\documentclass[onecolumn]{aastex61}

\hypersetup{linkcolor=red,citecolor=blue,filecolor=cyan,urlcolor=magenta}

\usepackage[all]{hypcap}
\usepackage{amsmath}
\usepackage{aas_macros}
\usepackage{graphics}
\usepackage{longtable} 
\usepackage{amsmath}
\usepackage{natbib}
\usepackage{tablefootnote}

\newcommand{\myemail}{adams@psi.edu}
\newcommand{\kepler}{\emph{Kepler}}
\newcommand{\ktwo}{\emph{K2}}
\newcommand{\tess}{\emph{TESS}}
\newcommand{\numusp}{74} 
\newcommand{\numcompanions}{30} 
\newcommand{\numdetected}{104} 
\newcommand{\newusp}{33} 
\newcommand{\newcompanions}{13} 
\newcommand{\newmultisystems}{4} 
\newcommand{\nummultisystems}{24} 
\newcommand{\multirate}{32} 
\newcommand{\numvalidusp}{21} 
\newcommand{\numvalidcomp}{15} 
\newcommand{\numvalid}{36} 
\newcommand{\newlyvalid}{10} 
\newcommand{\newlyvalidusp}{7} 
\newcommand{\newlyvalidcomp}{3} 
\newcommand{\numhighfpp}{17} 
\newcommand{\numfp}{5} 
\newcommand{\numstringent}{22} 
\newcommand{\numstringentmissed}{12} 
\newcommand{\lowsnr}{7} 
\newcommand{\marginalsnrusp}{14} 
\newcommand{\marginalsnrcomp}{6} 
\newcommand{\smallerthanearth}{32} 

\revised{\today}
\submitjournal{AAS}

\begin{document}


\title{Ultra Short Period Planets in \ktwo\ III: Neighbors are Common with \newcompanions\ New Multi-Planet Systems and \newlyvalid\ Newly Validated Planets in Campaigns 0-8, 10}


\author[0000-0002-9131-5969]{Elisabeth R. Adams}
\affil{Planetary Science Institute, 1700 E. Ft. Lowell, Suite 106, Tucson, AZ 85719, USA}
\email{\myemail}

\author[0000-0002-9495-9700]{Brian Jackson}
\affil{Department of Physics, Boise State University, 1910 University Drive, Boise ID 83725, USA}

\author{Samantha Johnson}
\affil{Department of Physics, University of North Carolina at Chapel Hill, Chapel Hill, North Carolina 27599, USA
Triangle Universities Nuclear Laboratory, Duke University, Durham, North Carolina 27708, USA}

\author{David R. Ciardi}
\affil{NASA Exoplanet Science Institute, 770 South Wilson Avenue, Pasadena, CA 91125, USA}

\author[0000-0001-9662-3496]{William D. Cochran}
\affil{Center for Planetary Systems Habitability and McDonald Observatory, The University of Texas at Austin, Austin, TX 78712, USA}

\author[0000-0002-7714-6310]{Michael Endl}
\affil{McDonald Observatory, The University of Texas at Austin, Austin, TX 78712, USA}

\author[0000-0002-0885-7215]{Mark E. Everett}
\affil{NSF's Optical Infrared Astronomy Research Laboratory, 950 North Cherry Avenue Tucson, AZ 85719, USA}
 
\author[0000-0001-9800-6248]{Elise Furlan}
\affil{NASA Exoplanet Science Institute, Caltech/IPAC, 770 South Wilson Avenue, Pasadena, CA 91125, USA}

\author[0000-0002-2532-2853]{Steve B. Howell}
\affil{NASA Ames Research Center, Moffett Field, CA 94035, USA}

\author{Prasanna Jayanthi}
\affil{Department of Physics, Boise State University, 1910 University Drive, Boise ID 83725, USA}

\author{Phillip J. MacQueen}
\affil{McDonald Observatory, The University of Texas at Austin, Austin, TX 78712, USA}

\author[0000-0001-7233-7508]{Rachel A. Matson}
\affil{U.S. Naval Observatory, 3450 Massachusetts Avenue NW, Washington, D.C. 20392, USA}

\author{Ciera Partyka-Worley}
\affil{Department of Physics, Boise State University, 1910 University Drive, Boise ID 83725, USA}

\author{Joshua Schlieder}
\affil{NASA Goddard Space Flight Center, Greenbelt, MD, USA}

\author[0000-0003-1038-9702]{Nicholas J. Scott}
\affil{NASA Ames Research Center, Moffett Field, CA 94035, USA}

\author{Sevio M. Stanton}
\affil{Department of Physics, Boise State University, 1910 University Drive, Boise ID 83725, USA}

\author{Carl Ziegler}
\affil{Dunlap Institute for Astronomy and Astrophysics, University of Toronto, 50 St. George Street, Toronto, Ontario M5S 3H4, Canada}


\begin{abstract}

Using the EVEREST photometry pipeline, we have identified \numusp\ candidate ultra-short-period planets (orbital period $P<1$ d) in the first half of the \ktwo\ data (Campaigns 0-8 and 10). Of these, \newusp\ candidates have not previously been reported. A systematic search for additional transiting planets found \newcompanions\ new multi-planet systems containing a USP, doubling the number known and representing a third (\multirate\%) of USPs in our sample from \ktwo. We also identified \numcompanions\ companions, which have periods from 1.4 to 31 days (median 5.5 d). A third (\numvalid\ of \numdetected) of the candidate USPs and companions have been statistically validated or confirmed in this work, \newlyvalid\ for the first time, including \newlyvalidusp\ USPs. Almost all candidates, and all validated planets, are small (radii $R_p \le 3~R_{\oplus}$) with a median radius of $R_p=1.1~R_{\oplus}$; the validated and confirmed USP candidates have radii between $0.4~R_{\oplus}$ and $2.4~R_{\oplus}$ and periods from $P=0.18$ to $0.96$ d. The lack of candidate (a) ultra-hot-Jupiters ($R_p > 10~R_{\oplus}$) and (b) short-period desert ($3\le R_p \le 10~R_{\oplus}$) planets suggests that both populations are rare, although our survey may have missed some of the very deepest transits. These results also provide strong evidence that we have not reached a lower limit on the distribution of planetary radius values for planets at close proximity to a star, and suggest that additional improvements in photometry techniques would yield yet more ultra-short-period planets. The large fraction of USPs in known multi-planet systems supports origins models that involve dynamical interactions with exterior planets coupled to tidal decay of the USP orbits.

\end{abstract}


\section{Background}
\label{sec:Introduction}

\begin{figure}
\includegraphics[width=0.9\textwidth]{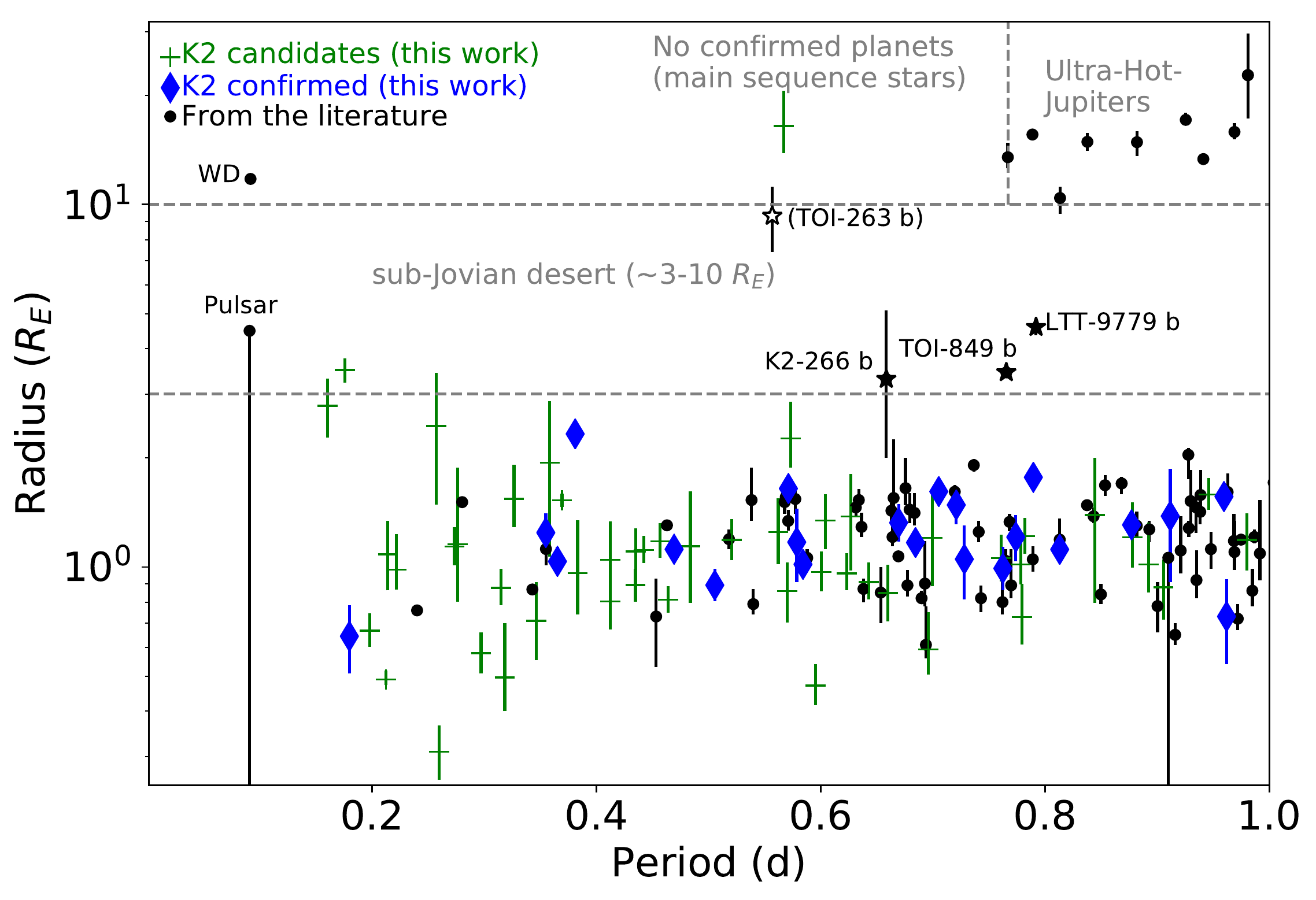}
\caption{Ultra-short-period planets ($P < 1\,$day). All confirmed (blue diamond) and candidate (green plus sign) USPs from this work for \ktwo\ C0-8 and 10 are shown, along with confirmed planets from the literature (black circles). The few known} candidate (open black star) and confirmed (closed black star) planets in the sub-Jovian desert (roughly $3 < R_{\rm p} < 10\, {\rm R_{\oplus}}$ and $P<1.6\,$d) are also shown. The only confirmed planets with $P<0.15$ d are around pulsars or white dwarfs. Data compiled from \url{exoplanetarchive.ipac.caltech.edu/}, \url{exoplanet.eu}, and this work.
\label{fig:period_radius}
\end{figure}

With about two hundred planets known with orbital periods less than a day (\autoref{fig:period_radius}), it seems that planets can occupy every available nook and cranny of their stellar systems. These ultra-short-period planets (USPs) lie at the edge of orbital stability and have defied easy explanation, although it seems unlikely that they formed where we find them \citep{2016CeMDA.126..227J}. Empirically, USPs around main sequence stars can be divided into three populations by size. (1) Small USPs ($\le$ 2-3 Earth radii, or ${\rm {\rm R_{\oplus}}}$) have been found with stable orbits down to orbital period $P=4.2\, {\rm hr}$ or $0.17 \, {\rm d}$ \citep[KOI-1843.03,][]{2013ApJ...773L..15R}. (2) Large USPs, or ultra-hot-Jupiters ($R_{\rm p} \ge 10\, {\rm {\rm R_{\oplus}}}$) have been confirmed with periods as short as $P=19\, {\rm hr}$ or $0.767\,{\rm d}$ \citep[NGTS-10 b,][]{2020MNRAS.493..126M}, near where Roche lobe overflow may begin \citep{2013ApJ...773L..15R}. (3) Very few intermediate-sized planets ($R_p=~3-10~R_{\oplus}$) have been confirmed, a gap known as the short-period or sub-Jovian desert that extends out to $P=2-3$ d \citep[e.g.,][]{2014ApJ...792....1L, 2016A&A...589A..75M}. Although the physical mechanisms that shape the desert are still uncertain, the location of the gap corresponds to the radius values expected for planets that are too large to be predominantly rocky. Just three confirmed planets are known in the inner desert ($P<1.5~$d): LTT 9779b ($R_{\rm p} = 4.59\pm0.23 {\rm R_{\oplus}}$, $M_{\rm p} = 29.32^{+0.78}_{-0.81} {\rm M_{\oplus}}$, $P = 0.79$ d,  \citealp{2019ESS.....410307J});  TOI-849 b ($R_p = 3.44_{-0.115}^{+0.157}\ {\rm R_{\oplus}}$, $M_p=40.8^{+2.4}_{-2.5} {\rm M_{\oplus}}$, $P = 0.7655$ d,  \citealp{2020Natur.583...39A}); and K2-266 b from \ktwo\ Campaign 14 ($R_{\rm p} = 3.3\pm1.8 {\rm R_{\oplus}}$, $P = 0.658$ d,  \citealp{2018AJ....156..245R}).

Of the three leading classes of theory to explain the existence of ultra-short-period planets -- in-situ formation, migration powered by tides raised on the host stars, or tidal migration coupled with multi-body interactions -- the first is unlikely due to the extreme temperatures at such short orbital periods \citep{1998AREPS..26...53B} and likely truncation of the protoplanetary disk much farther out \citep{2017ApJ...842...40L}. However, the roles of tidal migration in the planets' origins are the subjects of active research \citep[e.g.][]{2018ARA&A..56..175D}. Whatever its origins, once a planet occupies a short-period orbit, the host star's influence may remove it in a few billion years, with two chief mechanisms theorized. For volatile-rich planets, high atmospheric temperatures and intense stellar irradiation can lead to atmospheric loss \citep{2019AREPS..47...67O}. Strong tidal forces lead to orbital decay, driving the planet into unstable Roche lobe overflow, with complete disruption of the atmosphere occurring in just a few orbits \citep{2017MNRAS.465..149J}. 

The long-term stability of USPs is highly uncertain \citep{2015MNRAS.446.3676A}. \citet{2019AJ....158..190H} determined that the population of hot-Jupiter host stars is younger than either the population all field stars or the population of all stars known to host planets. This result suggests that close, giant planets quickly in-spiral or are otherwise destroyed, while their stars are still on the main sequence. Moreover, at least two small USPs around main-sequence stars have been discovered which appear to be actively disintegrating, as seen by distinctive features in their transit light curves \citep{2012ApJ...752....1R, 2015ApJ...812..112S}. The environment in which USPs find themselves is clearly perilous, but at the same time, hundreds of USPs -- which may orbit 0.1-1\% of all Sun-like stars \citep{SanchisOjeda2014} -- have been found and require explanation.

Studying ultra-short-period planets, both individually and as a population, thus presents unique opportunities to observe the end-state of planet evolution. Of particular use for testing theoretical models are the relative abundance of USPs of different masses and orbital periods, and the architecture (sizes, separations, and mutual inclinations) of multi-planet systems that contain a USP. Systems hosting USPs are also ideal for radial-velocity follow-up (assuming the star is bright enough) to determine masses, long-term timing measurements to watch for orbital decay \citep[e.g. WASP-12 b,][]{2017AJ....154....4P, 2019MNRAS.tmp.2246B}, and ultimately direct measurements of atmospheric escape using JWST or other next-generation telescopes.

Fortunately, USPs are a particularly advantageous group to observe. Their short orbital periods make it easy to observe multiple planetary orbits, allowing for a more rapid determination of precise planetary parameters. Even very small USPs are relatively easy to detect in space-based datasets such as those provided by \kepler, \ktwo, and \tess, which span several months to several years of continuous observations and therefore contain hundreds to thousands of individual transits. Some USPs have transit depths of less than 100 parts-per-million (ppm), and the typical depth is only a few hundred ppm, which means that stacking multiple transits together is often required for transit light curve fits. USPs also have larger radial velocity (RV) signatures than more distant planets of the same size, facilitating mass estimation. Kepler-78~b is one of the smallest exoplanets with a measured mass, about 1.7 ${\rm M_{\oplus}}$ \citep{2013Natur.503..381H, 2013Natur.503..377P}, and the planet's extremely short period, 8.5 hours, allowed its radial velocity (RV) signal to be disentangled from the star's longer-term activity cycle. Other measurements of planetary properties are easier at ultra-short periods: an 8 ${\rm M_{E}}$ \citep{2012ApJ...759...19E} member of a five-planet system, 55 Cnc e has $P=17$~hours and such a high brightness temperature ($\sim$1,800 K) that its eclipse (the occultation of the planet by the star) was detected in IR by the Spitzer Space Telescope \citep{2016Natur.532..207D}, the smallest planet for which that's been possible. \citet{2016MNRAS.455.2018D} reported that the planet's eclipse exhibits year-to-year IR variability, a result recently corroborated by \citet{2018AJ....155..221T} and potentially signaling active volcanism. Given their observational prominence, it is important to understand whether small USPs are representative of all small exoplanets or whether they constitute a unique species. They are critical tracers of theories of planet formation and evolution. 

In this study, we continue the work of the Short-Period Planet Group (SuPerPiG) collaboration \citep{2013ApJ...779..165J, 2016AJ....152...47A,2017AJ....153...82A} to discover, corroborate, and characterize USPs using data from the \ktwo\ Mission. We analyzed observations from Campaigns 0-8 and 10 of \ktwo\ (ignoring the microlensing campaign in C9) using the improved EVEREST 2.0 pipeline \citep{2016AJ....152..100L, 2018AJ....156...99L} and compared the results to different analyses using the k2sff pipeline \citep{2014PASP..126..948V} and the EVEREST 1.0 pipeline \citep{2019ApJS..244...11K}. In this work, using the EVEREST 2.0 photometry has resulted in twice as many USPs smaller than Earth as \citet{2018AJ....155..136M} (see Section \ref{sec:othersurveys}), and roughly six times as many sub-Earths and twice as many 1-2$~ R_{\oplus}$ planets as our previous USP survey \citep{2016AJ....152...47A} (see Section \ref{sec:survey_completeness}), which both used k2sff photometry. To estimate the false-positive probability (FPP) for each candidate, we use statistical validation by employing the \emph{vespa} model \citep{2012ApJ...761....6M}. We find that, of \numusp\ total USP candidates, \numvalidusp\ have FPPs $<1\%$, suggesting they are bonafide planets. An additional \numvalidcomp\ out of \numcompanions\ companion planets in the same systems were also validated. (Note: although many works use the 1\% validation threshold, including \cite{2015ApJ...809...25M, 2018AJ....155..127H, 2018AJ....156...78L}, a few use a more stringent 0.1\% cutoff, e.g. \cite{2018AJ....155..136M}; if we were to use the same threshold, only \numstringent\ of \numdetected\ candidates would have been validated by this work, and that would not include \numstringentmissed\ planets that have been reported confirmed or validated in other works.) At the other extreme, \numhighfpp\ candidates have FPP $>50\%$ and \numfp\ have FPP $>90\%$, suggesting that they are unlikely to be planets. The faintest object we validated has \kepler\ magnitude ${\rm Kep}=17.3$, and most of the host stars for our candidate planets are sufficiently bright for ground-based imaging to provide effective imaging constraints, while about half have magnitudes suitable for ground-based spectroscopy (${\rm Kep}<14$). 

The plan for this paper is as follows: in Section \ref{sec:search}, we detail our transit search procedure. Our systematic search for additional transiting companions is described in Section \ref{sec:multi_search}. In Section \ref{sec:obs}, we present our ground-based follow-up observations. In Section \ref{sec:validation}, we discuss our statistical validation results. We discuss individual candidates in Section \ref{sec:candidates}. In Section \ref{sec:survey_completeness}, we estimate our survey's completeness. Finally, in Section \ref{sec:context}, we explore the properties of the \ktwo\ population of ultra-short-period planets as a whole and examine the implications for planet formation theories, with a focus on the USP systems hosting multiple planets. 

\begin{longrotatetable}
\begin{deluxetable}{lccclllclrrrrcl}
\tablecaption{Candidate Host Star Parameters\label{table:stars}} 
\tabletypesize{\scriptsize}
\tablewidth{0pt}
\tablehead{
\colhead{Candidate} & \colhead{C.}&
\colhead{RA}& \colhead{Dec} & \colhead{$m_{Kep}$} & 
\colhead{$R_\star$ ($R_{\odot}$)} & \colhead{$T_{\rm eff}$} &
\colhead{[Fe/H]} & \colhead{log(g)} & \colhead{$u_1$\tablenotemark{a}} &
\colhead{$u_2$} & \colhead{$u_3$} & \colhead{$u_4$} & 
\colhead{Name} & \colhead{Source\tablenotemark{b}}
}
\startdata
$201085153$	&C10	&$180.122852$	&$-7.012986$	&$17.33$	&$0.38\pm0.15$	&$3945\pm386$	&$-0.2\pm0.35$	&$4.9\pm0.2$	&$0.66$	&$0.394$	&$-0.364$	&$0.137$	&	&E\\
$201089381$	&C10	&$181.393564$	&$-6.894076$	&$14.28$	&$0.77\pm0.14$	&$5709\pm170$	&$-0.8\pm0.42$	&$4.5\pm0.1$	&$0.202$	&$1.06$	&$-0.798$	&$0.224$	&	&E\\
$201110617$	&C10	&$182.700187$	&$-6.294205$	&$12.95$	&$0.62\pm0.02$	&$4460\pm50$	&$-0.33\pm0.08$	&$4.7\pm0.1$	&$0.62$	&$-0.489$	&$1.13$	&$-0.435$	&K2-156	&M18\\
$201130233$	&C10	&$183.751337$	&$-5.781991$	&$12.6$	&$0.88\pm0.04$	&$5456\pm50$	&$0.13\pm0.08$	&$4.5\pm0.1$	&$0.612$	&$-0.325$	&$0.944$	&$-0.446$	&K2-157	&M18\\
$201214691$	&C1	&$174.871463$	&$-3.755974$	&$14.23$	&$0.78\pm0.05$	&$5122\pm82$	&$-0.049\pm0.25$	&$4.5\pm0.04$	&$0.612$	&$-0.453$	&$1.16$	&$-0.519$	&	&E,B\\
$201230694$	&C1	&$171.09001$	&$-3.501374$	&$15.18$	&$0.48\pm0.04$	&$4086\pm80$	&$-0.1\pm0.18$	&$4.8\pm0.05$	&$0.614$	&$0.18$	&$0.033$	&$0.004$	&	&E\\
$201231940$	&C1	&$174.120612$	&$-3.482449$	&$18.11$	&$0.27\pm0.02$	&$3927\pm39$	&$-0.56\pm0.05$	&$5\pm0.1$	&$0.881$	&$0.123$	&$-0.324$	&$0.147$	&	&E\\
$201239401$	&C1	&$173.052956$	&$-3.370242$	&$15.04$	&$0.33\pm0.06$	&$3782\pm90$	&$-0.13\pm0.2$	&$5\pm0.02$	&$0.551$	&$0.731$	&$-0.694$	&$0.226$	&	&E\\
$201264302$	&C1	&$169.598013$	&$-2.991577$	&$13.88$	&$0.19\pm0$	&$3875\pm157$	&$-0.8\pm0.2$	&$4.2\pm0.2$	&$0.791$	&$0.147$	&$-0.226$	&$0.127$	&	&M\\
$201427007$	&C10	&$184.602205$	&$-0.551349$	&$14.22$	&$0.94\pm0.09$	&$5633\pm111$	&$-0.053\pm0.24$	&$4.4\pm0.07$	&$0.517$	&$0.007$	&$0.532$	&$-0.299$	&	&E\\
$201438507$	&C10	&$182.714439$	&$-0.383279$	&$14.07$	&$0.89\pm0.1$	&$5532\pm113$	&$-0.13\pm0.32$	&$4.5\pm0.07$	&$0.514$	&$-0.024$	&$0.597$	&$-0.328$	&	&E\\
$201461352$	&C1	&$175.686546$	&$-0.0553$	&$17.15$	&$0.15\pm0$	&$3092\pm37$	&$-0.036\pm0.14$	&$5.2\pm0.02$	&$0.432$	&$0.793$	&$-0.518$	&$0.108$	&	&E\\
$201469738$	&C1	&$177.738562$	&$0.068657$	&$15.82$	&$0.48\pm0.05$	&$4050\pm166$	&$-0.011\pm0.2$	&$4.8\pm0.04$	&$0.587$	&$0.21$	&$0.041$	&$-0.008$	&	&E\\
$201549425$	&C1	&$176.938421$	&$1.278706$	&$15.47$	&$0.5\pm0.08$	&$4415\pm178$	&$-0.55\pm0.3$	&$4.8\pm0.1$	&$0.591$	&$-0.203$	&$0.693$	&$-0.27$	&	&E\\
$201595106$	&C10	&$183.720347$	&$1.96828$	&$11.68$	&$0.93\pm0.06$	&$5820\pm20$	&$-0.01\pm0.03$	&$4.6\pm0.1$	&$0.493$	&$0.135$	&$0.341$	&$-0.226$	&	&M,M18\\
$201609593$	&C1	&$173.066309$	&$2.192621$	&$14.71$	&$0.29\pm0.01$	&$3630\pm36$	&$0.007\pm0.05$	&$5\pm0.04$	&$0.465$	&$0.758$	&$-0.542$	&$0.131$	&	&E\\
$201637175$	&C1	&$169.482818$	&$2.61907$	&$14.93$	&$0.58\pm0.03$	&$3879\pm90$	&$0.032\pm0.12$	&$4.7\pm0.04$	&$0.553$	&$0.285$	&$0.021$	&$-0.030$	&K2-22A	&Dr17\\
$201650711$	&C1	&$172.044052$	&$2.826891$	&$12.25$	&$0.34\pm0.05$	&$4310\pm92$	&$-0.82\pm0.05$	&$4.2\pm0.3$	&$0.581$	&$-0.272$	&$0.826$	&$-0.323$	&	&M\\
$201687804$	&C1	&$175.012744$	&$3.434154$	&$13.13$	&$0.63\pm0.06$	&$4483\pm184$	&$-0.01\pm0.09$	&$4.7\pm0.04$	&$0.674$	&$-0.682$	&$1.34$	&$-0.491$	&	&E\\
$201717159$	&C1	&$178.436482$	&$3.932202$	&$15.98$	&$0.46\pm0.07$	&$4092\pm198$	&$-0.15\pm0.2$	&$4.8\pm0.06$	&$0.635$	&$0.164$	&$0.021$	&$0.0108$	&	&E\\
$201796374$	&C1	&$171.527163$	&$5.301112$	&$16.37$	&$0.26\pm0.02$	&$3587\pm71$	&$-0.057\pm0.08$	&$5\pm0.02$	&$0.46$	&$0.793$	&$-0.588$	&$0.148$	&	&E\\
$201923289$	&C1	&$174.920526$	&$7.825885$	&$13.69$	&$0.85\pm0.07$	&$5350\pm127$	&$-0.021\pm0.15$	&$4.5\pm0.03$	&$0.591$	&$-0.315$	&$0.968$	&$-0.46$	&	&E\\
$203533312$	&C2	&$243.955378$	&$-25.818471$	&$12.16$	&$1.3\pm0.1$	&$6620\pm86$	&$0.09\pm0.05$	&$4.2\pm0.1$	&$0.412$	&$0.691$	&$-0.591$	&$0.186$	&	&M,I20\\
$205002291$	&C2	&$247.749587$	&$-19.875631$	&$14.27$	&$0.31\pm0.06$	&$3685\pm108$	&$-0.002\pm0.1$	&$5\pm0.04$	&$0.482$	&$0.744$	&$-0.559$	&$0.145$	&	&E\\
$205152172$	&C2	&$245.180008$	&$-19.14414$	&$13.49$	&$0.53\pm0.09$	&$4202\pm242$	&$-0.04\pm0.2$	&$4.8\pm0.04$	&$0.613$	&$-0.11$	&$0.485$	&$-0.149$	&	&E\\
$206024342$	&C3	&$331.277219$	&$-14.121665$	&$13.05$	&$1.1\pm0.36$	&$5724\pm72$	&$-0.22\pm0.06$	&$4.5\pm0.1$	&$0.431$	&$0.291$	&$0.181$	&$-0.168$	&K2-299	&M\\
$206042996$	&C3	&$337.513575$	&$-13.610113$	&$16.06$	&$0.5\pm0.04$	&$4114\pm99$	&$0.006\pm0.12$	&$4.8\pm0.05$	&$0.592$	&$0.059$	&$0.264$	&$-0.081$	&K2-301	&E\\
$206103150$	&C3	&$331.203044$	&$-12.018893$	&$11.76$	&$1.1\pm0.04$	&$5565\pm60$	&$0.43\pm0.04$	&$4.3\pm0.07$	&$0.68$	&$-0.457$	&$1.06$	&$-0.479$	&WASP-47	&S15,M13\\
$206151047$	&C3	&$333.71501$	&$-10.769168$	&$13.43$	&$1.2\pm0.55$	&$5971\pm67$	&$-0.25\pm0.06$	&$4.6\pm0.09$	&$0.367$	&$0.576$	&$-0.24$	&$0.006$	&	&M\\
$206215704$	&C3	&$335.094822$	&$-9.509531$	&$15.6$	&$0.25\pm0.01$	&$3297\pm73$	&$-0.33\pm0.093$	&$4.9\pm0.08$	&$0.487$	&$0.846$	&$-0.718$	&$0.203$	&K2-302	&D19\\
$206298289$	&C3	&$337.357688$	&$-8.266702$	&$14.69$	&$0.33\pm0.03$	&$3875\pm157$	&$-1\pm0.1$	&$4.4\pm0.2$	&$0.92$	&$0.098$	&$-0.369$	&$0.189$	&	&M\\
$206417197$	&C3	&$337.133466$	&$-6.347505$	&$13.35$	&$0.75\pm0.04$	&$5245\pm55$	&$-0.15\pm0.05$	&$4.6\pm0.07$	&$0.566$	&$-0.29$	&$0.975$	&$-0.468$	&	&M\\
$210605073$	&C4	&$65.62329$	&$17.037875$	&$17.89$	&$1.6_{-0.24}^{+0.42}$	&$7918\pm500$	&$0\pm0$	&$0\pm0$	&$0.417$	&$0.766$	&$-0.825$	&$0.291$	&	&Inf.,I\\
$210707130$	&C4	&$59.464394$	&$18.465254$	&$12.1$	&$0.68\pm0.03$	&$4427\pm48$	&$-0.36\pm0.05$	&$4.3\pm0.1$	&$0.646$	&$-0.658$	&$1.37$	&$-0.525$	&K2-85	&M,Dr17\\
$210801536$	&C4	&$52.818936$	&$19.891303$	&$13.18$	&$0.85_{-0.09}^{+0.28}$	&$5304\pm259$	&$-0.22\pm0.25$	&$3.8\pm0.8$	&$0.523$	&$-0.118$	&$0.752$	&$-0.385$	&	&E,I\\
$210848775$	&C4	&$55.615947$	&$20.609693$	&$16.08$	&$0.44\pm0.08$	&$4054\pm217$	&$-0.19\pm0.2$	&$4.8\pm0.06$	&$0.666$	&$0.223$	&$-0.119$	&$0.061$	&	&E\\
$210931967$	&C4	&$59.771022$	&$21.895214$	&$15.44$	&$0.26\pm0.05$	&$3625\pm197$	&$-0.19\pm0.28$	&$5\pm0.07$	&$0.53$	&$0.772$	&$-0.694$	&$0.209$	&	&E\\
$211305568$	&C5	&$133.122327$	&$9.962734$	&$13.85$	&$0.45\pm0.03$	&$3612\pm84$	&$-0.18\pm0.09$	&$4.8\pm0.06$	&$0.524$	&$0.699$	&$-0.551$	&$0.148$	&	&Dr17\\
$211336288$	&C5	&$133.423618$	&$10.564473$	&$14.61$	&$0.59\pm0.03$	&$3996\pm79$	&$-0.075\pm0.084$	&$4.7\pm0.03$	&$0.591$	&$0.266$	&$-0.056$	&$0.030$	&	&Dr17\\
$211357309$	&C5	&$133.23263$	&$10.944721$	&$13.15$	&$0.46\pm0.03$	&$4240\pm223$	&$-1\pm0$	&$4.6\pm0.1$	&$0.674$	&$0.074$	&$0.042$	&$0.014$	&	&M,Dr17\\
$211562654$	&C5	&$125.00716$	&$14.019461$	&$12.75$	&$0.87\pm0.04$	&$5482\pm50$	&$0.09\pm0.08$	&$4.6\pm0.1$	&$0.597$	&$-0.275$	&$0.889$	&$-0.429$	&K2-183	&M18\\
$211566738$	&C5	&$128.32327$	&$14.078237$	&$14.93$	&$0.47\pm0.05$	&$4091\pm155$	&$-0.009\pm0.16$	&$4.8\pm0.04$	&$0.59$	&$0.144$	&$0.135$	&$-0.037$	&	&E\\
$211828293$	&C5	&$123.468508$	&$17.718615$	&$14.71$	&$0.3\pm0.05$	&$3647\pm87$	&$0.017\pm0.1$	&$5\pm0.06$	&$0.472$	&$0.725$	&$-0.501$	&$0.118$	&	&E\\
$212092746$	&C5	&$127.644804$	&$21.881783$	&$16.16$	&$0.3\pm0.04$	&$3669\pm114$	&$0.013\pm0.15$	&$5\pm0.03$	&$0.477$	&$0.733$	&$-0.527$	&$0.13$	&	&E\\
$212157262$	&C5	&$132.523609$	&$23.192601$	&$12.86$	&$0.83\pm0.11$	&$5438\pm63$	&$0\pm0.1$	&$4.6\pm0.08$	&$0.584$	&$-0.268$	&$0.901$	&$-0.438$	&K2-187	&M\\
$212303338$	&C6	&$202.9588318$	&$-17.4307041$	&$9.96$	&$0.8\pm0.05$	&$5148\pm155$	&$0.005\pm0.1$	&$4.6\pm0.03$	&$0.626$	&$-0.487$	&$1.19$	&$-0.528$	&	&E\\
$212425103$	&C6	&$204.3695831$	&$-14.3488894$	&$14.49$	&$0.83\pm0.06$	&$5490\pm133$	&$-0.19\pm0.25$	&$4.5\pm0.06$	&$0.507$	&$-0.029$	&$0.622$	&$-0.342$	&	&E,K19\\
$212443973$	&C6	&$205.0088959$	&$-13.9320383$	&$14.95$	&$0.34\pm0.03$	&$3423\pm84$	&$0.2\pm0.084$	&$4.9\pm0.1$	&$0.496$	&$0.415$	&$-0.004$	&$-0.081$	&	&Dr17\\
$212470904$	&C6	&$205.8204651$	&$-13.3571358$	&$13.02$	&$0.72\pm0.04$	&$4820\pm124$	&$0.18\pm0.08$	&$4.6\pm0.2$	&$0.685$	&$-0.71$	&$1.41$	&$-0.557$	&	&M\\
$212532636$	&C6	&$205.7254181$	&$-12.0385857$	&$14.05$	&$0.59\pm0.04$	&$4485\pm73$	&$-0.3\pm0.15$	&$4.7\pm0.04$	&$0.62$	&$-0.531$	&$1.2$	&$-0.465$	&	&E\\
$212610063$	&C6	&$207.1204529$	&$-10.3622961$	&$14.36$	&$0.77\pm0.09$	&$5146\pm185$	&$-0.19\pm0.3$	&$4.6\pm0.06$	&$0.575$	&$-0.358$	&$1.07$	&$-0.498$	&	&E\\
$212624936$	&C6	&$210.6377716$	&$-10.0237617$	&$13.87$	&$0.93\pm0.2$	&$5765\pm170$	&$-0.39\pm0.25$	&$4.4\pm0.07$	&$0.347$	&$0.576$	&$-0.186$	&$-0.023$	&	&E\\
$213715787$	&C7	&$293.8331909$	&$-28.4978333$	&$13.36$	&$0.31\pm0.08$	&$3775\pm186$	&$-0.1\pm0.15$	&$5\pm0.07$	&$0.528$	&$0.76$	&$-0.698$	&$0.222$	&K2-147	&E\\
$214189767$	&C7	&$285.0127869$	&$-27.1352367$	&$15.42$	&$0.53\pm0.04$	&$4148\pm123$	&$0.043\pm0.16$	&$4.8\pm0.03$	&$0.595$	&$-0.051$	&$0.432$	&$-0.138$	&	&E\\
$214539781$	&C7	&$287.9277344$	&$-26.2420712$	&$14.63$	&$0.41\pm0.04$	&$3856\pm122$	&$0.071\pm0.12$	&$4.9\pm0.03$	&$0.531$	&$0.431$	&$-0.166$	&$0.024$	&	&E\\
$219752140$	&C7	&$288.7794495$	&$-15.9167862$	&$12.55$	&$1_{-0.12}^{+0.24}$	&$5708\pm111$	&$-0.068\pm0.15$	&$4.2\pm0.2$	&$0.479$	&$0.161$	&$0.322$	&$-0.215$	&	&E,I\\
$220250254$	&C8	&$17.6409359$	&$1.578266$	&$11.51$	&$0.89\pm0.15$	&$5500\pm45$	&$0\pm0.02$	&$4.6\pm0.07$	&$0.573$	&$-0.217$	&$0.834$	&$-0.416$	&K2-210	&M\\
$220256496$	&C8	&$21.1059532$	&$1.704954$	&$12.87$	&$0.79\pm0.08$	&$5360\pm51$	&$-0.04\pm0.03$	&$4.7\pm0.06$	&$0.579$	&$-0.281$	&$0.933$	&$-0.449$	&K2-211	&M\\
$220272424$	&C8	&$10.5750151$	&$2.048125$	&$13.84$	&$0.52\pm0.07$	&$4183\pm128$	&$-0.044\pm0.15$	&$4.8\pm0.05$	&$0.61$	&$-0.057$	&$0.408$	&$-0.124$	&	&E\\
$220383386$	&C8	&$8.7396841$	&$4.3814678$	&$8.95$	&$0.83\pm0.03$	&$5438\pm63$	&$-0.1\pm0.1$	&$4.6\pm0.1$	&$0.542$	&$-0.137$	&$0.748$	&$-0.383$	&HD 3167	&M,V16\\
$220440058$	&C8	&$15.725172$	&$5.4905128$	&$16.27$	&$0.28\pm0.08$	&$3682\pm217$	&$-0.18\pm0.25$	&$5\pm0.07$	&$0.544$	&$0.774$	&$-0.734$	&$0.232$	&	&E\\
$220492298$	&C8	&$16.7614613$	&$6.588829$	&$13.79$	&$0.89\pm0.09$	&$5620\pm31$	&$0.28\pm0.03$	&$4.5\pm0.08$	&$0.629$	&$-0.309$	&$0.887$	&$-0.422$	&	&M\\
$220554126$	&C8	&$15.5148172$	&$7.9936891$	&$15.79$	&$0.45\pm0.06$	&$4073\pm201$	&$-0.19\pm0.2$	&$4.8\pm0.06$	&$0.658$	&$0.186$	&$-0.049$	&$0.037$	&	&E\\
$220554210$	&C8	&$13.4320173$	&$7.995316$	&$13.72$	&$1\pm0.04$	&$5499\pm109$	&$-0.23\pm0.16$	&$4.5\pm0.04$	&$0.49$	&$0.025$	&$0.561$	&$-0.321$	&K2-282	&H16,K19\\
$220674823$	&C8	&$13.0797796$	&$10.7946987$	&$11.96$	&$0.95\pm0.05$	&$5613\pm39$	&$0.01\pm0.01$	&$4.6\pm0.07$	&$0.549$	&$-0.104$	&$0.673$	&$-0.354$	&K2-106	&M,S17\\
$220687583$	&C8	&$16.4097919$	&$11.1298952$	&$13.64$	&$0.81\pm0.04$	&$5259\pm148$	&$-0.15\pm0.16$	&$4.5\pm0.04$	&$0.565$	&$-0.278$	&$0.956$	&$-0.46$	&	&E,I\\
$228721452$	&C10	&$185.306152$	&$-10.282037$	&$11.32$	&$1\pm0.06$	&$5859\pm50$	&$0.19\pm0.08$	&$4.5\pm0.1$	&$0.55$	&$-0.001$	&$0.48$	&$-0.272$	&K2-223	&M18\\
$228732031$	&C10	&$182.751564$	&$-9.765218$	&$11.94$	&$0.81\pm0.03$	&$5200\pm100$	&$-0.02\pm0.08$	&$4.6\pm0.1$	&$0.612$	&$-0.43$	&$1.13$	&$-0.511$	&K2-131	&Da17\\
$228739208$	&C10	&$188.441461$	&$-9.419237$	&$12.11$	&$1\pm0.26$	&$5791\pm139$	&$-0.21\pm0.32$	&$4.4\pm0.2$	&$0.413$	&$0.372$	&$0.062$	&$-0.118$	&	&E\\
$228801451$	&C10	&$186.873264$	&$-6.721861$	&$10.96$	&$0.79\pm0.01$	&$5315\pm35$	&$-0.09\pm0.02$	&$4.6\pm0.01$	&$0.568$	&$-0.26$	&$0.916$	&$-0.443$	&K2-229	&L18\\
$228813918$	&C10	&$186.870725$	&$-6.195224$	&$14.53$	&$0.29\pm0.06$	&$3697\pm109$	&$-0.059\pm0.18$	&$5\pm0.06$	&$0.493$	&$0.788$	&$-0.66$	&$0.191$	&K2-137	&H16\\
$228814754$	&C10	&$186.267537$	&$-6.162735$	&$14.57$	&$0.21\pm0.09$	&$3521\pm467$	&$-0.22\pm0.6$	&$5.1\pm0.1$	&$0.503$	&$0.781$	&$-0.64$	&$0.174$	&	&E\\
$228836835$	&C10	&$189.575383$	&$-5.244647$	&$14.93$	&$0.32\pm0.07$	&$3782\pm152$	&$-0.18\pm0.25$	&$5\pm0.07$	&$0.576$	&$0.714$	&$-0.712$	&$0.238$	&	&E\\
$228877886$	&C10	&$186.533953$	&$-3.65254$	&$14.37$	&$0.73\pm0.04$	&$4837\pm117$	&$-0.18\pm0.35$	&$4.6\pm2$	&$0.619$	&$-0.578$	&$1.34$	&$-0.566$	&	&E,I\\
\enddata
\tablenotetext{a}{$u_1$, $u_2$, $u_3$, $u_4$: Limb-darkening parameters from \citet{2011AA...529A..75C}}       
\tablenotetext{b}{Source of stellar parameter values: E = used EPIC parameters; M = $T_{eff}$, $log(g)$, and [Fe/H] from McDonald observations; I = $R_\star$ calculated using isochrones from \citep{2008ApJS..178...89D}; B = $R_\star$ calculated using relations from \citet{Boyajian2012}; M13 = \citet{2013AA...558A.106M}; B15 = \citet{2015ApJ...812L..18B}; H16 = \citet{2016ApJS..224....2H}; V16 = \citet{2016ApJ...829L...9V}; Da17 = \citet{2017AJ....154..226D}; Dr17 = \citet{2017AJ....154..207D}; S17 = \citet{2017AJ....153..271S}; M18 = \citet{2018AJ....155..136M}; Dr19 = \citet{2019AJ....158...87D}; K19 = \citet{2019ApJS..244...11K}; P = inferred from photometry using \citet{2012ApJS..199...30P}; see Section \ref{sec:uhj_cand}.}
\end{deluxetable}
\end{longrotatetable}

\section{Photometry and transit search}
\label{sec:search}

The candidate search method used in this paper is similar to prior SuPerPiG papers \citep{2016AJ....152...47A, 2017AJ....153...82A}, with one key difference: the choice of photometry pipeline. As a test of the precision of the EVEREST pipeline, we searched both k2sff and EVEREST 2.0 photometry, using data from Campaigns C0-8 and 10. EVEREST 2.0 claims 10-20\% better photometric precision than EVEREST 1.0, which in turn claimed to be 25\% better than k2sff -- and for the mostly small USP population, these differences are highly significant in terms of the number of planets detected. 

The discovery method employed is as follows: we retrieved the EVEREST 2.0 (hereafter referred to as EVEREST) photometry \citep{2016AJ....152..100L, 2018AJ....156...99L} from MAST\footnote{\url{http://archive.stsci.edu/k2/hlsp/everest/search.php}} and the k2sff photometry \citep{2014PASP..126..948V} from MAST\footnote{\url{http://archive.stsci.edu/k2/hlsp/k2sff/search.php}} for Campaigns 0-8 and 10. For each of the 10,000-30,000 light curves per campaign, we subtracted a median boxcar filter with a window size of 1 day. A window size of a day is well-suited to USPs and other short period planets, with transit durations of 1-3 hours. It might not be appropriate for long period planets where the transit duration could be an appreciable fraction of a day, and could result in lower yields, particularly of shallow transits. The performance of our pipeline at periods beyond a few days is left for future research.

We estimated the overall scatter, $\sigma$, using 1.4826 times the median absolute deviation and masked out individual points that lay $\ge10 \sigma$ from the campaign median value. The choice to cut points more than 10$\sigma$ out of transit was made to facilitate small planet discovery, and had the desired side effect of eliminating many eclipsing binary blends before the vetting process, whose inclusion would have increased the workload on the survey by a factor of several. However, we may also have inadvertently eliminated some of the deepest ultra-hot Jupiter candidates. Many deep transits were nonetheless identified as periodic candidates, since a substantial fraction of the points in transit during ingress and egress are within 10$\sigma$ of the baseline, which causes the light curve to show up with the bottom removed. In such cases, we re-analyzed the light curve with a much higher noise clipping (50$\sigma$) so that we could fit the transit (e.g., the 4.5 d companion EPIC 206103150/HD 3167 b). Almost all those candidates were then subsequently eliminated upon finding clear signs of phase variability or odd/even transit depth differences that are associated with eclipsing binaries. The largest viable candidate in this work has a depth of 6400 ppm (EPIC 210605073).

We next searched for transit signals with orbital periods between 3 and 72 hours using the EEBLS algorithm \citep{2002A&A...391..369K}. Although the goal of this work is to discover planets with $P \leq 24$ hr, we gave a first examination to candidates at longer periods to identify potential  aliased signals, which are particularly common at twice the true orbital period. Issues with aliasing USPs to longer periods has been a problem since the identification of the very first USP, 55 Cnc e \citep{2010ApJ...722..937D}.

To maximize the yield of USPs, which are intrinsically uncommon \citep[e.g.,][]{SanchisOjeda2014}, and since many candidates are easily recovered at a lower signal-to-noise threshold, we relaxed the discovery threshold to $\ge7 \sigma$, as opposed to $\ge10 \sigma$ in e.g. \citet{2017AJ....153...82A}. We found \numdetected\ candidates with transit-like signals and $P<1$ d that met that transit depth threshold in at least one pipeline, and subjected each one to further vetting by hand by some of the co-authors of this paper, with each candidate examined by at least two people. During this vetting step, transits were binned at low multiples of the nominal orbital period to search for masquerading binary stars, and the transit depth, duration, and odd-even differences were also examined for discrepancies.

For the surviving candidates, we fit the transit light curve using Markov chain Monte Carlo (MCMC) methods. In the first iteration of this search using the k2sff pipeline, we used the algorithm from \citet{2002ApJ...580L.171M} as implemented by the Python packages Batman \citep{2015PASP..127.1161K} and PyLightCurve (\url{https://github.com/ucl-exoplanets/pylightcurve}); in those fits, which are retained in this publication only for comparison to the EVEREST results, we used 100,000 links in each chain, thinning the sample by a factor of 10 and then discarding as burn-in the first 1000 iterations. For the bulk of our analysis using EVEREST, we fit the light curves using the PyLightcurve package, which is based on Emcee, an affine-invariant ensemble sampler \citep{2013PASP..125..306F}. We used 600,000 iterations, 1500 walkers, and discarded the first half (300,000 iterations) as burn-in. (No significant difference was seen in the fit results between the two methods; the switch was made for reasons of software compatibility and run-time improvements.)

We fixed the eccentricity for all fits at zero, since the orbits for all USPs are expected to be tidally circularized \citep{2008ApJ...678.1396J}; whether any of the more distant companions may have non-zero eccentricity is left for future work. The limb-darkening coefficients were calculated using the stellar parameters in \autoref{table:stars} using coefficients from \citet{2011AA...529A..75C}. During fitting, the inclination was allowed to range freely above 90 degrees, since transit photometry cannot distinguish between prograde and retrograde orbits. For a few companions with low signal-to-noise ratios (SNR), the quality of the light curve was poor enough that we fixed the inclination (to $i=90^{\circ}$) and $a/R_*$ (to a reasonable guess scaled from the periods of both planets and $a/R_*$ values of the USP) and fit only for $R_P/R_*$ and $T_{mid}$; those fits are noted in \autoref{table:planets}. The best-fit model parameters in \autoref{table:planets} are calculated using a confidence interval of 68\% of the posterior values, where the value reported is at 50\% of the distribution. For parameters that are derived from the fitted parameters, namely the radius in $R_{\oplus}$ (which depends on $R_{\odot}$) and the transit duration (which depends on $a/R_*$, $i$, and $R_p/R_*$), the errors are calculated by sampling from the appropriate chains for the dependent variables. For the planet radius, two half-Gaussian distributions were constructed around the asymmetric error on the stellar radius; we generated 150,000 points for both half-distributions and randomly assigned one point to each link in the radius ratio chain. For the transit duration, a random value from each of the inclination, semimajor axis, and radius ratio chains was drawn to create a duration chain of equal length, using the fixed value for the few cases where those values were not fitted. 

To check convergence, we use the autocorrelation statistic test in \emph{Emcee}, checking that the correlation length is always much less than the length of the chain, and found values between 1 and 15, indicating that the 300,000-link chains were likely well-converged. We also visually inspected the traces for each parameter and the correlations between parameters (the so-called triangle plot).

We examined additional information on each system, including the EVEREST data validation (DV) reports (which include photometry and images of the aperture) and stellar observations that had either been reported to the ExoFOP\footnote{\url{https://exofop.ipac.caltech.edu/}} or were made for the purpose of this work (see Section \ref{sec:spectra}). If more than one transiting candidate was detected in the system, we reported fits for each planet candidate (see Section \ref{sec:multi_search}).

Although our final results are based on the EVEREST light curves, we found it useful to compare those results to the fits to the k2sff curves, since they were derived from an independent pipeline with different aperture sizes and calibration choices; background blends are likely to be blended by different amounts, while the parameters for true planets should remain constant. Out of 93 signals originally flagged for follow-up, thirty-five were only detected in EVEREST, while five were only detected in k2sff. For the majority of signals which were detected in both pipelines, we compared the planetary properties and flagged any significant discrepancies. For example, if the fitted depths disagreed by more than 50\%, that was considered a clear sign of an eclipsing binary that has a different blended fraction in each pipeline. When the period differed by more than a few minutes, it was typically a sign of a spurious detection picking up on noise or other non-transiting features. Of the five signals detected only in the k2sff pipeline, all were clearly tagged as false positives during the initial vetting due to features in the folded EVEREST light curves; among them was EPIC 211995325, previously reported as a planet using k2sff data by \citet{2016AJ....152...47A}. Meanwhile, most of the EVEREST-only signals (30 of 35) remained viable candidates after vetting, and are therefore more likely to be real planets that were just below the k2sff detectability threshold. We conclude that the EVEREST photometry is preferable for small planet searches, and we use it in all analyses and results presented in this paper going forward, with one exception noted in Section \ref{everest_issue}. (However, we note that eight of the EVEREST-only candidates have high false-positive probabilities, as discussed in Section \ref{sec:highfpp}.)


\begin{longrotatetable}
\begin{deluxetable}{llllllllllccc}
\tablecaption{Transit Model Fit Results\label{table:planets}} 
\tabletypesize{\scriptsize}
\tablewidth{0pt}
\tablehead{
\colhead{EPIC}& \colhead{Pl.\tablenotemark{a}} & \colhead{Period (d)} &
\colhead{$T_0$ (HJD-2450000)} & \colhead{$R_p/R_*$} &
\colhead{$R_{\rm p} (R_{\oplus})$} & \colhead{$a/R_*$} &
\colhead{$i(^\circ)$} & \colhead{Ind. FPP} & \colhead{FPP} &
\colhead{SNR} & \colhead{Status \tablenotemark{b}} &
\colhead{Prior\tablenotemark{c}}
}
\startdata
$201085153$	&p1	&$0.25722\pm0.000029$	&$7627.05811_{-0.00083}^{+0.00094}$	&$0.0584_{-0.0024}^{+0.004}$	&$2.5\pm1$	&$2.4_{-0.4}^{+0.3}$	&$90.7_{-10}^{+20}$	&$0.11$	&$0.11$	&26	&	&\\
$201085153$	&p2	&$2.259796\pm0.00026$	&$7642.94739_{-0.017}^{+0.011}$	&$0.0341_{-0.0055}^{+0.0041}$	&$1.3\pm0.6$	&$9$	&$90$	&$0.13$	&$0.0087$	&4.1	&	&\\
$201089381$	&p1	&$0.160419\pm0.0000083$	&$7604.12593\pm0.00027$	&$0.033_{-0.00077}^{+0.0013}$	&$2.8\pm0.5$	&$2.8_{-0.3}^{+0.1}$	&$91.7_{-9}^{+10}$	&NA\tablenotemark{d}	&--	&40	&	&\\
$201110617$	&p1	&$0.813143\pm0.000077$	&$7641.67248_{-0.0012}^{+0.0014}$	&$0.0162_{-0.0006}^{+0.0013}$	&$1.1_{-0.06}^{+0.09}$	&$4.4_{-0.8}^{+0.4}$	&$89.8_{-7}^{+8}$	&1.6e-8	&1.6e-8	&26	&Conf.	&Dr17\\
$201130233$	&p1	&$0.365261\pm0.00017$	&$7638.7224_{-0.0019}^{+0.0021}$	&$0.0107_{-0.00063}^{+0.0011}$	&$1_{-0.08}^{+0.1}$	&$2.2\pm0.4$	&$89.7\pm20$	&$0.0046$	&$0.0046$	&15	&Conf.	&M18\\
$201214691$	&p1	&$0.600446\pm0.000068$	&$6880.68578_{-0.0029}^{+0.0034}$	&$0.0112_{-0.00098}^{+0.0016}$	&$0.97\pm0.1$	&$3.3_{-0.7}^{+0.9}$	&$91.1\pm10$	&$0.035$	&$0.035$	&9.5	&	&\\
$201230694$	&p1	&$0.273413\pm0.000061$	&$6853.61225_{-0.0016}^{+0.0018}$	&$0.0212_{-0.0012}^{+0.0024}$	&$1.1\pm0.1$	&$2.2_{-0.4}^{+0.6}$	&$91.5\pm20$	&$0.067$	&$0.067$	&16	&	&\\
$201231940$	&p1	&$0.57309\pm0.000045$	&$6862.0756_{-0.0014}^{+0.0016}$	&$0.0691_{-0.0062}^{+0.025}$	&$2.3_{-0.4}^{+0.6}$	&$5.8_{-2}^{+7}$	&$90.2_{-5}^{+6}$	&$0.76$	&$0.76$	&14	&	&V16a\\
$201239401$	&p1	&$0.905626\pm0.00002$	&$6811.84172_{-0.0013}^{+0.0015}$	&$0.0239_{-0.00087}^{+0.0017}$	&$0.88\pm0.2$	&$5.2_{-1}^{+0.6}$	&$89.3_{-7}^{+9}$	&$0.53$	&$0.53$	&24	&	&V16a\\
$201239401$	&p2	&$6.783592\pm0.00019$	&$6863.39748_{-0.0059}^{+0.0089}$	&$0.0207_{-0.002}^{+0.003}$	&$0.77\pm0.2$	&$19\pm6$	&$90.3\pm2$	&$0.77$	&$0.051$	&5.8	&	&\\
$201264302$	&p1	&$0.212203\pm0.0000078$	&$6828.10358_{-0.00041}^{+0.00041}$	&$0.0233_{-0.00055}^{+0.0015}$	&$0.49_{-0.02}^{+0.03}$	&$3_{-0.4}^{+0.3}$	&$91.2_{-9}^{+10}$	&$0.13$	&$0.13$	&39	&	&V16a\\
$201427007$	&p1	&$0.72091\pm0.00014$	&$7625.7943_{-0.0023}^{+0.0027}$	&$0.0142_{-0.00067}^{+0.0013}$	&$1.5\pm0.2$	&$3.1_{-0.6}^{+0.3}$	&$88.5\pm10$	&$0.00024$	&$0.00024$	&16	&Conf.	&\\
$201438507$	&p1	&$0.326564\pm0.000034$	&$7611.32309_{-0.0013}^{+0.0014}$	&$0.0145_{-0.00096}^{+0.0038}$	&$1.5_{-0.3}^{+0.4}$	&$3.5_{-0.8}^{+3}$	&$89.3_{-7}^{+10}$	&$0.46$	&$0.46$	&13	&	&\\
$201438507$	&p2	&$2.536236\pm0.00031$	&$7609.54335_{-0.0074}^{+0.0065}$	&$0.0112_{-0.0019}^{+0.0017}$	&$1.1\pm0.2$	&$16$	&$90$	&$0.5$	&$0.033$	&3.4	&	&\\
$201461352$	&p1	&$0.696099\pm0.000057$	&$6858.64842_{-0.002}^{+0.0029}$	&$0.0331_{-0.0036}^{+0.011}$	&$0.59_{-0.09}^{+0.2}$	&$5.9_{-2}^{+6}$	&$90.1_{-5}^{+6}$	&$0.85$	&$0.85$	&8.7	&	&\\
$201469738$	&p1	&$0.276097\pm0.000052$	&$6830.54448_{-0.0024}^{+0.0016}$	&$0.0167_{-0.0013}^{+0.016}$	&$1.2_{-0.4}^{+0.7}$	&$2.6_{-0.6}^{+10}$	&$90_{-6}^{+9}$	&$0.67$	&$0.67$	&7.2	&	&\\
$201549425$	&p1	&$0.383255\pm0.000052$	&$6865.80605_{-0.0022}^{+0.0033}$	&$0.0155_{-0.0016}^{+0.0069}$	&$0.96_{-0.2}^{+0.4}$	&$3.1_{-0.6}^{+5}$	&$89.8_{-7}^{+9}$	&$0.39$	&$0.39$	&8.9	&	&\\
$201595106$	&p1	&$0.877214\pm0.00006$	&$7608.48962_{-0.0012}^{+0.0014}$	&$0.0126_{-0.00036}^{+0.00098}$	&$1.3\pm0.1$	&$5.8_{-1}^{+0.7}$	&$89.2_{-5}^{+8}$	&$0.0076$	&$0.0076$	&24	&Conf.	&M18\\
$201609593$	&p1	&$0.318254\pm0.000041$	&$6814.04212_{-0.0019}^{+0.0022}$	&$0.0138_{-0.0016}^{+0.0068}$	&$0.5_{-0.1}^{+0.2}$	&$3_{-0.7}^{+5}$	&$89.6_{-7}^{+9}$	&$0.72$	&$0.72$	&8.9	&	&\\
$201609593$	&p2	&$1.392535\pm0.00053$	&$6856.48324_{-0.0075}^{+0.0066}$	&$0.011_{-0.0019}^{+0.0015}$	&$0.34_{-0.07}^{+0.05}$	&$9$	&$90$	&2.3e-8	&2.3e-8	&3.8	&	&\\
$201637175$	&p1	&$0.381073\pm0.000024$	&$6846.94318_{-0.00036}^{+0.00031}$	&$0.0366_{-0.00083}^{+0.0014}$	&$2.3\pm0.1$	&$3_{-0.2}^{+0.08}$	&$90.6_{-7}^{+8}$	&$1$	&$1$	&37	&Conf.	&S15\\
$201650711$	&p1	&$0.259661\pm0.000058$	&$6828.61939_{-0.001}^{+0.0012}$	&$0.00824_{-0.00056}^{+0.00093}$	&$0.31_{-0.05}^{+0.06}$	&$2.9_{-0.6}^{+0.7}$	&$89\pm10$	&$0.49$	&$0.49$	&18	&	&\\
$201650711$	&p2	&$5.544059\pm0.000046$	&$6853.70238\pm0.004$	&$0.0106_{-0.00068}^{+0.0014}$	&$0.41_{-0.07}^{+0.08}$	&$20_{-5}^{+4}$	&$90.2\pm2$	&$0.029$	&$0.0019$	&9.4	&	&K19\\
$201687804$	&p1	&$0.297142\pm0.000095$	&$6844.48164_{-0.0015}^{+0.0023}$	&$0.00833_{-0.00056}^{+0.00094}$	&$0.58_{-0.07}^{+0.08}$	&$2.1_{-0.3}^{+0.4}$	&$89.8\pm20$	&$0.074$	&$0.074$	&14	&	&\\
$201717159$	&p1	&$0.604066\pm0.000039$	&$6813.86171_{-0.0016}^{+0.0018}$	&$0.0261_{-0.0014}^{+0.0028}$	&$1.3\pm0.2$	&$4.3_{-1}^{+0.9}$	&$90.5_{-9}^{+10}$	&$0.024$	&$0.024$	&17	&	&V16a\\
$201796374$	&p1	&$0.412222\pm0.000045$	&$6844.67348\pm0.0012$	&$0.0338_{-0.0027}^{+0.011}$	&$1_{-0.2}^{+0.3}$	&$5.5_{-1}^{+6}$	&$89.7\pm5$	&$0.29$	&$0.29$	&16	&	&\\
$201923289$	&p1	&$0.782113\pm0.000033$	&$6861.99499_{-0.0022}^{+0.0025}$	&$0.0128_{-0.00066}^{+0.0013}$	&$1.2\pm0.1$	&$4_{-0.9}^{+0.7}$	&$91.5\pm10$	&$0.054$	&$0.054$	&16	&	&V16a\\
$203533312$	&p1	&$0.175661\pm0.00002$	&$6964.02466\pm0.00025$	&$0.0237_{-0.0002}^{+0.00056}$	&$3.5\pm0.3$	&$1.6_{-0.09}^{+0.04}$	&$89.6_{-9}^{+20}$	&NA\tablenotemark{d}	&--	&54	&	&A16\\
$205002291$	&p1	&$0.857941\pm0.000092$	&$6895.58103_{-0.0013}^{+0.0024}$	&$0.0241_{-0.0013}^{+0.0026}$	&$0.83\pm0.2$	&$6\pm1$	&$90.6\pm7$	&$0.97$	&$0.97$	&16	&FP	&\\
$205152172$	&p1	&$0.980088\pm0.000061$	&$6895.59262_{-0.00092}^{+0.0011}$	&$0.02_{-0.00053}^{+0.0016}$	&$1.2\pm0.2$	&$6.6_{-1}^{+0.5}$	&$89.6\pm5$	&$0.24$	&$0.24$	&28	&	&V16a\\
$205152172$	&p2	&$3.224517\pm0.00028$	&$6931.20494_{-0.0025}^{+0.0022}$	&$0.0166_{-0.0014}^{+0.0028}$	&$0.99\pm0.2$	&$25_{-7}^{+10}$	&$90_{-1}^{+2}$	&$0.25$	&$0.017$	&10	&	&\\
$206024342$	&p1	&$0.911724\pm0.000076$	&$7016.52991_{-0.0018}^{+0.0027}$	&$0.0117_{-0.00056}^{+0.0011}$	&$1.4\pm0.5$	&$4.7_{-1}^{+0.7}$	&$91.7\pm9$	&6.5e-5	&4.4e-6	&11	&Conf.	&V16a\\
$206024342$	&p2	&$4.508035\pm0.00042$	&$7033.78077_{-0.0025}^{+0.0033}$	&$0.0149_{-0.00086}^{+0.0014}$	&$1.7\pm0.6$	&$17_{-4}^{+2}$	&$89.8_{-2}^{+3}$	&$0.0063$	&$0.00042$	&12	&Conf.	&V16a\\
$206024342$	&p3	&$14.647531\pm0.0000041$	&$6985.47506_{-0.0025}^{+0.0038}$	&$0.0217_{-0.00073}^{+0.0016}$	&$2.6\pm0.9$	&$29_{-6}^{+2}$	&$90_{-1}^{+2}$	&1.9e-6	&1.9e-6	&15	&Conf.	&K19\\
$206024342$	&p4	&$27.718603\pm0.1$	&$7017.0414_{-0.018}^{+0.039}$	&$0.0111_{-0.0015}^{+0.0023}$	&$1.3\pm0.5$	&$28_{-9}^{+10}$	&$90_{-1}^{+2}$	&6.8e-5	&4.6e-6	&3.6	&Conf.	&\\
$206042996$	&p1	&$0.354884\pm0.000063$	&$7007.06827_{-0.0017}^{+0.002}$	&$0.0223_{-0.0013}^{+0.0026}$	&$1.2_{-0.1}^{+0.2}$	&$2.6\pm0.5$	&$89.9_{-10}^{+20}$	&$0.029$	&$0.0019$	&14	&Conf.	&V16a\\
$206042996$	&p2	&$5.2986\pm0.0006$	&$6989.0462_{-0.0047}^{+0.0053}$	&$0.0303_{-0.0025}^{+0.0041}$	&$1.7\pm0.2$	&$16_{-4}^{+3}$	&$90.5\pm3$	&$0.015$	&$0.00098$	&12	&Conf.	&K19\\
$206103150$	&p1	&$0.789655\pm0.00004$	&$6990.82197_{-0.00097}^{+0.0013}$	&$0.0139_{-0.00034}^{+0.00076}$	&$1.8_{-0.09}^{+0.1}$	&$3.2_{-0.4}^{+0.2}$	&$89.6_{-9}^{+10}$	&1.2e-7	&8e-9	&20	&Conf.	&B15\\
$206103150$	&p2	&$9.028052\pm0.000065$	&$6997.34398_{-0.00088}^{+0.0013}$	&$0.026_{-0.00033}^{+0.00072}$	&$3.3\pm0.1$	&$17_{-1}^{+0.3}$	&$89.5_{-0.9}^{+2}$	&$0.00013$	&$8.5e-6$	&39	&Conf.	&B15\\
$206151047$	&p1	&$0.358442\pm0.0000085$	&$7015.6764_{-0.00087}^{+0.0013}$	&$0.0138_{-0.00076}^{+0.0021}$	&$1.9\pm0.9$	&$4.4_{-0.9}^{+2}$	&$91\pm8$	&$0.57$	&$0.57$	&18	&	&V16a\\
$206215704$	&p1	&$0.623232\pm0.00003$	&$7039.60634_{-0.0012}^{+0.0014}$	&$0.0346_{-0.0025}^{+0.0051}$	&$0.96\pm0.1$	&$7.3_{-2}^{+3}$	&$90.1_{-4}^{+6}$	&$0.44$	&$0.018$	&14	&Conf.	&V16a\\
$206215704$	&p2	&$2.255436\pm0.000091$	&$7010.0072_{-0.0016}^{+0.0018}$	&$0.0383_{-0.0024}^{+0.0036}$	&$1_{-0.09}^{+0.1}$	&$17_{-4}^{+3}$	&$89.9_{-2}^{+3}$	&$0.085$	&$0.0034$	&13	&Conf.	&K19\\
$206215704$	&p3	&$2.909219\pm0.000035$	&$7043.29037_{-0.0017}^{+0.0019}$	&$0.0398_{-0.0025}^{+0.0051}$	&$1.1\pm0.1$	&$24\pm7$	&$90\pm2$	&$0.27$	&$0.011$	&12	&Conf.	&\\
$206298289$	&p1	&$0.434847\pm0.000046$	&$6987.59112_{-0.00078}^{+0.00089}$	&$0.0247_{-0.00094}^{+0.0016}$	&$0.89_{-0.09}^{+0.1}$	&$4.2_{-0.8}^{+0.4}$	&$88.5_{-7}^{+10}$	&$0.037$	&$0.037$	&26	&	&V16a\\
$206298289$	&p2	&$3.215738\pm0.000078$	&$6992.50972_{-0.0028}^{+0.0032}$	&$0.0232_{-0.0018}^{+0.0028}$	&$0.85\pm0.1$	&$20_{-6}^{+5}$	&$90\pm2$	&$0.41$	&$0.027$	&7.8	&	&\\
$206417197$	&p1	&$0.442112\pm0.000039$	&$7022.51794_{-0.0016}^{+0.0014}$	&$0.0133_{-0.00059}^{+0.0013}$	&$1.1_{-0.09}^{+0.1}$	&$3_{-0.6}^{+0.4}$	&$88.5_{-9}^{+10}$	&$0.11$	&$0.11$	&19	&	&V16a\\
$210605073$	&p1	&$0.567003\pm0.000025$	&$7130.83584_{-0.00052}^{+0.00059}$	&$0.0903_{-0.0021}^{+0.0047}$	&$16_{-3}^{+4}$	&$5.2_{-0.9}^{+0.3}$	&$90.2_{-7}^{+8}$	&$0.72$	&$0.72$	&39	&	&A16\\
$210707130$	&p1	&$0.684559\pm0.00001$	&$7102.33296_{-0.00042}^{+0.00056}$	&$0.0155_{-0.00032}^{+0.00088}$	&$1.2_{-0.07}^{+0.08}$	&$6.1_{-0.7}^{+0.4}$	&$89.4_{-4}^{+6}$	&$0.00017$	&$0.00017$	&37	&Conf.	&Dr17\\
$210801536$	&p1	&$0.892592\pm0.000036$	&$7130.80967_{-0.0025}^{+0.0037}$	&$0.0105_{-0.0008}^{+0.0012}$	&$1_{-0.2}^{+0.3}$	&$4.3_{-1}^{+0.8}$	&$90.8_{-8}^{+10}$	&$0.026$	&$0.026$	&9.6	&	&K19\\
$210801536$	&p2	&$5.044155\pm0.00022$	&$7102.84382_{-0.0063}^{+0.0094}$	&$0.0126_{-0.0017}^{+0.002}$	&$1.2_{-0.2}^{+0.4}$	&$16_{-5}^{+8}$	&$90.2_{-2}^{+3}$	&$0.39$	&$0.39$	&4.4	&	&\\
$210848775$	&p1	&$0.214079\pm0.00003$	&$7104.33168_{-0.0015}^{+0.0018}$	&$0.0224_{-0.0021}^{+0.0031}$	&$1.1_{-0.2}^{+0.3}$	&$2.2_{-0.4}^{+0.9}$	&$90.8_{-10}^{+20}$	&$0.28$	&$0.28$	&11	&	&\\
$210931967$	&p1	&$0.346547\pm0.000042$	&$7124.31547_{-0.0011}^{+0.0012}$	&$0.0232_{-0.0015}^{+0.006}$	&$0.71\pm0.2$	&$4.9_{-1}^{+4}$	&$89.3_{-5}^{+7}$	&$0.68$	&$0.68$	&18	&	&\\
$211305568$	&p1	&$0.197843\pm9.4e-5$	&$7173.01294_{-0.00084}^{+0.0011}$	&$0.0134_{-0.00072}^{+0.0014}$	&$0.67_{-0.07}^{+0.08}$	&$2.1_{-0.4}^{+0.6}$	&$91.6_{-10}^{+20}$	&$0.59$	&$0.59$	&15	&	&Dr17\\
$211305568$	&p2	&$11.54962\pm1.2e-10$	&$7203.98574_{-0.0032}^{+0.0021}$	&$0.0306_{-0.0042}^{+0.0063}$	&$1.5_{-0.2}^{+0.3}$	&$72_{-30}^{+50}$	&$90.1_{-0.5}^{+0.6}$	&$1$	&$0.067$	&4.6	&	&Dr17\\
$211336288$	&p1	&$0.221814\pm0.000062$	&$7143.81329_{-0.0014}^{+0.0021}$	&$0.0147_{-0.0011}^{+0.0039}$	&$0.98_{-0.1}^{+0.2}$	&$2.1_{-0.4}^{+2}$	&$89.4_{-10}^{+20}$	&$0.78$	&$0.78$	&12	&	&Dr17\\
$211357309$	&p1	&$0.463974\pm0.0000011$	&$7189.86394_{-0.00079}^{+0.0009}$	&$0.0159_{-0.00055}^{+0.0011}$	&$0.81\pm0.07$	&$4.3_{-0.7}^{+0.5}$	&$90.7_{-7}^{+9}$	&$0.36$	&$0.36$	&26	&	&A16\\
$211357309$	&p2	&$11.623464\pm0.0000039$	&$7174.22919_{-0.0013}^{+0.0015}$	&$0.0248_{-0.0014}^{+0.0028}$	&$1.3_{-0.1}^{+0.2}$	&$98_{-30}^{+20}$	&$90\pm0.4$	&$0.32$	&$0.021$	&8.8	&	&\\
$211562654$	&p1	&$0.469285\pm0.000064$	&$7196.91237_{-0.0017}^{+0.0019}$	&$0.0117_{-0.00049}^{+0.00098}$	&$1.1_{-0.08}^{+0.1}$	&$2.5_{-0.4}^{+0.3}$	&$90.1_{-10}^{+20}$	&$0.00042$	&2.8e-5	&13	&Conf.	&M18\\
$211562654$	&p2	&$10.787992\pm0.00033$	&$7212.47321_{-0.0023}^{+0.0026}$	&$0.0242_{-0.00065}^{+0.0014}$	&$2.3_{-0.1}^{+0.2}$	&$22_{-3}^{+1}$	&$90.5\pm2$	&$0.025$	&$0.0017$	&18	&Conf.	&M18\\
$211562654$	&p3	&$22.620344\pm0.000067$	&$7212.0276\pm0.0036$	&$0.0244_{-0.0011}^{+0.0021}$	&$2.3\pm0.2$	&$49_{-10}^{+5}$	&$89.9_{-0.7}^{+0.8}$	&$0.028$	&$0.0011$	&11	&Conf.	&M18\\
$211566738$	&p1	&$0.760983\pm0.00004$	&$7206.50131_{-0.0016}^{+0.0021}$	&$0.0198_{-0.0014}^{+0.0028}$	&$1.1_{-0.1}^{+0.2}$	&$5.7_{-1}^{+2}$	&$89.9_{-6}^{+7}$	&$0.35$	&$0.35$	&14	&	&\\
$211828293$	&p1	&$0.699739\pm0.000035$	&$7211.67533\pm0.0015$	&$0.0303_{-0.0041}^{+0.015}$	&$1.2_{-0.3}^{+0.4}$	&$8.8_{-1}^{+20}$	&$89.9_{-2}^{+3}$	&$0.48$	&$0.48$	&12	&	&\\
$212092746$	&p1	&$0.562193\pm0.000034$	&$7158.44518_{-0.0011}^{+0.0013}$	&$0.0375_{-0.0049}^{+0.0082}$	&$1.2_{-0.2}^{+0.3}$	&$7.8_{-2}^{+6}$	&$90.1_{-4}^{+5}$	&$0.48$	&$0.48$	&14	&	&\\
$212157262$	&p1	&$0.77401\pm0.000083$	&$7184.43637_{-0.0018}^{+0.0016}$	&$0.0132_{-0.00053}^{+0.0011}$	&$1.2\pm0.2$	&$3.9_{-0.8}^{+0.4}$	&$89.7_{-8}^{+10}$	&1.5e-6	&5.8e-8	&14	&Conf.	&M18\\
$212157262$	&p2	&$2.871512\pm0.00012$	&$7144.63842_{-0.0024}^{+0.0027}$	&$0.0154_{-0.00067}^{+0.0015}$	&$1.4\pm0.2$	&$9.2_{-2}^{+0.8}$	&$90.4_{-4}^{+5}$	&$0.00034$	&1.4e-5	&14	&Conf.	&M18\\
$212157262$	&p3	&$7.149584\pm0.00012$	&$7203.52085_{-0.0013}^{+0.0019}$	&$0.0259_{-0.00069}^{+0.0015}$	&$2.4\pm0.3$	&$18_{-3}^{+0.9}$	&$90.3_{-1}^{+2}$	&3.6e-5	&1.4e-6	&19	&Conf.	&M18\\
$212157262$	&p4	&$13.606629\pm0.0000036$	&$7155.08495_{-0.0027}^{+0.0031}$	&$0.0206_{-0.00077}^{+0.0017}$	&$1.9\pm0.3$	&$28_{-6}^{+2}$	&$90.1_{-2}^{+1}$	&$1.6e-11$	&$6.5e-13$	&13	&Conf.	&M18\\
$212303338$	&p1	&$0.595491\pm0.000055$	&$7264.23209_{-0.0025}^{+0.0028}$	&$0.00539_{-0.00048}^{+0.00073}$	&$0.47_{-0.06}^{+0.07}$	&$4\pm1$	&$89.4_{-9}^{+10}$	&$0.8$	&$0.8$	&9	&	&M18\\
$212425103$	&p1	&$0.945974\pm0.000011$	&$7226.68722\pm0.0019$	&$0.0172_{-0.00078}^{+0.0016}$	&$1.6_{-0.1}^{+0.2}$	&$4.9_{-1}^{+0.6}$	&$90.2_{-8}^{+9}$	&$0.012$	&$0.012$	&18	&	&K19\\
$212443973$	&p1	&$0.779407\pm0.000014$	&$7239.5629_{-0.0019}^{+0.0021}$	&$0.018_{-0.0013}^{+0.0048}$	&$0.73_{-0.1}^{+0.2}$	&$7.4_{-2}^{+6}$	&$90.9\pm5$	&$0.59$	&$0.59$	&9.2	&	&Dr17\\
$212470904$	&p1	&$0.642926\pm0.000019$	&$7280.92925_{-0.0014}^{+0.0019}$	&$0.0115_{-0.00093}^{+0.0015}$	&$0.91_{-0.09}^{+0.1}$	&$5.5_{-1}^{+2}$	&$89.6_{-6}^{+7}$	&$0.16$	&$0.16$	&14	&	&\\
$212532636$	&p1	&$0.314842\pm0.000024$	&$7242.75694_{-0.0015}^{+0.0013}$	&$0.0132_{-0.00081}^{+0.0018}$	&$0.88_{-0.09}^{+0.1}$	&$2.9_{-0.6}^{+0.7}$	&$89.7\pm10$	&$0.078$	&$0.078$	&18	&	&\\
$212610063$	&p1	&$0.43514\pm0.000049$	&$7272.1437_{-0.0018}^{+0.0021}$	&$0.0131_{-0.00093}^{+0.0014}$	&$1.1\pm0.2$	&$3\pm0.6$	&$88.1_{-10}^{+20}$	&$0.037$	&$0.037$	&13	&	&\\
$212624936$	&p1	&$0.578657\pm0.000016$	&$7232.53108_{-0.0022}^{+0.0025}$	&$0.0114_{-0.00071}^{+0.0011}$	&$1.2\pm0.3$	&$3_{-0.6}^{+0.4}$	&$90.3\pm10$	&$0.0041$	&$0.0041$	&13	&Conf.	&\\
$212624936$	&p2	&$11.811437\pm0.00011$	&$7233.78707_{-0.0033}^{+0.0029}$	&$0.0226_{-0.0011}^{+0.0022}$	&$2.3\pm0.5$	&$41_{-10}^{+5}$	&$90_{-0.9}^{+1}$	&$0.048$	&$0.0032$	&10	&Conf.	&\\
$213715787$	&p1	&$0.961918\pm0.000013$	&$7327.91683_{-0.001}^{+0.00089}$	&$0.0213_{-0.00072}^{+0.0014}$	&$0.73\pm0.2$	&$7.9_{-2}^{+0.8}$	&$89.9_{-4}^{+5}$	&$2.7e-5$	&$2.7e-5$	&25	&Conf.	&H18\\
$214189767$	&p1	&$0.520638\pm0.000091$	&$7327.11571_{-0.0028}^{+0.0033}$	&$0.0205_{-0.0017}^{+0.0026}$	&$1.2_{-0.1}^{+0.2}$	&$3.1_{-0.7}^{+0.9}$	&$90.6\pm10$	&$0.93$	&$0.93$	&8	&	&\\
$214539781$	&p1	&$0.437366\pm0.000039$	&$7309.36511_{-0.0016}^{+0.0014}$	&$0.0169_{-0.0011}^{+0.0023}$	&$0.78\pm0.1$	&$4.1_{-0.9}^{+2}$	&$89.4_{-7}^{+10}$	&$0.97$	&$0.97$	&15	&FP	&\\
$219752140$	&p1	&$0.877836\pm0.00021$	&$7371.26939_{-0.0037}^{+0.0048}$	&$0.0104_{-0.00099}^{+0.0015}$	&$1.2_{-0.2}^{+0.3}$	&$4.6_{-1}^{+2}$	&$89.5_{-8}^{+10}$	&$0.96$	&$0.96$	&7.9	&	&\\
$220250254$	&p1	&$0.57023\pm0.000032$	&$7465.80382_{-0.0012}^{+0.0023}$	&$0.00874_{-0.00061}^{+0.001}$	&$0.86\pm0.2$	&$4.4\pm1$	&$89.2_{-8}^{+10}$	&$0.036$	&$0.036$	&16	&Conf.	&M18\\
$220256496$	&p1	&$0.669532\pm0.000019$	&$7395.82322\pm0.0016$	&$0.0149_{-0.00048}^{+0.0013}$	&$1.3\pm0.2$	&$3.7_{-0.8}^{+0.5}$	&$89.6\pm10$	&$0.0029$	&$0.0029$	&21	&Conf.	&M18\\
$220256496$	&p2	&$2.601267\pm0.00014$	&$7470.75432\pm0.0033$	&$0.0145_{-0.00091}^{+0.0018}$	&$1.3\pm0.2$	&$11_{-3}^{+2}$	&$89.8_{-3}^{+4}$	&$0.0064$	&$0.00043$	&8.7	&Conf.	&\\
$220272424$	&p1	&$0.659967\pm0.000064$	&$7452.27858_{-0.0017}^{+0.0023}$	&$0.0145_{-0.0011}^{+0.0022}$	&$0.85_{-0.1}^{+0.2}$	&$5.1_{-1}^{+2}$	&$89.5_{-7}^{+9}$	&$0.2$	&$0.2$	&9.3	&	&\\
$220383386$	&p1	&$0.959622\pm0.000044$	&$7410.69248_{-0.00091}^{+0.001}$	&$0.0171_{-0.00042}^{+0.00092}$	&$1.6_{-0.08}^{+0.09}$	&$4.7_{-0.7}^{+0.2}$	&$90.2_{-6}^{+7}$	&$0.0003$	&$2e-5$	&21	&Conf.	&V16b\\
$220383386$	&p2	&$29.875221\pm0.0004$	&$7454.69902_{-0.0033}^{+0.0038}$	&$0.0251_{-0.00087}^{+0.0017}$	&$2.3_{-0.1}^{+0.2}$	&$39_{-6}^{+2}$	&$90_{-0.8}^{+0.9}$	&$0.14$	&$0.0095$	&14	&Conf.	&V16b\\
$220440058$	&p1	&$0.626683\pm0.000026$	&$7428.38519_{-0.0013}^{+0.0019}$	&$0.0442_{-0.0025}^{+0.0049}$	&$1.4\pm0.4$	&$5.2\pm1$	&$89.4_{-7}^{+9}$	&$0.06$	&$0.06$	&17	&	&\\
$220440058$	&p2	&$3.835187\pm0.000098$	&$7401.85905_{-0.0024}^{+0.0027}$	&$0.053_{-0.0057}^{+0.014}$	&$1.7_{-0.5}^{+0.6}$	&$39$	&$88.9_{-0.2}^{+2}$	&$0.64$	&$0.043$	&7	&	&\\
$220492298$	&p1	&$0.762406\pm0.00015$	&$7431.67734_{-0.0033}^{+0.0037}$	&$0.01_{-0.00072}^{+0.0012}$	&$0.99\pm0.1$	&$3.2_{-0.7}^{+0.6}$	&$92.6\pm10$	&$0.00069$	&$0.00069$	&11	&Conf.	&\\
$220554126$	&p1	&$0.412326\pm0.00007$	&$7430.58836_{-0.0029}^{+0.0033}$	&$0.0161_{-0.0014}^{+0.0021}$	&$0.8\pm0.1$	&$2.5_{-0.5}^{+0.7}$	&$90.7_{-10}^{+20}$	&$0.098$	&$0.098$	&8.8	&	&\\
$220554210$	&p1	&$0.70531\pm0.000037$	&$7466.1131_{-0.0022}^{+0.0019}$	&$0.0145_{-0.00065}^{+0.0011}$	&$1.6\pm0.1$	&$3_{-0.5}^{+0.3}$	&$91.3\pm10$	&$1.1e-9$	&$7.5e-11$	&21	&Conf.	&\\
$220554210$	&p2	&$4.169783\pm0.000089$	&$7457.54493_{-0.0022}^{+0.0025}$	&$0.0227_{-0.00073}^{+0.0016}$	&$2.5_{-0.1}^{+0.2}$	&$11_{-2}^{+0.8}$	&$90.1_{-3}^{+4}$	&$1.8e-7$	&$1.2e-8$	&18	&Conf.	&K19\\
$220554210$	&p3	&$27.263595\pm0.0009$	&$7426.94269_{-0.0052}^{+0.01}$	&$0.0233_{-0.0014}^{+0.003}$	&$2.6_{-0.2}^{+0.3}$	&$42_{-10}^{+5}$	&$90.1\pm1$	&$0.00013$	&$5.4e-6$	&8.6	&Conf.	&\\
$220674823$	&p1	&$0.571212\pm0.000038$	&$7417.43601_{-0.00079}^{+0.00091}$	&$0.0157_{-0.00031}^{+0.00069}$	&$1.6\pm0.1$	&$2.9_{-0.3}^{+0.1}$	&$89.7\pm10$	&$6.3e-5$	&$4.2e-6$	&36	&Conf.	&A17\\
$220674823$	&p2	&$13.334615\pm0.0001$	&$7405.70999_{-0.0032}^{+0.0028}$	&$0.0209_{-0.00055}^{+0.0015}$	&$2.2\pm0.2$	&$26_{-5}^{+2}$	&$90.2\pm1$	&$5.4e-8$	&$3.6e-9$	&19	&Conf.	&A17\\
$220687583$	&p1	&$0.45693\pm0.000073$	&$7466.50516_{-0.002}^{+0.0023}$	&$0.013_{-0.00087}^{+0.0017}$	&$1.2\pm0.1$	&$3.4_{-0.8}^{+0.9}$	&$89.2\pm10$	&$0.013$	&$0.013$	&11	&	&\\
$228721452$	&p1	&$0.50557\pm0.000088$	&$7637.66396_{-0.002}^{+0.0023}$	&$0.00801_{-0.00053}^{+0.0008}$	&$0.89_{-0.09}^{+0.1}$	&$3.3_{-0.7}^{+0.8}$	&$89.7\pm10$	&$0.029$	&$0.0019$	&10	&Conf.	&\\
$228721452$	&p2	&$4.562546\pm0.00014$	&$7628.61208_{-0.0025}^{+0.0029}$	&$0.0129_{-0.00042}^{+0.0012}$	&$1.5\pm0.1$	&$13_{-3}^{+1}$	&$89.6_{-3}^{+4}$	&$0.0025$	&$0.0025$	&15	&Conf.	&M18\\
$228732031$	&p1	&$0.369311\pm0.00003$	&$7605.09575_{-0.00025}^{+0.00091}$	&$0.0172_{-0.00027}^{+0.00041}$	&$1.5_{-0.06}^{+0.07}$	&$3\pm0.2$	&$89.7_{-6}^{+7}$	&$0.45$	&$0.45$	&32	&Conf.	&Da17\\
$228739208$	&p1	&$0.483787\pm0.000033$	&$7607.43487_{-0.0019}^{+0.0021}$	&$0.00918_{-0.001}^{+0.004}$	&$1.1_{-0.3}^{+0.5}$	&$4.2_{-0.9}^{+7}$	&$90.8_{-6}^{+7}$	&$0.99$	&$0.99$	&9.5	&	&\\
$228801451$	&p1	&$0.584216\pm0.000022$	&$7646.56655_{-0.00077}^{+0.00046}$	&$0.0118_{-0.0002}^{+0.00023}$	&$1\pm0.02$	&$4.2_{-0.2}^{+0.3}$	&$89.7\pm4$	&$0.00012$	&$0.00012$	&29	&Conf.	&M18\\
$228813918$	&p1	&$0.179719\pm0.000028$	&$7639.27571_{-0.00068}^{+0.00077}$	&$0.02_{-0.00081}^{+0.0016}$	&$0.64\pm0.1$	&$2.1_{-0.3}^{+0.4}$	&$93.1_{-20}^{+10}$	&$0.0011$	&$0.0011$	&25	&Conf.	&S18\\
$228814754$	&p1	&$0.844537\pm0.000025$	&$7650.61924_{-0.00047}^{+0.00071}$	&$0.0593_{-0.0027}^{+0.0053}$	&$1.4\pm0.6$	&$11\pm2$	&$89.9\pm3$	&$0.39$	&$0.39$	&25	&	&\\
$228836835$	&p1	&$0.728113\pm0.000063$	&$7585.45061_{-0.0012}^{+0.0014}$	&$0.0299_{-0.0018}^{+0.003}$	&$1.1_{-0.2}^{+0.3}$	&$6.4_{-2}^{+1}$	&$89.7_{-5}^{+7}$	&$0.0058$	&$0.0058$	&16	&Conf.	&\\
$228877886$	&p1	&$0.778286\pm0.00011$	&$7642.77006_{-0.0029}^{+0.0038}$	&$0.0125_{-0.0011}^{+0.0019}$	&$1\pm0.1$	&$3.9_{-1}^{+0.9}$	&$89.6\pm10$	&$0.015$	&$0.015$	&8.7	&	&\\
\enddata
\tablenotetext{a}{All known planets in a system are numbered in order of distance from the star, and may differ from the b/c/d/e letters assigned to confirmed planets in order of discovery, and from the 01, 02, etc. designations of \ktwo\ canidadates.}
\tablenotetext{b}{Conf.=Statistically validated or confirmed by radial velocity mass measurement here or in prior work. FP=Likely false positive. Note that candidates with SNR$<10$ are not identified as either confirmed or false positives regardless of FPP.}
\tablenotetext{c}{Candidate or confirmed planet was first noted in a prior paper: B15 = \citet{2015ApJ...812L..18B}, S15: \citet{2015ApJ...812L..11S}, V16a = \citet{2016ApJS..222...14V}, V16b = \citet{2016ApJ...829L...9V}, A16 = \citet{2016AJ....152...47A}, A17 = \citet{2017AJ....153...82A}, Dr17 = \citet{2017AJ....154..207D}, Da17 = \citet{2017AJ....154..226D}, M18 = \citet{2018AJ....155..136M}, H18 = \citet{2018AJ....155..127H}, S18 = \citet{2018MNRAS.474.5523S}, K19 = \citet{2019ApJS..244...11K} }
\tablenotetext{d}{See Section \ref{sec:desert}}
\end{deluxetable}
\end{longrotatetable}

\section{Systematic Search for Additional Transiting Planets}
\label{sec:multi_search}

\subsection{The importance of multi-planet systems}

Multi-planet systems are important for several reasons. First, the system architecture (orbits, sizes, and compositions) is key to unlocking the formation history of these systems. Second, the mere presence of one or more additional planets boosts the likelihood that all candidates are planets \citep{2012ApJ...750..112L} We follow the revised analysis of \citet{2019AJ....157..143B} and assign likelihood boost factors of 10 for two-candidate systems and 15 for three or more candidates. We do not apply a validation boost, however, for candidates detected with SNR$ \leq 10$; see Section \ref{sec:validation} for a discussion of the effect of low SNR on validation of both single and multi-planet systems. In \autoref{table:planets}, we report the individual and boosted FPP, where applicable, for each candidate.

\subsubsection{Identification of potential multi-planet systems has often been piecemeal}

Previous work has turned up numerous cases of ultra-short-period planets in multi-planet systems, often identified because the companion planet(s) had obvious transits in the photometry \citep[e.g.][]{2017AJ....153...82A}. It is also common, however, for multiple candidates at different periods to be reported individually in separate works, with no combined system analysis performed, or even consistent candidate numbering (e.g. EPIC 206215704/K2-302; see Section \ref{sec:newcomps}). Motivated by these discrepancies, for the first time we have systematically searched every candidate USP in \ktwo\ Campaigns 0-8 and 10 for evidence of transiting companions.

We note that for completeness, a full census of USPs in multi-planet systems would require searching all known transiting planet candidates at any period for heretofore missed USPs, using the targeted short-period search methods described in this paper. Complementary programs to identify non-transiting, longer-period companions are also recommended, such as long-term radial velocity monitoring, through which planets have been found around HD 3167 \citep{2017AJ....154..122C} and WASP-47 \citep{2016A&A...586A..93N}. Both projects are, however, well beyond the scope of the present work.

To conduct our search, we first compiled values from the literature for all of the known companions to our \numusp\ USP candidates. For each candidate that has already been identified in the literature, starting with the USP (which we designate p1, to avoid confusion with other designations in the literature), we folded the transit data at the known period and fit for the transit light curve (as in Section \ref{sec:search}). 

Overlapping transits are common in multi-planet systems, particularly for USPs with frequent transits. We found, however, that the impact of overlapping exterior transits on the parameters derived for the USP is minimal, since the fraction of USP transits that are overlapped is typically small. Overlap is more important for the outer planets, particularly those with only a handful of transits. Thus, we subtracted the best fit model for the USP from the light curve before searching for the next potential candidate planet. For systems with companions that have already been identified, we used the reported orbital periods to progressively search for, fit, and remove each of the companion transit signals. After exhausting the list of known companion periods, we searched for new periodic signals, both between known orbital periods (if there are already multiple candidates identified) and beyond the longest period identified (we formally searched out to half the length of the time series). For simplicity, we used the same EEBLS routine for longer-period planets as we used to search for USPs, which may affect recovery efficiencies, though with a median $P=4.5$ d, most companions have short enough transit durations that the 1-day smoothing window will not have a huge effect. We also dropped the initial SNR detection threshold for periodic signals and gave everything above 3$\sigma$ a visual examination, since the SNR of a periodic signal can be difficult to calculate for a multiple planet system with many overlapping periodic peaks. Some candidates which originally appeared at low SNR in the power spectra were actually detected at moderate SNR once we had fit for the SNR of the transit light curve with the other signals removed. We found \lowsnr\ companions with SNR=3-7, and another \marginalsnrcomp\ between SNR=7-10. None of the low and marginal SNR candidates are listed as validated; see discussion in Section \ref{sec:validation}. In this manner, we identified \nummultisystems\ systems with \numcompanions\ candidates, in addition to our \numusp\ USP candidates. \newcompanions\ of these systems are newly identified as hosting multiple candidates, and of these \newmultisystems\ are entirely new systems (both the USP and companion are new).

Both the USPs and their companions tend to be small (see \autoref{fig:period_radius}). Of the \nummultisystems\ candidate multi-planet systems we identified using EVEREST photometry, we documented only three cases where both the USP and the companion(s) would have been detectable using k2sff photometry: EPIC 206103150 (a.k.a. WASP-47), three transiting planets \citep{2015ApJ...812L..18B}; EPIC 220383386 (a.k.a. HD 3167), two transiting planets \citep{2016ApJ...829L...9V} \citep[as well as one non-transiting planet,][]{2017AJ....154..122C}; and EPIC 220674823 (a.k.a. K2-106), two transiting planets \citep{2017AJ....153...82A}.

\section{Ground-based Observations}
\label{sec:obs}

\subsection{Spectroscopy}
\label{sec:spectra}

Accurate planetary radius values are only as good as the stellar parameters used to derive the stellar radius. To better determine the stellar parameters, we obtained reconnaissance precision spectra for some systems with the Tull Coud\'e spectrograph \citep{1995PASP..107..251T} at the Harlan J. Smith 2.7-m telescope at McDonald Observatory. The exposure times ranged from 100 to 4800 s, resulting in signal-to-noise ratios (SNR) from 25 to 60 per resolution element at 5650\AA. We determined stellar parameters for the host stars with the spectral fitting tool {\it Kea} \citep{2016PASP..128i4502E}, which compares high-resolution, low-SNR spectra of stars to a massive grid of synthetic stellar spectral models in order to determine the fundamental stellar parameters of the \kepler\ target stars. For typical Kea uncertainties and the range in $T_{\rm eff}$ over which it derives reliable stellar parameters, see \citet{2016PASP..128i4502E}. We also determined absolute radial velocities by cross-correlating the spectra with known RV standard stars. The stellar parameters from each observation and the derived values of  $T_{\rm eff}$, $log(g)$, and [Fe/H] are shown in \autoref{table:spectra}. When multiple observations were taken to constrain the stellar RV, we assigned the star to have the average value of the parameters for $T_{\rm eff}$, $log(g)$, and [Fe/H] reported in \autoref{table:stars} for the purpose of our analyses. A few observational targets showed significant variations in radial velocity values over time that likely indicate stellar-mass companions, as noted in \autoref{table:stars}, and are not included in the \numusp\ systems we retained as candidates. 

\begin{longrotatetable}
\begin{deluxetable}{cccrclrrrrc}
\tablecaption{McDonald Spectroscopic Observations\label{table:spectra}}
\tabletypesize{\scriptsize}
\tablewidth{0pt}
\tablehead{
\colhead{Target} & \colhead{UT}	& \colhead{HJD} & \colhead{T$_{exp}$[s]} &
\colhead{SNR} & \colhead{T$_{eff}$[K]} & \colhead{[Fe/H]} &
\colhead{log(g)} & \colhead{rot[km/s]} & \colhead{RV[km/s]} &
\colhead{Note}\tablenotemark{a}
}
\startdata
201264302	& 2017-04-07 05:32:06.9	& 2457850.75592	 & 3477.8	& 24	& $3875	\pm 157$    & $ -0.90	\pm 0.10$	& $4.33	\pm 0.21$	& $ 9.00	\pm 4.65$	& $ 50.20	 \pm 0.80 $	& \\
201264302	& 2017-05-18 03:27:50.2	& 2457891.66756	 & 3600.0	& 31	& $3875	\pm 157$ 	& $ -0.80	\pm 0.20$	& $4.17	\pm 0.17$	& $ 9.00	\pm 3.84$	& $ 49.70	 \pm 1.00$	& M\\
201595106	& 2017-01-19 12:31:35.0 & 2457773.0273	 & 486.1	& 33	& $5820	\pm 20$ 	& $ -0.01	\pm 0.03$	& $4.62	\pm 0.12$	& $ 3.62	\pm 0.18$	& $0.59\pm0.25$         & \\
201606542	& 2016-03-15 07:31:09.2	& 2457462.81892	 & 597.1	& 35	& $5540	\pm 108$ 	& $ 0.03	\pm 0.04$	& $4.81	\pm 0.12$	& $ 13.23	\pm 0.53$	& $13.11\pm0.61$	& S \\
201650711	& 2017-02-17 05:15:59.1	& 2457801.73285	 & 1412.0	& 48	& $4280	\pm 153$ 	& $ -0.83	\pm 0.07$	& $4.25	\pm 0.48$	& $ 2.77	\pm 0.43$	& $4.88\pm0.58$	&  \\
201650711	& 2016-03-16 07:28:38.6	& 2457463.81728	 & 1659.2	& 43	& $4340	\pm 103$ 	& $ -0.81	\pm 0.08$	& $4.25	\pm 0.42$	& $ 2.31	\pm 0.40$	& $ 5.52	 \pm 0.43$	&  \\
202094740	& 2016-09-10 10:39:49.1	& 2457641.94531	 & 536.7	& 41	& $6250	\pm 341$ 	& $ -0.20	\pm 0.12$	& $4.92	\pm 0.08$	& $ 120.00  \pm 0.00$	& $ 49.02	 \pm 1.48$	& S,F  \\ 
202094740	& 2017-02-17 04:52:40.7	& 2457801.70827	 & 225.0	& 36	& $6300	\pm 129$ 	& $ -0.14	\pm 0.08$	& $4.06	\pm 0.36$	& $ 50.00	\pm 0.00$	& $-14.64\pm6.29$	& S,F \\
203533312	& 2016-03-15 11:07:20.5	& 2457462.97775	 & 2161.2	& 38	& $6620	\pm 86$ 	& $ 0.09	\pm 0.05$	& $4.19	\pm 0.12$	& $ 23.85	\pm 0.83$	& $ 18.56	 \pm 1.50$	& R\\
203533312	& 2016-03-16 11:33:30.0	& 2457463.99681	 & 2299.6	& 37	& $6320	\pm 111$ 	& $-0.06	\pm 0.06$	& $4.12	\pm 0.26$	& $ 25.08	\pm 0.834$	& $19.98\pm4.43$	& R \\
206024342	& 2017-10-17 03:59:20.6	& 2458043.68139	 & 2059.5	& 38	& $5760	\pm 60$ 	& $ -0.24	\pm 0.03$	& $4.62	\pm 0.16$	& $ 3.80	\pm 0.20$	& $ 52.98	 \pm 0.24$	&  \\
206024342	& 2017-11-11 01:24:33.1	& 2458068.56905	 & 1619.1	& 37	& $5688	\pm 132$ 	& $ -0.20	\pm 0.12$	& $4.33	\pm 0.17$	& $ 2.90	\pm 0.40$	& $ 53.06	 \pm 0.35$	& \\ 
206151047	& 2016-09-10 07:16:53.3	& 2457641.81914	 & 1756.7	& 33	& $5880	\pm 73$ 	& $ -0.30	\pm 0.04$	& $4.56	\pm 0.16$	& $ 4.00	\pm 0.23$	& $ -16.86 \pm.25$      &  \\
206151047	& 2017-11-11 02:00:34.4	& 2458068.59742	 & 2141.2	& 34	& $6062	\pm 113$ 	& $ -0.20	\pm 0.12$	& $4.58	\pm 0.08$	& $ 3.90	\pm 0.30$	& $-17.01\pm0.42$	    &  \\
206169375	& 2016-09-10 07:54:54.5	& 2457641.84179	 & 1076.6	& 39	& $6180	\pm 73$ 	& $ -0.27	\pm 0.03$	& $4.44	\pm 0.06$	& $ 5.38	\pm 0.21$	& $ 14.59	 \pm 0.30$	&  \\
206169375	& 2017-10-15 04:05:09.3	& 2458041.67913	 & 779.3	& 37	& $6100	\pm 141$ 	& $ -0.25	\pm 0.04$	& $4.38	\pm 0.16$	& $ 6.20	\pm 0.40$	& $14.93\pm0.42$	    &  \\
206298289	& 2017-10-17 04:43:18.3	& 2458043.72144	 & 3600.0	& 18	& $3875	\pm 157$ 	& $ -1.00	\pm 0.10$	& $4.42	\pm 0.20$	& $ 8.50	\pm 2.90$	& $-26.69\pm1.26$	    & M\\
206417197	& 2016-09-10 06:31:47.4	& 2457641.79023	 & 2153.1	& 39	& $5240	\pm 60$ 	& $ -0.20	\pm 0.04$	& $4.62	\pm 0.07$	& $ 1.77	\pm 0.28$	& $ 14.23	 \pm 0.27$	&  \\
206417197	& 2017-11-11 02:48:00.2	& 2458068.63202	 & 2351.5	& 38	& $5250	\pm 94$ 	& $ -0.10	\pm 0.10$	& $4.67	\pm 0.11$	& $ 2.00	\pm 0.20$	& $14.09\pm0.33$	    &  \\
210414957	& 2015-12-17 09:15:41.1	& 2457373.91896	 & 4800.0	& 32	& $5838	\pm 102$ 	& $ 0.30	\pm 0.07$	& $4.12	\pm 0.14$	& $ 8.25	\pm 0.28$	& $ 43.76	 \pm 0.09$	&  \\
210707130	& 2017-02-17 04:27:48.1	& 2457801.69147	 & 902.6	& 44	& $4440	\pm 93$ 	& $ -0.36	\pm 0.09$	& $4.38	\pm 0.33$	& $ 1.92	\pm 0.38$	& $-4.26\pm 0.56$	    &  \\
210707130	& 2015-12-18 03:50:12.9	& 2457374.67776	 & 2189.7	& 47	& $4462	\pm 84$ 	& $ -0.28	\pm 0.10$	& $4.17	\pm 0.18$	& $ 1.92	\pm 0.34$	& $ -4.12	 \pm 0.08$	&  \\
210707130	& 2016-09-10 08:27:53.7	& 2457641.86788	 & 2335.0	& 51	& $4380	\pm 73$ 	& $ -0.43	\pm 0.09$	& $4.31	\pm 0.16$	& $ 1.38	\pm 0.27$	& $ -3.14	 \pm 0.30$	&  \\
210754505	& 2015-12-18 06:30:20.8	& 2457374.79139	 & 2586.7	& 36	& $5875	\pm 103$ 	& $ 0.04	\pm 0.05$	& $4.04	\pm 0.19$	& $ 8.75	\pm 0.22$	& $ -1.82	 \pm 0.07$	&  \\
210954046	& 2015-12-18 08:05:17.3	& 2457374.84919	 & 1200.0	& 50	& $6188	\pm 298$ 	& $ 0.00	\pm 0.16$	& $3.00	\pm 0.58$	& $ 53.33	\pm 3.55$	& $ 18.83	 \pm 0.49$	& F\\
210954046	& 2015-12-17 03:25:23.3	& 2457373.65486	 & 1200.0	& 38	& $6062	\pm 274$ 	& $ 0.00	\pm 0.16$	& $3.00	\pm 0.58$	& $ 55.00	\pm 2.89$	& $ 40.57	 \pm 1.05$	& F\\
210961508	& 2015-12-19 03:08:23.2	& 2457375.65275	 & 2895.7	& 42	& $4925	\pm 45$ 	& $ -0.06	\pm 0.07$	& $3.42	\pm 0.08$	& $ 2.42	\pm 0.15$	& $ -58.53   \pm 0.05$	& E\\
210961508	& 2017-11-11 06:33:04.3	& 2458068.78856	 & 1735.3	& 40	& $5000	\pm 80$ 	& $ 0.00	\pm 0.10$	& $3.92	\pm 0.20$	& $ 2.30	\pm 0.30$	& $-58.67\pm0.31$	    & E\\
211152484	& 2015-12-18 07:28:11.5	& 2457374.82461	 & 1407.3	& 44	& $6188	\pm 67$ 	& $ -0.12	\pm 0.06$	& $4.12	\pm 0.15$	& $ 8.42	\pm 0.15$	& $ -49.55   \pm 0.05$	&  \\
211357309	& 2017-03-21 05:04:38.0	& 2457833.7298	 & 2499.8	& 50	& $4240	\pm 223$ 	& $ -1.00	\pm 0.00$	& $4.62	\pm 0.12$	& $ 4.23	\pm 0.39$	& $18.09\pm0.74$        & \\
212157262	& 2017-11-11 11:19:00.4	& 2458068.98210	 & 1649.2	& 35	& $5438	\pm 63$ 	& $ 0.00	\pm 0.10$	& $4.58	\pm 0.08$	& $ 2.20	\pm 0.20$	& $-8.52\pm0.23$    	&  \\
212432685	& 2017-05-18 05:59:41.1	& 2457891.76177	 & 1196.5	& 35	& $6180	\pm 136$ 	& $ -0.36	\pm 0.04$	& $4.50	\pm 0.18$	& $ 8.00	\pm 0.00$	& $ 102.50   \pm 0.60$	& A\\
212470904	& 2018-05-24 05:00:23.0	& 2458262.7259	 & 2151.5	& 25	& $4820	\pm 124$ 	& $ 0.18	\pm 0.08$	& $4.62	\pm 0.22$	& $ 2.85	\pm 0.34$	& $-11.71\pm0.60$       & \\
212470904	& 2019-04-28 06:43:15.0	& 2458601.7998	 & 2422.9	& 42	& $4800	\pm 63$ 	& $ -0.16	\pm 0.08$	& $4.56	\pm 0.16$	& $ 1.46	\pm 0.18$	& $-11.71\pm0.60$       & \\
213703832	& 2017-06-18 08:14:48.8	& 2457922.86992	 & 3600.0	& 27	& $4750	\pm 94$ 	& $ -0.30	\pm 0.25$	& $2.50	\pm 0.40$	& $ 13.00	\pm 0.00$	& $ 151.00   \pm 0.50$	& G,R\\
214547683	& 2018-07-21 06:12:05.0	& 2458320.7856	 & 3729.2	& 32	& $4860	\pm 24$ 	& $ -0.12	\pm 0.08$	& $3.38	\pm 0.12$	& $ 3.31	\pm 0.24$	& $-20.53\pm0.39$       & B\\
214984368	& 2016-09-09 03:27:21.6	& 2457640.66333	 & 2945.0	& 29	& $4160	\pm 60$ 	& $ -0.70	\pm 0.05$	& $1.12	\pm 0.12$	& $ 4.92	\pm 0.18$	& $ 81.40	 \pm 0.22$	& G\\
218952024	& 2017-10-15 02:31:39.1	& 2458041.61074	 & 1093.0	& 25	& $4125	\pm 82$ 	& $ -0.60	\pm 0.10$	& $1.67	\pm 0.21$	& $ 5.50	\pm 0.20$	& $50.4 \pm 0.7$	                & G,S\\
220198551	& 2016-09-08 11:07:19.7	& 2457639.97939	 & 1915.9	& 33	& $5000	\pm 63$ 	& $ 0.01	\pm 0.07$	& $4.50	\pm 0.18$	& $ 1.46	\pm 0.18$	& $ -11.25   \pm 0.17$	&  \\
220250254	& 2016-08-15 09:30:35.7	& 2457615.90196	 & 397.4	& 29	& $5500	\pm 45$ 	& $ 0.00	\pm 0.02$	& $4.62	\pm 0.07$	& $ 2.00	\pm 0.25$	& $ -1.15	 \pm 0.17$	&  \\
220250254	& 2016-10-11 07:12:25.0	& 2457672.80795	 & 332.9	& 41	& $5500	\pm 89$ 	& $-0.09	\pm 0.02$	& $4.50	\pm 0.10$	& $ 2.54	\pm 0.22$	& $-1.37\pm0.25$	    &  \\
220256496	& 2016-09-08 10:26:01.9	& 2457639.95073	 & 1937.7	& 38	& $5360	\pm 51$ 	& $-0.04	\pm 0.03$	& $4.69	\pm 0.06$	& $ 1.77	\pm 0.28$	& $ -14.60   \pm 0.13$	&  \\
220256496	& 2016-12-16 02:16:39.3	& 2457738.60416	 & 1175.9	& 33	& $5340	\pm 51$ 	& $-0.04	\pm 0.03$	& $4.81	\pm 0.06$	& $2.15     \pm 0.25$	& $-14.41\pm0.22$	    &  \\
220383386	& 2017-11-11 03:56:05.8	& 2458068.66911	 & 122.5	& 72	& $5438	\pm 63$ 	& $ -0.10	\pm 0.10$	& $4.58	\pm 0.15$	& $ 2.00	\pm 0.20$	& $19.29\pm0.16$	    &  \\
220492298	& 2017-10-15 06:00:03.1	& 2458041.76168	 & 1020.0	& 17	& $5680	\pm 37$ 	& $ 0.29	\pm 0.03$	& $4.56	\pm 0.12$	& $ 3.40	\pm 0.10$	& $ -19.38   \pm 0.29$	& \\ 
220492298	& 2017-10-18 07:41:14.7	& 2458044.84403	 & 3112.9	& 36	& $5560	\pm 51$ 	& $ 0.26	\pm 0.05$	& $4.44	\pm 0.12$	& $ 3.50	\pm 0.20$	& $ -19.47   \pm 0.26$	&  \\
220674823	& 2016-08-15 09:46:55.0	& 2457615.91451	 & 598.2	& 29	& $5580	\pm 86$ 	& $ 0.04	\pm 0.03$	& $4.62	\pm 0.16$	& $ 2.62	\pm 0.18$	& $ -15.86   \pm 0.13$	&  \\
220674823	& 2016-09-08 09:17:37.4	& 2457639.90160	 & 1618.3	& 59	& $5600	\pm 55$ 	& $ 0.01	\pm 0.02$	& $4.50	\pm 0.10$	& $ 2.23	\pm 0.23$	& $ -16.01   \pm 0.10$	&  \\
220674823	& 2016-10-11 07:27:13.7	& 2457672.81868	 & 408.9	& 39	& $5660	\pm 60$ 	& $ -0.01	\pm 0.02$	& $4.62	\pm 0.07$	& $ 3.08	\pm 0.21$	& $ -15.80   \pm 0.28$	&  \\
\enddata
\label{table:spectra}
\tablenotetext{a}{A: asymmetric CCF, possible blend. E: somewhat evolved. F: false positive (large RV shifts). G: giant star. R: fast rotator. M: M dwarf has unreliable [Fe/H]. S: spectral binary with double-peaked cross-correlation function (CCF). B: identified as binary star from imaging or phase curve variations}
\end{deluxetable}
\end{longrotatetable}


\subsection{Stellar Radius Estimation}

For systems with previous analyses, we used the literature value for $R_\star$ with the smallest error bar. Otherwise, we used our observationally derived values from \autoref{table:spectra}, where available, and then the values from the Ecliptic Plane Input Catalog, or EPIC \citep{2016ApJS..224....2H}, for $T_{\rm eff}$, $log(g)$, and [Fe/H]. To calculate $R_\star$, we used the isochrones from the Dartmouth models assuming a stellar age of 5 Gyr \citep{2008ApJS..178...89D}, individual stellar ages being largely unavailable. As a check, we compared the isochrones values for $R_\star$ (which are based on a numerical grid) to the values from the equation-based models of \citet{Boyajian2012}. Both approaches agreed well within error, all but five with $\Delta R_{\star}<0.2 R_{\odot}$, and all within $<3\sigma$. We did not use the isochrones values in one case, EPIC 201214691, where isochrones produced $R_\star=38\pm14 R_{\odot}$, which, although technically within 3$\sigma$, is an extremely inflated radius for otherwise unremarkable stellar parameters, with $T_{\rm eff}=5122\pm82$ K, Fe/H$=-0.049\pm0.25$, and $log(g) = 4.5\pm0.04$. For this system we used the value $R_\star= 0.78\pm0.05 R_{\odot}$ from \citet{Boyajian2012} instead. All stellar radius sources are noted in \autoref{table:stars}.

\subsection{High-Resolution Imaging Constraints}
\label{sec:imaging}

\begin{figure}
\includegraphics[width=0.5\textwidth]{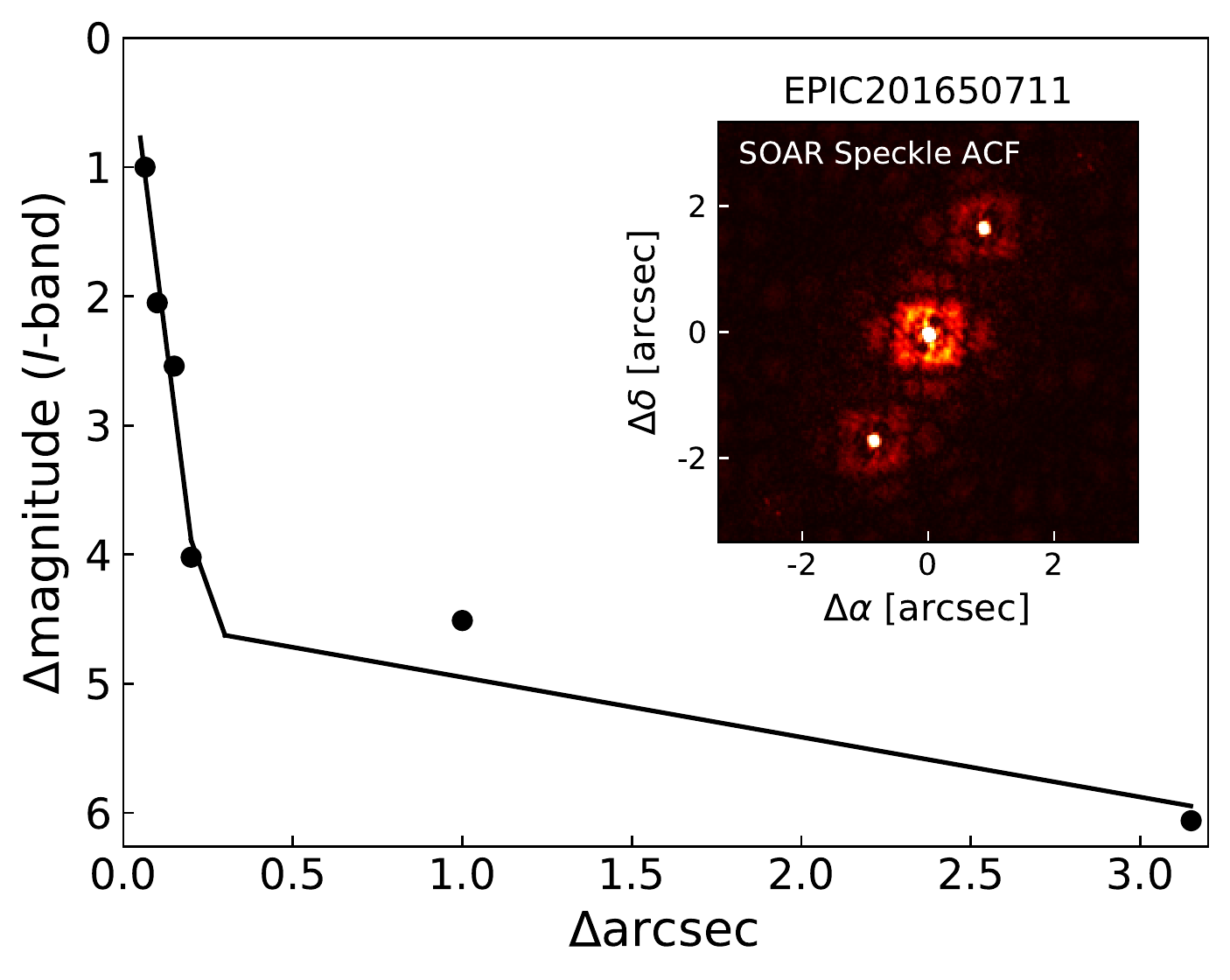}
\caption{The $5\sigma$ detection sensitivity to nearby companions for the SOAR speckle observation of EPIC 201650711 as a function of separation from the target star and magnitude difference with respect to the target. Inset is the speckle auto-correlation function from the observation. A nearby star at 1.79$\arcsec$ separation is mirrored in the image. To determine the star's true quadrant, a shift-and-add routine is performed resulting in a position angle of 333 degrees.}
\label{fig:201650711}
\end{figure}

High resolution images are a critical component of planetary validation efforts \citep[e.g.,][]{2012AJ....144...42A, 2013AJ....146....9A, 2014AJ....148...78D}. In particular, the contrast curves produced by imaging dramatically decrease the physical area in which an undetected stellar binary blend could lurk, and thus increase the a priori odds that a transit signal is planetary. If additional stars are detected, the observations may confirm a false positive \citep[if, for instance, the location of the star matches an observed shift in the photocenter during transit, e.g.][]{2010ApJ...713L.103B}. Not all companion stars signify false positives, however; if the stars have similar magnitudes and the transit is shallow enough, the signal may still be from a planetary-sized body regardless of which star it is around. However, it is critical to know both (a) which star is the host star, since the host star radius determines the planet's radius, and (b) by how much the companion star is diluting the transit. Calculating the dilution factor makes it possible to adjust the light-curve-derived parameters to provide the true planetary radius, which in turn is critical for accurate individual planetary and population analyses. We identified two systems with very close companions (within 4-5\arcsec) in \autoref{table:imaging_comps}. An example contrast curve is shown for EPIC 201650711, which also has a close companion, in \autoref{fig:201650711}.

We identified 28 systems for which imaging data is available using the ExoFOP website. (In a few well-studied cases, multiple contrast curves in similar wavelengths are available but not every curve was used, since additional curves were not needed to securely validate the planet.) Data came from six telescopes: WIYN, Keck, Gemini-South, Gemini-North, Palomar, and SOAR. Within these 28 systems, there are 46 planetary candidates (both USPs and companions), and we have newly validated six, while 27 were previously validated or confirmed, leaving 13 candidates in need of either validation or confirmation. 

\startlongtable
\begin{deluxetable}{cccc}
\tablewidth{0pt}
\tablecaption{Imaging Constraints Used In Validation\label{table:imaging}}
\tablehead{
\colhead{EPIC}  & \colhead{Telescope} & \colhead{Wavelength} & \colhead{Date }
}
\startdata
201089381	& GeminiS DSSI	& 692 (40) nm   & 2017-06-07\\	
201089381   & GeminiS DSSI  & 880 (50) nm   & 2017-06-07\\ 
201110617	& Gemini NIRI   & K             & 2017-01-28\\ 
201130233	& WIYN NESSI    & 832 (40) nm   & 2017-04-03\\ 
201595106	& WIYN NESSI    & 832 (40) nm   & 2017-03-18 \\ 
201637175	& GeminiN DSSI  & 692 (40) nm   & 2016-01-15\\
201637175   & GeminiN DSSI  & 880 (50) nm   & 2016-01-15\\ 
201650711	& SOAR HRCam    & I: 879 (289) nm & 2019-03-18 \\ 
205152172	& WIYN NESSI    & 832 (40) nm   & 2018-07-07 \\ 
206024342	& Keck2 NIRC2   & K             & 2015-08-07\\ 
206103150	& Keck2 NIRC2   & K             & 2015-08-21 \\ 
210707130	& Keck2 NIRC2   & K             & 2015-10-28\\ 
211562654	& Keck2 NIRC2   & K             & 2016-01-21\\ 
212157262	& Keck2 NIRC2   & K             & 2016-01-21\\ 
212303338	& WIYN NESSI    & 832 (40) nm   & 2018-02-01	\\ 
212470904	& WIYN DSSI     & 880 (54) nm   & 2016-04-21 \\ 
212532636	& WIYN DSSI    & 880 (54) nm   & 2016-04-21\\ 
220250254	& WIYN NESSI    & 832 (40) nm   & 2016-10-20\\ 
220256496	& WIYN NESSI    & 832 (40) nm   & 2016-10-21 \\ 
220383386	& GeminiS DSSI  & 880 (50) nm   & 2017-06-09 \\ 
220383386	& GeminiS DSSI  & 692 (40) nm   & 2017-06-09 \\ 
220383386	& WIYN NESSI    & 832 (40) nm   & 2016-11-11 \\ 
220492298	& WIYN NESSI    & 832 (40) nm   & 2016-10-20 \\ 
220554210	& Gemini NIRI   & K             & 2016-10-19\\
220554210	& Palomar PHARO-AO & K short    & 2016-10-19\\
220554210	& WIYN NESSI    & 832 (40) nm   & 2016-10-21 \\
220674823	& Keck2 NIRC2   & K             & 2016-10-16 \\ 
220687583	& WIYN NESSI    & 832 (40) nm   & 2016-11-15 \\ 
228721452	& WIYN NESSI    & 832 (40) nm   & 2017-03-11 \\ 
228732031	& GeminiN NIRI  & K             & 2017-02-19	 \\	
228732031   & WIYN NESSI    & 832 (40) nm   & 2017-04-05\\ 
228801451	& Palomar PHARO-AO & K short    & 2017-07-04\\ 
228813918	& WIYN NESSI    & 832 (40) nm   & 2017-05-12  \\ 
228814754	& GeminiS DSSI  & 692 (40) nm   & 2017-06-10 \\ 
228814754	& GeminiS DSSI  & 880 (50) nm   & 2017-06-10 \\
228836835	& GeminiS DSSI  & 692 (40) nm   & 2017-06-10\\	
228836835   & GeminiS DSSI  & 880 (50) nm   & 2017-06-10 
\enddata
\end{deluxetable}

\clearpage


\begin{deluxetable}{ccccc}
\tablewidth{0pt}
\tablecaption{Close Companions Detected Through Imaging\label{table:imaging_comps}}
\tablehead{
EPIC  &	Distance ($\arcsec$) & Wavelength & $\Delta M$ & Source }
\startdata
201650711	& 1.79    & I        & 1.2  & SOAR\\ 
212303338   & 0.113   & 880 nm   & 1.99 & WIYN \\
212303338   & 0.098   & 692 nm   & 2.45 & WIYN \\
\enddata
\end{deluxetable}


\subsubsection{WIYN and Gemini Speckle Imaging}

We observed targets using two similar speckle imaging instruments at three different telescopes.  The Differential Speckle Survey Instrument \citep[DSSI; ][]{2009AJ....137.5057H} was deployed as a visiting instrument at the WIYN 3.5m, Gemini North 8m and Gemini South 8m telescopes.  The NN-EXPLORE Exoplanet Stellar Speckle Imager \citep[NESSI; ][]{2018PASP..130e4502S} was installed at WIYN in late 2016 to replace DSSI at that facility. Both DSSI and NESSI operate by splitting the incoming light into blue and red channels and recording simultaneous images of each target with an Andor iXon Ultra 888 EMCCD camera on each channel.  For speckle imaging, we acquire a dataset consisting of a small number (1-10) of 1000-frame datacubes on each target.  The number of 1000-frame sets taken normally depends on target brightness and observing conditions. Each frame is 40 ms long at WIYN and 60 ms at the Gemini telescopes, exposure times intended to maximize signal-to-noise for the typical seeing at each site.  Along with datasets taken of science targets, we observed nearby single stars to serve as point source calibrators.  For each dataset, the fields of view are confined to 256x256 pixel subarray readouts of the full EMCCDs.  These narrow fields are sufficient to cover the typically $\sim$2 arcsec wide isoplanatic atmospheric patch centered on the target with a plate scale that critically samples the diffraction-limited resolution at each telescope.  At Gemini, the field-of-view is 2.8x2.8 arcsec with a plate scale of 0.011 arcsec/pixel and at WIYN it is 4.6x4.6 arcsec with a plate scale of 0.018 arcsec/pixel.  Filters with bandpass widths of 40-54 nm are used in each channel.  For these data, filters with central wavelengths at 692, 832 and 880 were used.  

A standardized data reduction pipeline was applied to all of these speckle data.  The basic reduction and production of high-level data products was previously detailed by \citet{2011AJ....142...19H}.  The high-level data pipeline products consist of a reconstructed, diffraction-limited resolution image of the target and surrounding field in each filter, relative photometry and astrometry of any companion sources detected near the target and a contrast curve that gives the detection limit for point sources as a function of angular separation from the target.

\subsubsection{SOAR Speckle Imaging}
\label{sec:soar}

We searched for stellar companions to EPIC 201650711 with speckle imaging using the HRCam instrument on the 4.1-m Southern Astrophysical Research (SOAR) telescope \citep{2018PASP..130c5002T} on 18 March 2019 UT. Observations were in Cousins I-band, a similar visible bandpass to \textit{Kepler}. The $5\sigma$ detection sensitivity and speckle auto-correlation functions from the observation are shown in \autoref{fig:201650711}. A 1.2 magnitude fainter companion was detected in the SOAR observation at a separation of 1.79 arcsec. This companion also appears in the Gaia DR2 catalog (GAIA DR2 source id 3812335125094701056, $g'=13.79$, $\Delta g' =1.35$) and has a comparable parallax and proper motion to the target star, indicating the two are likely associated; see Section \ref{sec:uncertainstars} for further discussion of this system.

\subsubsection{Keck and Palomar Imaging with Near-Infrared Adaptive Optics}

For high resolution imaging in the near-infrared, we utilized two adaptive optics systems: the first was on the Hale 5m telescope and the second was on the Keck-II 10m telescope.

The Palomar Observatory observations were made of EPIC 220554210 and EPIC 228801451 with the PHARO instrument \citep{2001PASP..113..105H} behind the natural guide star AO system P3K \citep{2013ApJ...776..130D} in 2016 and 2017 (see \autoref{table:imaging}).  The observations were taken in a standard 5-point quincunx dither pattern with steps of 5\arcsec\ in the narrow-band $K_{short}$ filter $(\lambda_o = 2.145)$.  Each dither position was observed three times, offset in position from each other by 0.5\arcsec\ for a total of 15 frames; with an integration time of 30 seconds per frame for EPIC 220554210 and 3 seconds per frame for EPIC 228801451. PHARO has a pixel scale of $0.025\arcsec$ per pixel for a total field of view of $\sim25\arcsec$. 

The Keck Observatory observations of five USP candidates (\autoref{table:imaging}) were made with the NIRC2 instrument on Keck-II behind the natural guide star AO system \citep{2000PASP..112..315W} in 2015 and 2016.  The observations were made in the standard 3-point dither pattern that is used with NIRC2 to avoid the left lower quadrant of the detector which is typically noisier than the other three quadrants. The dither pattern step size was $3\arcsec$ and was repeated twice, with each dither offset from the previous dither by $0.5\arcsec$.  NIRC2 was used in the narrow-angle mode with a full field of view of $\sim10\arcsec$ and a pixel scale of $0.0099442\arcsec$ per pixel.  The Keck observations were made in the $K$ filter $(\lambda_o = 2.196; \Delta\lambda = 0.336~\mu$m). 

The AO data were processed and analyzed with a custom set of IDL tools.  The science frames were flat-fielded and sky-subtracted.  The flat fields were generated from a median average of dark subtracted flats taken on-sky.  The flats were normalized such that the median value of the flats is unity.  The sky frames were generated from the median average of the dithered science frames; each science image was then sky-subtracted and flat-fielded.  The reduced science frames were combined into a single combined image using a intra-pixel interpolation that conserves flux, shifts the individual dithered frames by the appropriate fractional pixels, and median-coadds the frames.  The final resolution of the combined dithers was determined from the full-width half-maximum (FWHM) of the point spread function with typical final resolutions of 0.1\arcsec\ and 0.05\arcsec\ for the Palomar and Keck observations respectively.

To within the limits of the AO observations, no stellar companions were detected. The sensitivities of the final combined AO image were determined by injecting simulated sources azimuthally around the primary target every $20^\circ $ at separations of integer multiples of the central source's FWHM \citep[][Lund et. al 2020, submitted]{Furlan2017}. The brightness of each injected source was scaled until standard aperture photometry detected it with $5\sigma $ significance. The resulting brightness of the injected sources relative to the primary target set the contrast limits at that injection location. The final $5\sigma $ limit at each separation was determined from the average of all of the determined limits at that separation and the uncertainty on the limit was set by the RMS dispersion of the azimuthal slices at a given radial distance.

\section{Validation}
\label{sec:validation}

To validate a candidate as a planet, we used the \emph{vespa} software package \citep{2012ApJ...761....6M} to compare the relative likelihood that a transiting signal is caused by a planet vs.~a false positive (typically some kind of eclipsing binary star blend). Validation employs many kinds of observational constraints, including the photometric light curve for each candidate, the broadband photometric magnitudes of the star, the depth of a potential secondary transit (which must be consistent with a planetary occultation), spectroscopically derived stellar parameters ($T_{\rm eff}$, [Fe/H], $\log(g)$), and, where available, image contrast curves constraining the magnitude and distance of nearby but undetected stars that could contribute to a false positive (see Section \ref{sec:imaging}). The \emph{vespa} code provides an overall false-positive probability (FPP) for each candidate. For candidates with multiple candidates detected in the same system, a multi-candidate boost has been applied for targets with sufficient SNR. The threshold for applying this boost is conservatively defined, following \citet{2019AJ....157..143B}, to be SNR$>13$ for candidates with fewer than 10 transits and SNR$>11.9$ for candidates with 10 or more transits. The boosted FPP is noted separately from the individual FPP in \autoref{table:planets}.

Statistical validation may not be reliable for objects detected with low to moderate SNR and/or few transits \citep{2018AJ....155..210M, 2019AJ....157..143B}. While USPs have abundant transit numbers even in a single \ktwo\ campaign, the same is not always true of their longer period companions; 13 companions in our sample have fewer than 10 transits available. Another more nuanced issue with validation was noted by \citet{2019AJ....157..143B} and relates to the systematic false alarm rate (SFA), the rate at which spurious signals may be identified as transit-like by a given detection pipeline. Assumptions in the validation framework about the low SFA rate, which were developed using analyses of the \kepler\ main mission pipeline, may not directly translate to \ktwo\ data. In particular, in this survey the detection threshold has been pushed to lower limits (SNR=7) to identify more of the rare USP candidates. Performing a full SFA analysis falls outside the scope of the current work; it also is not possible to directly apply the approach of \citet{2019AJ....157..143B} to the single-pointing campaigns of \ktwo, e.g., permuting the order of multiple \kepler\ quarters. We did however perform a sanity check on our SFA rate using an inversion test of a single campaign, Campaign 2 (chosen for having average noise and number of transiting candidates). With inversion, the flux values are flipped around the nominal baseline level, so that valleys appear as peaks and vice versa, to create realistic noise without real signals. Of 13,394 stars in Campaign 2, 1,735 were identified in the inverted data by our pipeline as having periodic signals with $P<1$ d and SNR $>7$; most were immediately rejected for failing our initial depth and duration cuts, leaving 110 for visual inspection. Of those 110, almost all were easily rejected as non-planetary by the shape of the signal (most are sinusoids). Just 2 were passed along for further inspection and diagnostic tests, which they both failed (the pattern and amplitude of the out-of-transit variability did not match a planetary transit). Neither would have been identified as a candidate and sent to our pipeline for fitting and further tests. We conclude that our systematic false alarm rate is low enough that we may neglect it for the purposes of statistically validating planetary candidates. However, we have adopted a conservative stance toward candidates near the margins of validation and at low to moderate SNR, following the recommendations of \citet{2019AJ....157..143B}.

Planetary validation proceeds as follows. For each candidate, we calculate a false positive probability (FPP) using \emph{vespa}. If multiple candidates are detected in a system, each candidate with sufficient signal to noise (SNR$>13$ if $N<10$ transits and SNR$>11.9$ otherwise) provides a statistical boost to the FPPs of other candidates in the system, decreasing the FPP by a factor of 10 (if the system contains two qualifying planets) or 15 (three or more planets). A planet is declared validated if it has FPP $\le 1\%$, or a likely false positive if FPP $\ge 90\%$, following e.g. \citet{2015ApJ...809...25M}. Out of an abundance of caution, we do not declare candidates to be validated if they have SNR $<10$ even if their FPP is low, following \citet{2019AJ....157..143B}; this affects six candidates with low SNR (between 3-7): EPIC 206024342 p4, EPIC 201609593 p2, EPIC 201085153 p2, EPIC 220554210 p3, EPIC 220256496 p2, and EPIC 201650711 p2. We also have left as candidates three USPs with high FPPs but low SNR: EPIC 214189767, EPIC 219752140, and EPIC 228739208.

We validated \numvalid\ of \numdetected\ candidate planets, including \newlyvalid\ planets that had not been validated before. This number includes \newlyvalidusp\ newly validated USPs (EPIC 201427007 p1, EPIC 201595106 p1, EPIC 206024342 p1, EPIC 206042996 p1, EPIC 212624936 p1, EPIC 220492298 p1, and EPIC 228836835 p1) and \newlyvalidcomp\ companions (EPIC 206024342 p2, EPIC 206024342 p3, and EPIC 212624936 p2). Meanwhile, we found \numhighfpp\ candidates, all USPs, with better-than-even odds of being a false positive (FPP$>50\%$), of which \numfp\ have FPP$>90\%$. Most of the likely false positives lack any imaging or spectroscopic constraints, which are recommended where possible before finally ruling the candidates out, since ruling out the presence of nearby background stars can dramatically change a candidate's FPP.

\section{Discussion of Individual Candidates}
\label{sec:candidates}

We will now discuss the characteristics and dispositions of select individual candidates and classes of planets. 

\subsection{All validated planets are small}

We start with a few important categories where we did not find promising candidates: ultra-hot Jupiters and planets in the short-period desert ($R_p$ between 3-10 $R_\oplus$). Our \numusp\ USP candidates are almost uniformly small, and we did not validate any candidate with $R_{\rm p} > 2.4\, {\rm R_\oplus}$. 

\subsubsection{A single, faint ultra-hot Jupiter candidate}
\label{sec:uhj_cand}

\textbf{EPIC 210605073}, from Campaign 4, was previously reported as a candidate by \citet{2016AJ....152...47A}. There are no spectroscopically-derived stellar parameter values for $T_{\rm eff}$, [Fe/H], and $\log(g)$ because it is too faint for follow-up observations (${\rm Kep} = 17.887$). This system is our largest candidate and the only ultra-hot Jupiter. For its size, its period is short enough ($P=0.567\,$days) that its confirmation would be highly significant, since tidal effects are quite strong on giant planets at this distance; the shortest-period confirmed ultra-hot Jupiter, NGTS-10 b, has $P=0.767\,$days (\autoref{fig:period_radius}). For the sake of completeness, we attempted to validate the system despite the highly uncertain stellar parameters. We inferred a value of $T_{\rm eff}=7918$ K from the available stellar photometric magnitudes ($g=17.859$, $r =17.896$, $i=18.004$, $z=18.125$) and the corresponding temperatures in Table 1 from \citet{2012ApJS..199...30P}; the number we use is the median of the three estimates for $T_{\rm eff}$ using the differences in each band, using 500 K for the error bar. We arbitrarily assigned  $[Fe/H]=0.0\pm0.5$, and $\log(g)=4.0\pm0.5$. Even with these favorable guesses, \emph{vespa} returned a high FPP of 0.6, and this object remains an uncertain, though unlikely, candidate.

\subsubsection{Few plausible short-period desert candidates}
\label{sec:desert}

We do not find any compelling short-period desert candidates in this work. Just three candidates are within 1-$\sigma$ of the lower radius edge of the desert ($R_p>3~R_{\oplus}$): EPIC 201085153,  EPIC 201089381, and EPIC 203533312. 

\textbf{EPIC 201085153} ($R_p = 2.4\pm1~R_{\oplus}$, $P= 0.26 $ d, Kep=17.3) is a candidate planet in a two-candidate system (see also section \ref{sec:201085153}), but its planetary radius is low enough and its radius error high enough that it likely falls below the lower edge of the short-period desert. 

The only other plausible short-period desert candidates are \textbf{EPIC 201089381} ($R_p = 2.8\pm0.5~R_{\oplus}$, $P=0.16$ d, Kep=14.3) and \textbf{EPIC 203533312} ($R_p = 3.5\pm0.3~R_{\oplus}$, $P=0.176 $ d, Kep=12.2), which are also the two shortest-period planets in our sample. These two candidates are so close to their stars that the validation software fails to converge, with \emph{vespa} returning a ``Roche Lobe'' error, indicating that its orbital solutions are all necessarily unstable. Given that the location of the Roche limit depends very strongly on the stellar and planetary parameters, a closer examination of these systems is required to determine if these are legitimate planets that we have caught on the brink of disruption \citep{2016CeMDA.126..227J}, or (more likely) a pair of false positives.

\subsection{Shortest period planet that we validated}

We successfully re-validated \textbf{K2-137 b, aka EPIC 228813918}, at $P=0.1797\,$d, or 4.3 hr \citep{2018MNRAS.474.5523S}, with a low ${\rm FPP} = 0.001$. Despite its proximity to its star, it did not raise any ``Roche Lobe'' errors from \emph{vespa} as in Section \ref{sec:desert}. This planet is the second-shortest period of any confirmed planet around a main sequence star, after KOI 1843.03 at 4.2 hr \citep{2013ApJ...773L..15R}.

\begin{figure}
\includegraphics[width=\textwidth]{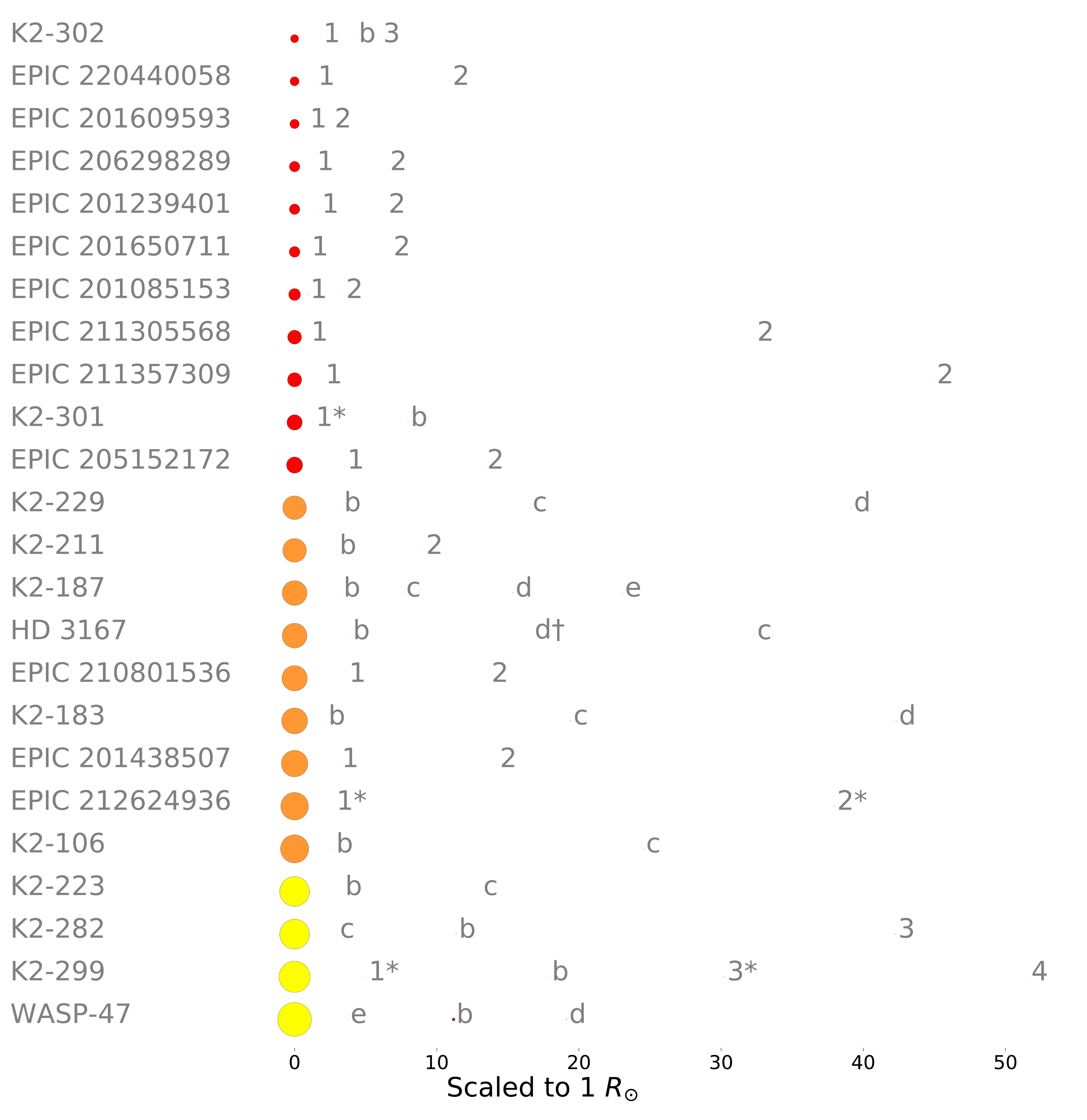}
\caption{All known USP multi-planet systems for Campaigns 0-8 and 10, plotted entirely to scale. All sizes and distances are in solar radii ($R_{\odot}$). Stars are color coded by spectral type using stellar radius as a proxy}: red=M ($R_\star\le0.7~R_{\odot}$), orange=K ($0.7<R_\star\le0.96~R_{\odot}$), and yellow=G ($0.96<R_\star\le1.15~R_{\odot}$). No USPs were found around F stars. Numbers denote candidates and letters denote confirmed or validated worlds. Planets with an asterisk (*) have been newly validated in this work but are pending an official letter designation. One non-transiting planet is marked with a dagger ($\dagger$) and plotted with zero radius at the correct separation from the star.
\label{fig:multi_plot}
\end{figure}

\subsection{Twice as many known multi-planet systems}
\label{sec:multis}

A plot of all \nummultisystems\ multi-planet systems in \ktwo\ Campaign C0-8 and 10, which are known to host at least one USP, is shown in \autoref{fig:multi_plot}. The extremely small semi-major axes for these systems make it possible to plot the stars and planets with the physical sizes and separations shown entirely to scale.

\begin{figure}
\includegraphics[width=\textwidth]{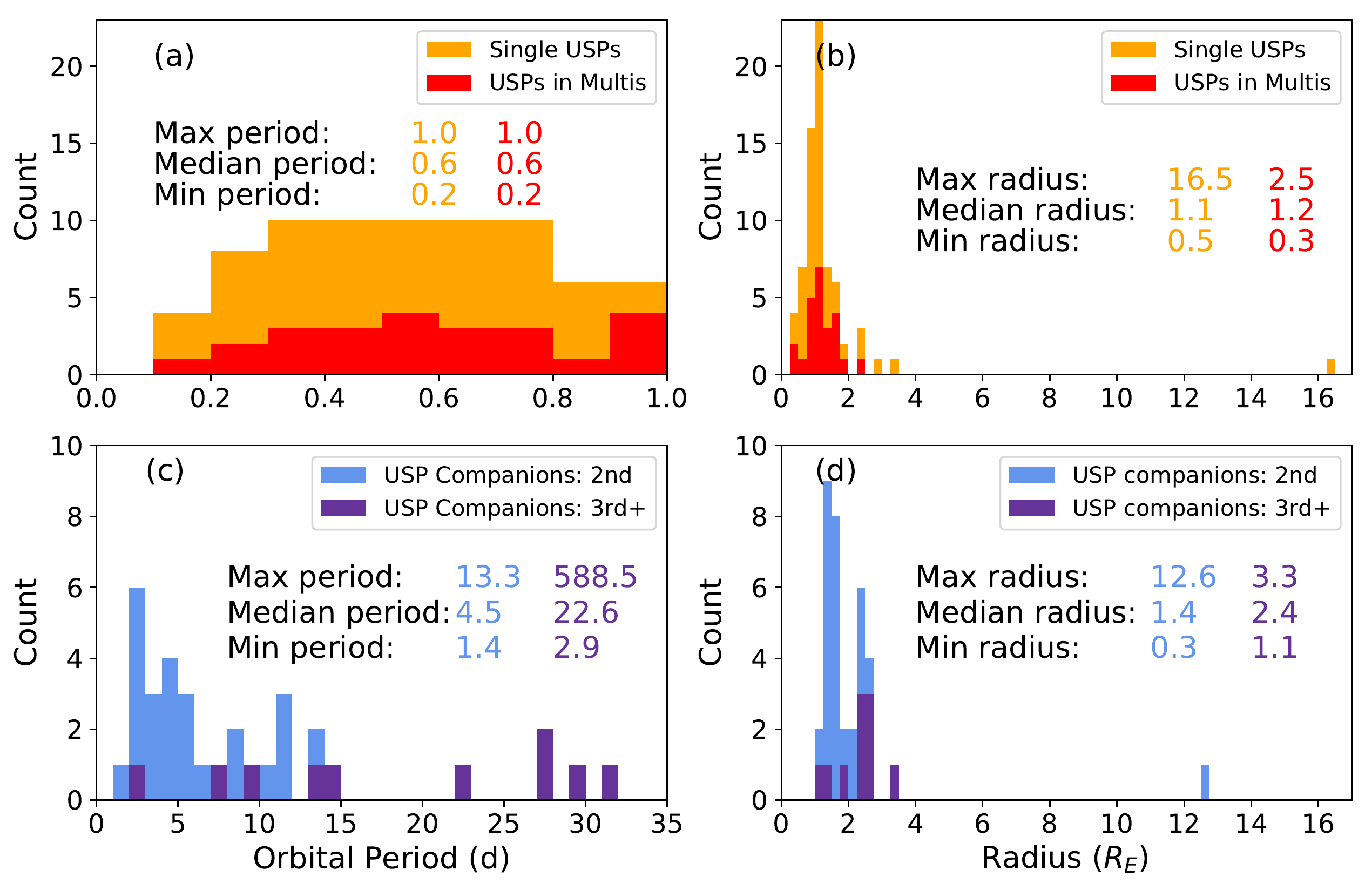}
\caption{Stacked histograms of properties of our population of 74 ultra-short-period planets and their 23 known companions. (a) Distribution of orbital periods for apparently single USPs (orange) and those in multi-planet systems (red). (b) Distribution of planetary radius for apparently single USPs (orange) and those in multi-planet systems (red). (c) Distribution of orbital periods for companion planets  second closest known planet (blue) and any additional companions (purple). Two companions that do not transit but have known periods from radial velocity, HD 3167 d and WASP-47 c, are included in the period statistics (though WASP-47 c is off the plot at 588 d); the most distant transiting companion is K2-229 d, at 31 d \citep{2018NatAs...2..393S}. (d) Distribution of planetary radius for companion planets to USPs, showing both the second closest known planet (blue) and any additional companions (purple). Non-transiting companions have no known radius values, and are not plotted. }
\label{fig:period_radius_hist}
\end{figure}

We find that our sample of ultra-short-period planets commonly have neighbors, with \nummultisystems\ out of \numusp\ (\multirate\%) candidates in this paper having additional transiting companions. We have found transiting companions with periods from 1.4 to 31 days, with a median of 4.5 d; two systems are also known to have radial-velocity companions, with the most distant one at 588 days (see \autoref{fig:period_radius_hist}). In half of the multi-candidate systems (\newcompanions) either the USP or the companion is new, and \newmultisystems\ represent entirely new candidate multi-planet systems.

With the new discoveries, the estimated occurrence rate of multi-planet systems for our \ktwo\ sample is comparable to the rate of all \kepler\ planets (with $R_p=0.5-10~R_{\oplus}$) at somewhat longer periods (P=3-300 d) \citep[24\% have 2+ planets,][]{2021AJ....161...16H}; a nearly identical observed rate was found by \citet{2019MNRAS.483.4479Z}, who placed the expected number of planets at $<N_p>=5.86$ planets per star ($P < 500$ d). For the shorter periods ($P<50$ d) that are more directly comparable to the USPs and companions in our survey, they found the expected number of planets per star to be 1.34 \citep{2019MNRAS.483.4479Z}, in close agreement with prior work by \citet[][$<N_p>=1.36$]{2011ApJ...742...38Y}. The number of planets per star in our survey is 1.41. Thus, at first glance the multiplicity of ultra-short-period planets is consistent with that of more distant planetary systems.

\subsubsection{Four Entirely New Multi-Planet Systems}

\textbf{EPIC 201609593}, from Campaign 1, has two new candidates ($P=0.318$ d and $P=1.39$ d). Both were detected with marginal SNR=9 and 4, respectively. The USP is among the smallest candidates in our sample ($R_p=0.5\, R_{\oplus}$) and has one of the shortest periods ($P=7.6$ hr), while the companion is the shortest-period companion to a USP candidate yet found. (No system has yet been found with two USPs.) Neither planet was validated due to their low SNR.

\textbf{EPIC 220440058}, from Campaign 8, has two new candidates ($P=0.627$ d and $P=3.836$ d). The USP was robustly detected with SNR=17, while the companion has SNR=7. No contrast curves are available, and both remain unvalidated as we do not apply a multiplicity boost due to the companion's marginal SNR.

\label{sec:201085153}
\textbf{EPIC 201085153}, from Campaign 10, has two new candidates ($P=0.257$ d and $P=2.26$ d). The USP was detected with high SNR=26, though the companion only has SNR=4. We note that further follow-up of this system is difficult since it is a faint (${\rm Kep}=17.3$, $J=15.8$) M dwarf ($T_{eff}=3945\pm386$ K).

\textbf{EPIC 201438507}, from Campaign 10, has two new candidates ($P=0.327$ d and $P=2.536$ d). Although the USP was detected with SNR=13, the companion has a very low SNR=3.4.  The USP has not been validated; no contrast curves are available, though the star is sufficiently bright for future observations (${\rm Kep} = 14.067$).

\subsubsection{New USP and known longer-period candidate newly validated}

\textbf{EPIC 212624936 (2 planets)}, from Campaign 6: \citet{2019ApJS..244...11K} reported a long-period candidate at $P=11.81$ d and SNR=10, and we found a new USP candidate at $P=0.57$ d and SNR=12.9. Both planets have been validated for the first time in this work.

\begin{figure}
\includegraphics[width=\textwidth]{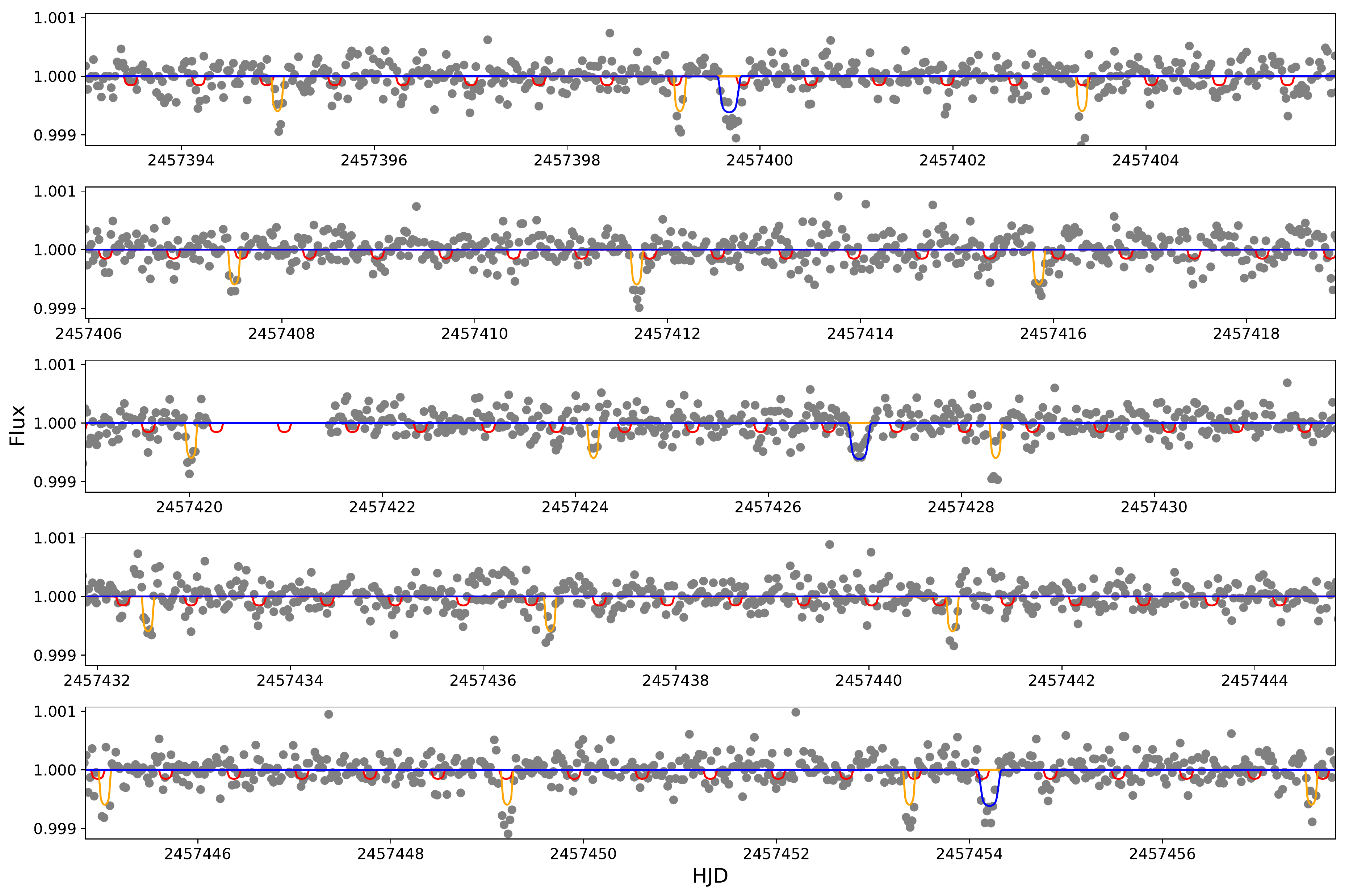}
\caption{K2-282, aka EPIC 220554210, with three known planets. The USP is planet b (red) and is too small to be detected in individual transits, unlike c (orange) and d (blue).}
\label{fig:k2282}
\end{figure}

\subsubsection{Seven new companions to previously reported USP candidates}
\label{sec:newcomps}

\textbf{EPIC 201239401}, from Campaign 1: the USP was reported by \citet{2016ApJS..222...14V} and has SNR=24, and we have newly identified a companion at $P=6.784$ d (low SNR=6). Due to the low SNR of the companion we have not applied a multiplicity boost, and both remain candidates.

\textbf{EPIC 205152172}, from Campaign 2: the USP \citep{2016ApJS..222...14V}, with SNR=28, has a newly identified companion at $P=3.225$ d (SNR=10). Both the USP and the companion have been validated after applying a multiplicity boost.

\label{sec:206215704}
\textbf{EPIC 206215704/K2-302}, from Campaign 3: An ultra-short-period planet with $P=0.623$ d was first reported by \citet{2016ApJS..222...14V}, while two candidate outer planets, but not the USP, were reported by \citet{2019ApJS..244...11K}. We have confirmed the validation of the planet at $P=2.25$ d, assigned the planetary designation K2-302 b by \citet{2019A&A...627A..66H}. However, we do not find the signal at 3.1 days, which is spurious and likely an alias ($3.1 = 5\times 0.62$ d, the USP period);  \textbf{instead we find a different} third planetary signal at $P=2.9$ d. The planets have SNR = 14, 13, and 12, ordered by distance from the star. We have verified that three separate transit signals are present in the data and fit each candidate separately. 
Even after applying a multiplicity boost, the USP and outermost companions narrowly missed validation, with FPP=0.017 and 0.011 respectively.

\textbf{EPIC 206298289}, from Campaign 3: the USP at $P=0.43$ d \citep{2016ApJS..222...14V} has a newly identified companion at $P=3.215$ d, with SNR=26 and 7.8, respectively. Due to the marginal SNR of the companion, we have not applied a multiplicity boost, and both remain candidates.

\textbf{EPIC 210801536}, from Campaign 4, has two candidates ($P=0.893$ d and $P=2.536$ d), with the USP previously reported by \citet{2019ApJS..244...11K}. Both candidates were detected at marginal SNR=9.6 and 4.4, respectively. There are no contrast curves available to provide constraints on background stars. The host star appeared to be a moderately evolved G or K star, with MAST values of $\log g=3.83\pm0.76$ and $R_\star = 2.0 \pm 1.7\, R_{\rm Sun}$; however, the isochrones value for the stellar radius is $0.85_{-0.09}^{+0.28}$, which we have used. Follow-up observations to improve the stellar parameters and search for nearby background stars are recommended, as the host star has magnitudes $Kep=13.176$ and $J=11.7$.

\textbf{EPIC 220256496/K2-221}, from Campaign 8, has two candidates. The USP ($P=0.66$ d) was reported by \citet{2018AJ....155..136M}, and we have found a new candidate with $P=2.601$ d. We validate the USP at SNR=21, but the companion has low SNR=8.7, and hence remains a candidate.

\label{sec:220554210}
\textbf{EPIC 220554210/K2-282}, from Campaign 8, has three planets. \citet{2019ApJS..244...11K} reported two planets, at $P=4.17$ and $P=0.705$ d \citep{2019ApJS..244...11K}, and \citet{2019A&A...627A..66H} confirmed the USP as K2-282 b. We did not identify the USP in our search because the dominant signal is from the planet at $P=4.17$ d. The larger planet did show up in our longer-period search out to $P=3$~d, at an alias at $P=2.08$ d or half its true period, but it was evidently not a USP and this system was set aside as outside our survey parameters. After finding this system during a review of the literature, we easily recovered the two known planets, and discovered a third signal at $P=27$ d, which has three transits during 78 days of data. A plot of the transits of each of these three systems is shown in \autoref{fig:k2282}. The planets have SNR=21, 18, and 8.6 respectively. All three planets were validated independently, even without a multiplicity boost.

\subsubsection{Three newly validated planets and an additional, low-SNR candidate for K2-299}

\textbf{EPIC 206024342/K2-299}: Three candidates have been previously reported in this system ($P=0.91$ d, $P=4.507$ d, and $P=14.637$ d) \citep{2016ApJ...829L...9V, 2019ApJS..244...11K}. The second planet ($P=4.5$ d) was previously validated as K2-299 b by \citet{2019ApJS..244...11K}; we have validated the other two for the first time. By distance from the sun, the candidates have SNR=11.2, 11.8, and 15. We have also identified a fourth transit-like signal at $P=27.7$ d, albeit with low SNR=3.6. Although the fourth candidate received a low FPP from \emph{vespa}, it remains a candidate due to its low SNR and long period, with only three transits detected.

\subsubsection{Two systems with companion candidates observed in Campaigns 5 and 18}
\label{sec:problematic}

Although the scope of this paper is confined to \ktwo\ Campaigns 0-8 and 10, two candidate planets were found in systems from Campaign 5 that were also observed in Campaign 18. We conditioned and searched the Campaign 18 data for just those two systems in the same manner as the rest of our data, recovering the known USP candidates at high SNR. In both cases, previous tentative detection of longer-period candidates were made more secure by the addition of the extra campaign of data. (Coincidentally, the companions both have periods near $P=11.6$ d, though the transit midtimes do not line up.)

\begin{figure}
\includegraphics[width=0.5\textwidth]{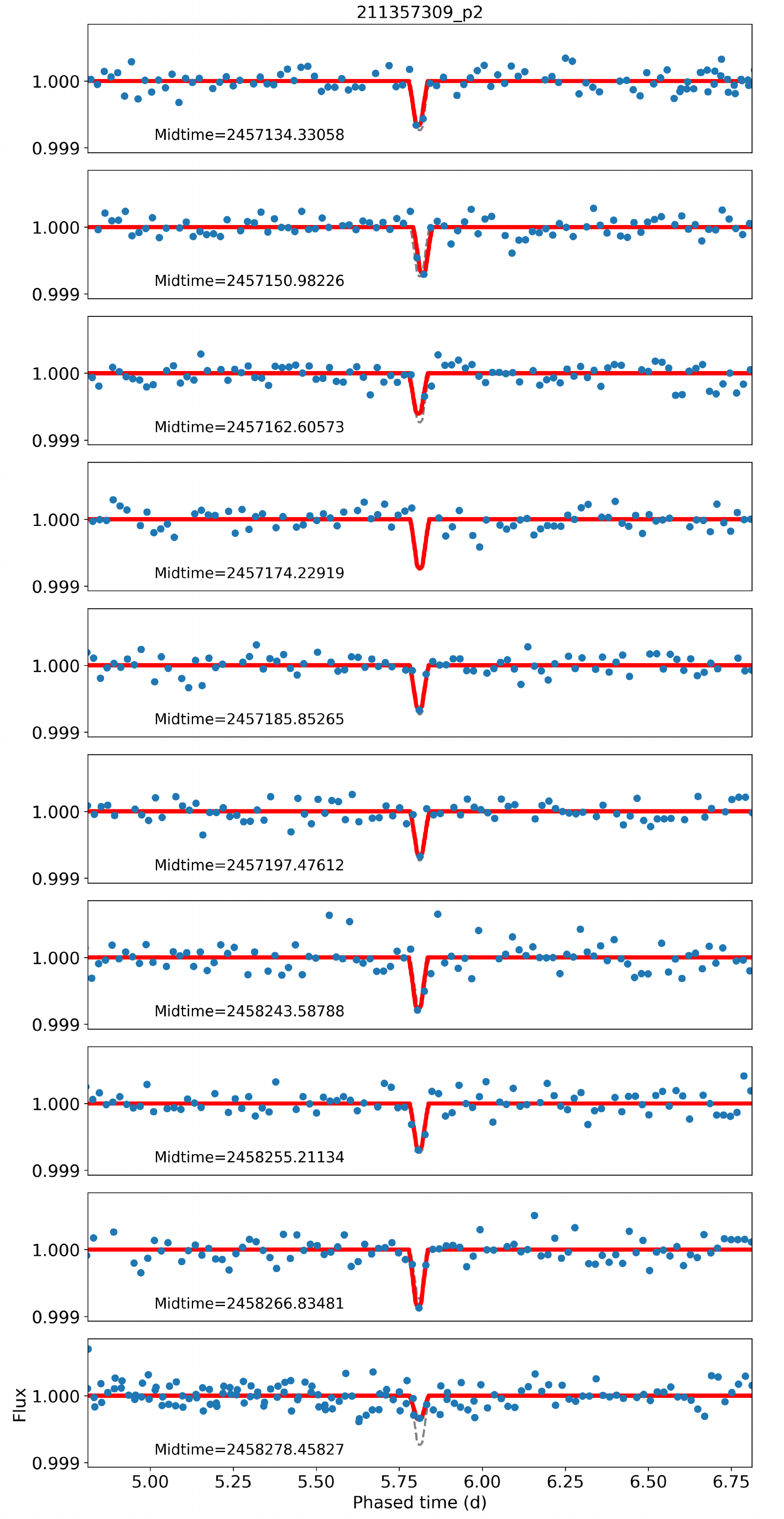}
\caption{Candidate second planet around EPIC 211357309, with $P=11.61$ d, combining data from Campaign 5 and 18. The best fits to each individual transit are shown in red, while the common fit to all transits is a grey dashed line.}
\label{fig:211357309}
\end{figure}

\textbf{EPIC 211357309}, from Campaign 5 and 18, has a known USP candidate at $P=0.46$ d with SNR=26 \citep{2016AJ....152...47A, 2017AJ....154..207D}. We also identified a new candidate, at $P=11.61$ d. We combined data from Campaigns 5 and 18 to search for the best period for both candidates, and found that the outer companion was detected during 9 transits across both campaigns, at SNR=8.8 (\autoref{fig:211357309}). Neither candidate is validated, since the SNR of the exterior candidate precludes a multi-planet boost.

\begin{figure}
\includegraphics[width=0.5\textwidth]{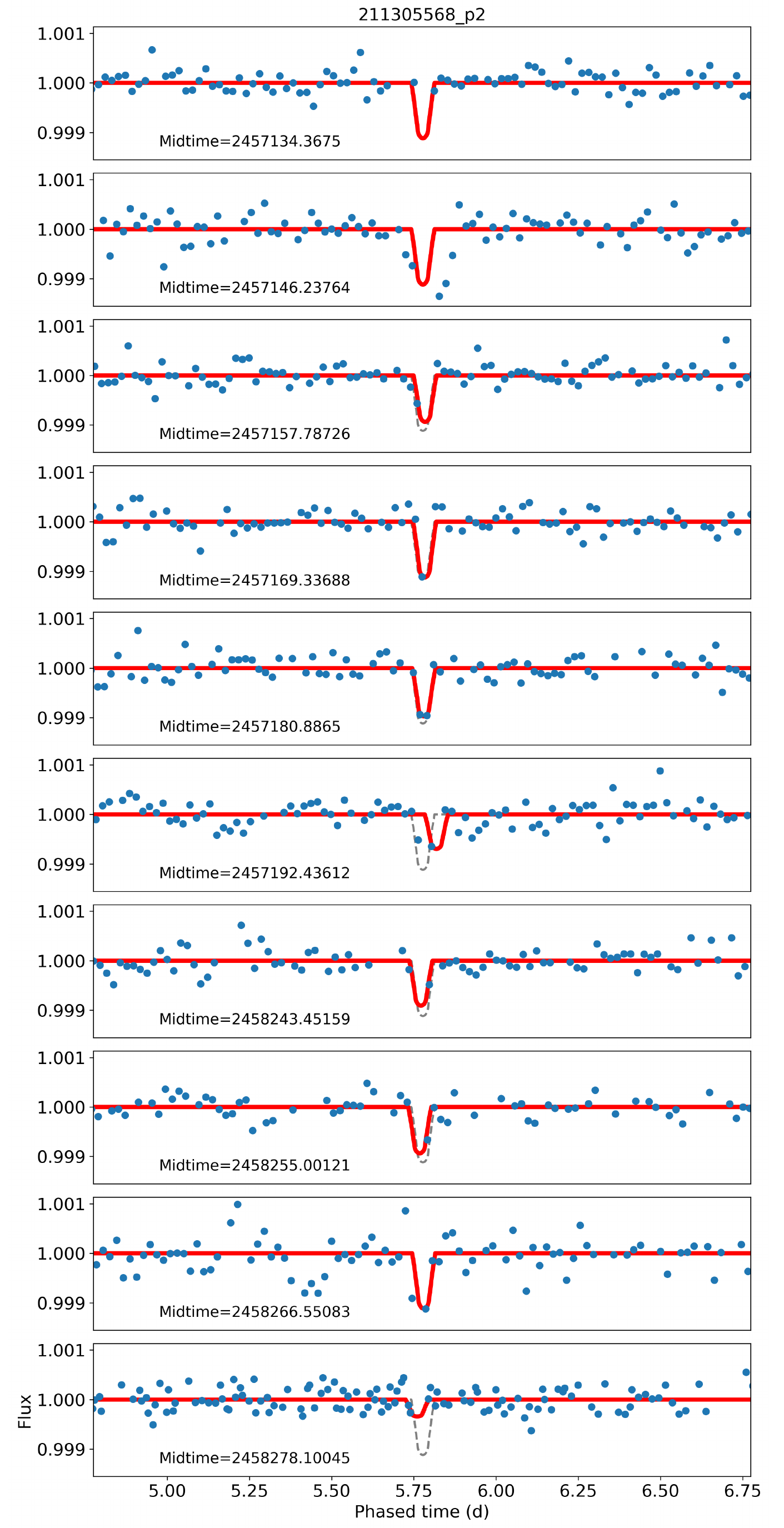}
\caption{Candidate second planet around EPIC 211305568, with $P=11.55$ d, combining data from Campaign 5 and 18. The best fits to each individual transit are shown in red, while the common fit to all transits is a grey dashed line.}
\label{fig:211305568}
\end{figure}

\textbf{EPIC 211305568}, also from Campaign 5 and 18, has two candidates identified by \citet{2017AJ....154..207D}, who calculated high false positive probabilities for both candidates. Our FPPs are similarly high, and due to the low SNR (4.6) of the outer companion, we did not 
apply a multi-planet boost to the USP (at SNR=15). There is a nearby faint stellar companion (4.8 arcsec, $\Delta J=2$, from 2MASS via ExoFOP website), which we have not corrected for since its dilution factor in visible wavelengths is unknown. The transits of the outer planet are show in \autoref{fig:211305568}.


\subsubsection{USP we originally missed in a system with a deeper transit of an exterior planet}
\label{sec:228721452}

\textbf{EPIC 228721452/K2-223} is a multi-planet system ($P_b=0.505$ d and $P_c=4.56$ d) identified by \citet{2018AJ....155..136M} and \citet{2018AJ....156...78L}. In our initial survey out to longer periods ($P<3$ d), we clearly identified a signal at $P=2.28$ d (an alias of half the true period of K2-223 c) with $SNR=15$. However, we did not  identify the USP  (SNR=10) until after reviewing the literature. We easily identified both planets and found fitted parameters are almost identical to those of \citet{2018AJ....155..136M} and \citet{ 2018AJ....156...78L}. We have validated both planets.

\subsection{Problematic Cases}

\subsubsection{Validated and confirmed candidates that we don't find and/or re-validate}
\label{sec:missedvalid}

\textbf{EPIC 228801451}, aka K2-229, from Campaign 10, has three candidates at $P=0.58$ d, $P=8.33$ d, and a single transit with $P=31\pm1$ d \citep{2018NatAs...2..393S, 2018AJ....156...78L}. Although we easily identified and validated the USP, neither of the longer period planets appears in our data. Our light curve of the USP also contains points that appear, in unbinned plots, to be artificially close to zero, due to some unknown processing problem in the smoothing step. Both of these issues are likely due to the star being active and variable; the K2-229 system required special processing to confirm in \citet{2018NatAs...2..393S}.

\textbf{EPIC 220250254}, aka K2-210, from Campaign 8, was validated as K2-210 b \citep{2018AJ....155..136M} with FPP=$2.31\times10^{-4}$; our FPP was high enough (FPP=0.036) that it remains a candidate, with a background eclipsing binary (BEB) the next-most likely hypothesis. It's not clear where the discrepancy lies, and a higher-contrast image than the WIYN curve in \autoref{table:imaging} or other follow-up is recommended.

\subsubsection{Two candidates with uncertain stellar host parameters}
\label{sec:uncertainstars}

\textbf{EPIC 201650711}, from Campaign 1: We originally flagged EPIC 201650711 p1 as a false positive in a single planet system with an FPP of 0.92. However, during the course of this work we reexamined the system after a second planet candidate was reported by \citet{2019ApJS..244...11K} at $P=5.54$ d (which did not note the presence of a USP candidate). We call the USP candidate ``p1'' and the longer-period candidate ``p2'' to avoid confusion, since both have been referred to as candidate 1 in the literature \citep{2016ApJS..222...14V, 2016AJ....152...47A,2019ApJS..244...11K}. We have verified that both sets of transits are present in the data (with SNR = 18 and 9.4, respectively). As noted in Section \ref{sec:soar}, there is a close, likely bound GAIA companion star 1.79\arcsec\ away. No significant odd-even differences were detected in the transit depths for either p1 or p2 (odd-even $\sigma=$ 0.01 and 0.06 respectively); no additional stellar companions were found with high-resolution imaging by SOAR (\autoref{fig:201650711}).

Unfortunately, we have not been able to determine which star hosts the candidates (or indeed if they are in the same system); an analysis of the centroid positions of the stars using the python package \emph{lightkurve} did not turn up any significant shifts during transit, not unexpected given the close proximity of the stars (less than half a \kepler\ pixel). The properties of the fainter star are largely unknown apart from having the same GAIA DR2 proper motion (${\rm RA}=80.1\pm0.3$ mas/yr,${\rm dec}=-29.1\pm0.3$ mas/yr) and parallax ($10.8\pm0.2$ mas) as the brighter star (all values within 1-2$\sigma$), indicating a likely bound companion. 

For either host star scenario, we estimated the effect of light curve dilution due to the other star on the calculated planetary radii. For transits around the brighter star (s1), the radius must be increased by $\sim15\%$, to $R_{p1,s1}=0.36 R_{\oplus}$ and $R_{p2,s1}=0.45 R_{\oplus}$. If the transits are around the fainter star (s2) instead, assuming the stellar properties are identical, both planets would be larger by a factor of 2, or $R_{p1,s2}=0.64 R_{\oplus}$ and $R_{p2,s2}=0.85 R_{\oplus}$. The radius values scale directly with the stellar radius value, so adjusting from the nominal radius of EPIC 201650711 \citep[$R_\star=0.34\pm0.05 R_{\rm Sun}$,][]{2019AJ....158...87D} for any reasonable main-sequence stellar radius would still yield planetary-sized values. 


More work needs to be done to determine which star(s) host the planets and what the stellar properties are. The relative orbital periods of the two planets, at $P=0.26$ d and $P=5.54$ d, are qualitatively consistent with most other observed USPs in multiplanet systems, as is the USP being the smaller planet (Section \ref{sec:size}). Assuming that the transit is around the central star, we re-ran the validation analysis including the SOAR contrast curve which was not available when we first attempted to validate. The contrast curve halved the FPP for the USP (though it remains unvalidated), though the companion remains unvalidated due to its marginal SNR. We note that the planetary radius values in \autoref{table:planets} have not been adjusted for dilution.

\textbf{EPIC 212303338}, from Campaign 6, has a USP candidate with $P=0.595$ d \citep{2018AJ....155..136M} at SNR=9. It also has a very close stellar companion (0.1\arcsec) in high-resolution images taken with WIYN in 2016, with $\Delta {\rm mag}$ = 2.45 at 692 nm (g') and $\Delta {\rm mag}$ = 1.99 at 880 nm (r') \citep{2018AJ....156...31M}. We first consider how much dilution the companion star introduces to the light curve. The candidate planet is small enough ($R_p = 0.46~R_{\oplus}$) and the stellar magnitudes similar enough that most scenarios are plausibly planetary. We consider two scenarios: (1) The candidate orbits the brighter star. In this case, the true radius should be increased from the value reported in \autoref{table:planets} by 5-7\%, to 0.5 $R_{\oplus}$. (2) The candidate orbits the fainter star. In this case, the radius should increase by a factor of about 3, to 1.5 $R_{\oplus}$, if we were to assume that the fainter star had the same radius as the bright star. The radius of the fainter star is also unknown; but if it is on the main sequence ($R_\star \approx 0.2-2~R_{\odot}$) the estimated radius will change by an additional factor of 0.25 to 2.5, using the brighter star radius of $0.8 R_{\odot}$; even in the maximal case, that would be $R_p \leq 4 R_{\oplus}$. We note that EPIC 212303338 p1 has a relatively high validation FPP (0.67), dominated by background eclipsing binary scenarios. We leave the final disposition of this candidate as undetermined. We also do not attempt to correct for dilution in the radius value reported in \autoref{table:planets} due to the uncertainty of which star the signal is around.


\subsubsection{EVEREST Photometry Issue for K2-131}
\label{everest_issue}

\textbf{EPIC 228732031, aka K2-131}, from Campaign 10, has a confirmed USP at $P=0.37$ d \citep{2017AJ....154..226D, 2018AJ....156...78L, 2018AJ....155..136M}. We were unable to validate this planet using the EVEREST light curve, and closer inspection revealed high-frequency noise that appeared as quasi-sinusoidal oscillations, likely negatively affecting \emph{vespa}'s score. Noting that our EVEREST best-fit period was 7.8 seconds longer than the published period (though only by 1.6$\sigma$), we re-did our analysis using the exact orbital period of \citet{2018AJ....155..136M}, which reduced, although did not entirely eliminate, the high-frequency noise, and also improved, but did not validate, the validation score (EVEREST FPP=0.45). When we instead used the k2sff light curve, in which no oscillations are visible, we easily re-validated the system with FPP=$2.6*10^{-4}$, demonstrating that the issue is with the photometry and not our validation framework. For consistency, the parameters for fits to the light curves derived from EVEREST photometry are the ones reported in the table \autoref{table:planets}. 

This was the only instance we noted in EVEREST photometry of such high frequency noise, though it is possible that other candidates with similar noise were rejected. Understanding the source of the noise would help understand whether this is a significant issue for completeness, but we leave that for future work.

\subsubsection{High false-positive probabilities}
\label{sec:highfpp}

Validation with \emph{vespa} produced \numhighfpp\ USPs that are more likely than not to be false positives, with false positive probabilities between $50-100\%$. Listed in decreasing order of FPP, the suspect USP candidates (above 90\%) are in the following systems: EPIC 228739208*, EPIC 205002291*, EPIC 214539781*, EPIC 219752140, EPIC 214189767; and from 50-90\%: EPIC 201461352, EPIC 212303338, EPIC 211336288*, EPIC 201231940*, EPIC 201609593, EPIC 210605073, EPIC 210931967*, EPIC 201469738, EPIC 211305568, EPIC 212443973, EPIC 206151047, and EPIC 201239401*. Three of these candidates (EPIC 201239401, 201609593, 211305568) have additional candidates in their system but do not receive a multi-planet boost due to low SNR of either the USP or the candidate. An additional three candidates in multi-candidate systems had individual FPPs $>$ 50\%, but which fell below 5\% after applying a multiplicity boost: EPIC 211305568 p2, EPIC 201239401 p2*, and EPIC 220440058 p2*.

Objects labeled with an asterisk (*) are candidates whose signals were only identified in our search of the EVEREST pipeline and not the k2sff pipeline, which may indicate that the signals are more likely to be spurious. This applies to three of the five candidates with FPP above 90\% and four of twelve single candidates with FPP from 50-90\%, as well as two potential companions in multi-planet systems.

One additional candidate we failed to validate was EPIC 201637175 aka K2-22 b, a disintegrating planet confirmed by radial velocity mass measurements elsewhere \citep{2015ApJ...812..112S}. We do not include this in our statistics as a true false positive, as we did not attempt to correct for the well-known light curve variability when fitting.


High-resolution imaging data is available for just one system with an elevated FPP (EPIC 212303338, FPP=0.74, which has a known stellar companion and is discussed in Section \ref{sec:uncertainstars}). For those targets on the elevated FPP list for which follow up observations are possible, they are recommended, since imaging contrast curves can make a huge difference in validation results, as noted previously for EPIC 201650711 where the FPP dropped in half once imaging was available. A more dramatic case was EPIC 213715787 aka K2-147 b, where we estimated the FPP both with and without the Suburu AO imaging constraints, and found that including them dropped the FPP from 0.6 to $3\times10^{-5}$.

\subsection{Comparisons to other surveys}
\label{sec:othersurveys}

Several recent surveys have discovered ultra-short-period planets in \ktwo\ Campaigns 0-10, usually in the context of searching for companions at all periods. Since these works used different photometry pipelines and/or search algorithms, it is worthwhile to compare our results to examine what lessons we can learn.

\subsubsection{Comparison to Mayo et al.~2018}

\citet{2018AJ....155..136M} completed a similar survey of \ktwo\ Campaigns 0-10 using their k2sff pipeline \citep{2014PASP..126..948V}. That work searched for candidates at all orbital periods, and found 14 of our \numusp\ USP candidates. A combination of the improved photometric sensitivity of the EVEREST pipeline over k2sff (as discussed in Section \ref{sec:search} and Subsection \ref{sec:comp_k2sff}) and our smoothing and detection algorithms, which have been tailored to detect short periodic signals, probably accounts for why they found only 1/5 of the candidates that we did. Our candidates also skew smaller: 43\% of our USP sample has $R_p<1 R_{\oplus}$ but only 17\% of theirs did (3/17).

They also reported 3 candidates that we do not have listed as planets, all of which we have identified as false positives in \autoref{table:fp}: EPIC 206169375, which has a close stellar companion and large odd/even differences visible in the light curve; and EPIC 212645891 and EPIC 229039390, two eclipsing binary stars with true periods at twice that identified by \citet{2018AJ....155..136M}.

\subsubsection{Comparison to Livingston et al.~2018}

\citet{2018AJ....156...78L} analyzed only Campaign 10 of \ktwo\ using their own photometric pipeline, and found 6 USP candidates: EPIC 201110617/K2-156, EPIC 201595106 (which we validate), EPIC 228721452 (two planets: P=0.5 and P=4.5 d), EPIC 228732031/K2-131 b, EPIC 228801451/K2-229 (two planets: P=0.58 and P=8.3 d), and EPIC 228836835 (which we validate). We identified 5 of the USP candidates, and discussed the one we missed, EPIC 228721452, in Section \ref{sec:228721452}. We found an additional 9 candidates that were not present in \citet{2018AJ....156...78L}, including EPIC 201438507 (two candidates), EPIC 228739208, EPIC 228814754, EPIC 228877886, EPIC 201427007 (which we validate), EPIC 201130233/K2-157, EPIC 228813918/K2-137, EPIC 201085153 (two candidates), and EPIC 201089381. 

\subsubsection{Comparison to Kruse et al.~2019}

\citet{2019ApJS..244...11K} used the earlier EVEREST 1.0 photometry to announce 818 planet candidates in C0-8 at all orbital periods (their Table 7). By design, their survey made conservative cuts ($10\sigma$) on odd-even depth differences, allowing likely eclipsing binaries to pass through. By comparing our set of candidates to theirs, we can offer a real-world assessment of the reliability of results from the EVEREST 1.0 catalog, particularly at the shortest periods where EVEREST 1.0 performed worst \citep{2019ApJS..244...11K}.

Removing our candidates from Campaign 10, we found 58 USP candidates in C0-8 (excluding EPIC 220554210, which we originally missed), of which \citet{2019ApJS..244...11K} detected 24 (33\%). Again, our sample skews smaller: \smallerthanearth\% of our sample had $R_p<1 R_{\oplus}$ but only 4\% of theirs did. They also reported 31 USP candidates that do not appear in our list (\autoref{table:planets}). We examined the folded light curves for all of these candidates, and found that almost all of the signals are either due to eclipsing binary blends (28) or to stellar variability (2), often at twice the period listed in \citet{2019ApJS..244...11K}. These objects are included in our table of rejected candidates in \autoref{table:fp}. Meanwhile, none of our candidates appear in their list of false positives or eclipsing binaries (their Tables 6 and 9). Just one of their candidates that we originally missed is genuine (EPIC 220554210) as discussed in Section \ref{sec:220554210}.

\begin{figure}
\includegraphics[width=0.5\textwidth]{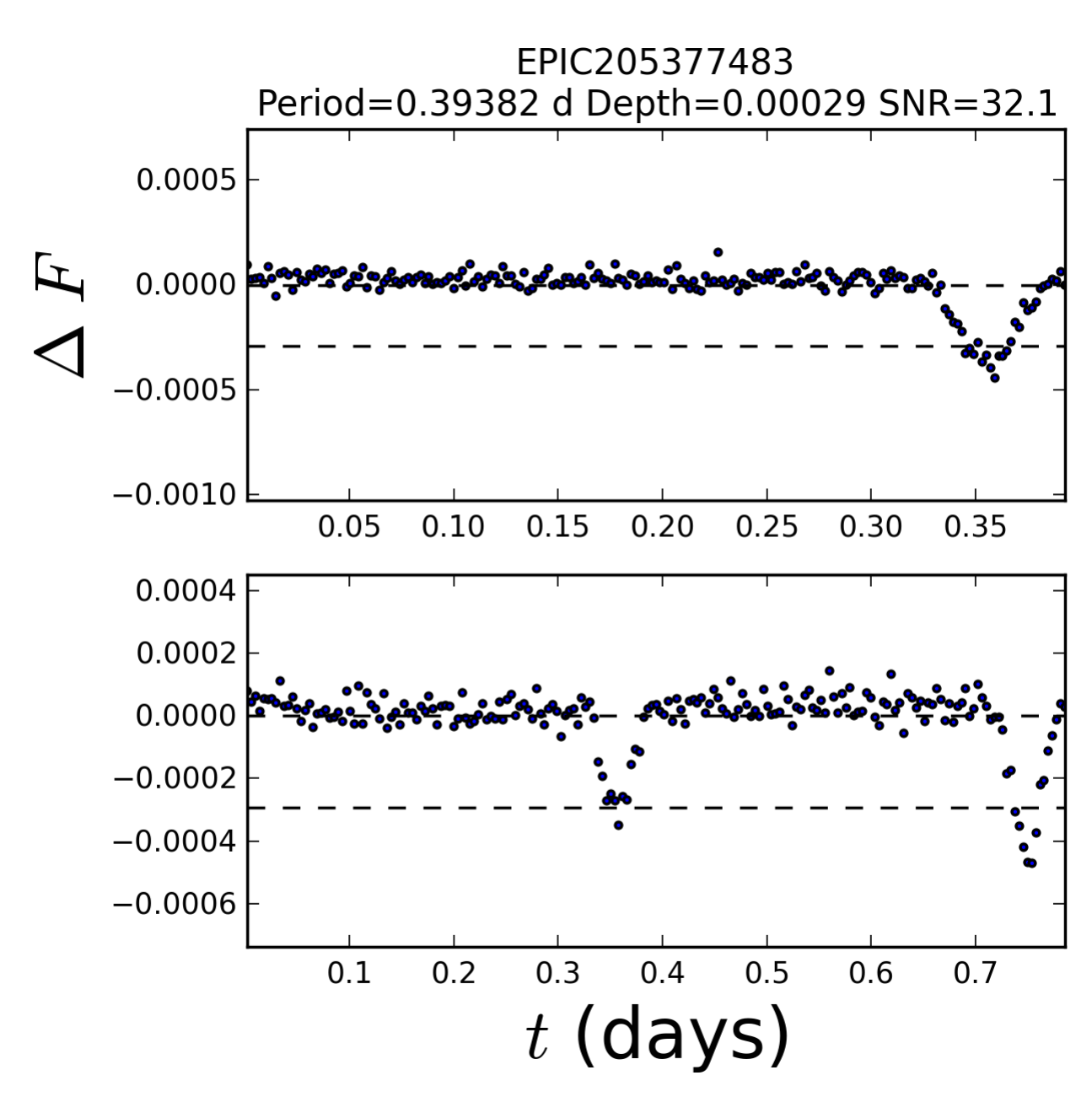}
\caption{Diagnostic plot showing the photometry for EPIC 205377483 folded at its nominal period (top) and twice that period (bottom). A potential transiting planet (at $P=0.39$ d) is revealed to be an eclipsing binary (at $P=0.79$ d) with telltale differences in transit depths between alternating transits.}
\label{fig:fold}
\end{figure}

Since the $P<1$ d sample of \citet{2019ApJS..244...11K} had so many false positives, we briefly examined some of their longer-period candidates ($1\le P \le 3$ d) to estimate what fraction might not be planets. For this paper, we did not vet any candidates with $P>1$ d (aside from companions to USPs), but our initial search went out to $P<3$ d. This was to guard against missing objects where the period had been aliased out of our search region, as discussed in Section \ref{sec:search}. Our initial diagnostic plots include the light curve folded to the nominal period, P, and $2\times P$. These plots are useful for quickly identifying eclipsing binary stars, which may exhibit different transit depths between odd and even transits, patterns of out-of-transit variability (often due to ellipsoidal variations in binary star systems), and other features that can be obscured when the data are folded to shorter periods than the true orbital period for the system. An example is shown in \autoref{fig:fold}, where a plausible transit of EPIC 205377483 at the originally identified period ($P=0.39$ d) is shown to be have unequal transit depths, an EB signature, when folded at twice the nominal period. By examining plots like this for the \citet{2019ApJS..244...11K} sample of 69 candidates with $1\le P<2$ d, we found 9 clear eclipsing binaries and 20 with some other kind of periodic signal, likely stellar variability; two more candidates we identified at somewhat different periods than they did. Repeating this exercise for candidates with $2\le P\le3$ d, we found that 15 of their 68 candidates in this range were likely non-planets (11 stellar noise/variability, 4 EBs) and 3 were candidates at different periods (including EPIC 206215704, discussed in Section \ref{sec:206215704}). We thus find that the short-period planet candidates reported in \citet{2019ApJS..244...11K} ($P<3$ d) include a significant fraction of variable and eclipsing binary stars mixed among the viable planetary candidates, with the percentage of false positives decreasing with longer orbital periods (47\% with $P\le1$ d, 42\% with $1\le P < 2$ d, and 22\% with $2\le P\le3$ d). Additional false positives are also expected after further vetting and follow-up observations, as in this work.

\subsection{Rejected USP candidates}

We investigated every reported USP candidate  in \ktwo\ Campaign 0-8 and 10 that we could identify in the literature. Some of these candidates \citep[including a few that we reported in prior work,][]{2016AJ....152...47A}, have turned out to be false positives, often after the advent of new spectroscopic or other observational constraints. We list the rejected candidates in \autoref{table:fp}. In addition to the candidates from \citet{2019ApJS..244...11K} and \citet{2018AJ....155..136M} we discussed above, we note one more rejected USP candidate:

\textbf{EPIC 203518244}, from Campaign 2, was reported as a USP at $P=0.84$ d using k2sff photometry by \citet{2016ApJS..222...14V}. At SNR=5.8 it was below the 10-$\sigma$ cutoff in  \citet{2016AJ....152...47A}. In this work, the EVEREST photometry clearly shows the transit to be a grazing eclipsing binary at $P=51.589$ hr or $P=2.15$ d, an alias (5/2) of the previously reported period.  

\begin{deluxetable*}{llll}
\tablewidth{0pt}
\tablecaption{Rejected Planet Candidates}
\tablehead{EPIC	&Reported P (d)	&Real P (d) 	&Notes 
}
\startdata
201577017   & 0.4775252 & 0.955	 & EB  \\
201606542   & 0.44437   & 0.8887 &   Spectroscopic binary (SB2, double peak) \\
202092559   & 0.6687    & 1.3374	&	 EB\\
202094740   & 0.6897    & 1.3794          &   Spectroscopic binary (SB2, single peak) \\
202139294   & 0.5217136 & 1.0434	&	 EB\\
202971774   & 0.2358684 & 0.47171 &	 EB\\
203518244   & 0.84      & 2.1495   &   EB \\
203633064   & 0.7099504 & 1.4199	&	 EB\\
204880394   & 0.2338108 & 0.71007	&	 EB\\
205377483   & 0.3938505 & 0.78763 &	 EB\\
206169375   & 0.36745   & --          &   Close AO companion + high odd/even difference \\
206315178   & 0.3176828 &	0.63532 &	 EB\\
210961508   & 0.349935  & 0.769     &   EB  \\
211158357   & 0.2990737 & 0.5981	&	 EB\\
211613886   & 0.9587591 & 1.9175	&	 EB\\
211685045   & 0.769057  & --          &   Variable star  \\
211843564   & 0.4520313 & 0.90399	&	 EB\\
211995325   & 0.279     & --          &  Pipeline depth mismatch \\
212066407   & 0.8218241 & 1.6436	&	 EB\\
212150006   & 0.89838    & --         &  Pipeline depth mismatch, close AO companion (0.5\arcsec)\\
212270970 &	0.7165461   & 1.4331	&	 EB\\
212362217 &	0.6962935   & 1.3928 &	 EB\\
212432685 &	0.531684    & 1.0634	&	 EB\\
212645891 &	0.32815384 & 0.65626 & EB\\
213123443 &	0.546127    & 1.0923	&	 EB\\
214984368 &	0.2633809   & 0.5268	&	 EB\\
215125108 &	0.738067    & 1.4761	&	 EB\\
215353525 &	0.9085915   & 1.8172	&	 EB\\
215381481 &	0.533393    & 1.0668	&	 EB\\
218195416 &	0.4951253   & 0.9903	&	 EB\\
218952024 &	0.56007     & 1.1201	&	 Spectroscopic binary (SB2, double peak)\\
219266595 &	0.944284    & 1.8886	&	 EB\\
219420915 &	0.5150196   & 1.02988	&	 EB\\
220198551 &	0.7988453   &	1.59758 &	 EB\\
220201207 &	0.4378443   & --	&	 Variable star\\
220282718 &	0.5551606   & 1.1103	&	 EB\\
220263105 & 1.44473     & --   & Variable star \\
220316844 &	0.6719977   &	1.34404 &	 EB \\
220569187   & 0.6045    & --          &  Pipeline depth mismatch \\
229039390 & 0.25605     & 0.5033 & EB, close AO comp. (0.6\arcsec) \\
\enddata
\tablecomments{EB = eclipsing binary from visual inspection of EVEREST photometry. Spectroscopic binary = binary found through follow-up spectroscopy, Section \ref{sec:spectra}. Close AO companion = nearby star likely contaminating the system, from \url{https://exofop.ipac.caltech.edu/}. Pipeline depth mismatch = EVEREST and k2sff depths differ by $>50\%$, indicating different blending fractions.}
\label{table:fp}
\end{deluxetable*}

\section{Survey Completeness}
\label{sec:survey_completeness}

\subsection{Detectability calculations}

\begin{figure}
\includegraphics[width=0.7\textwidth]{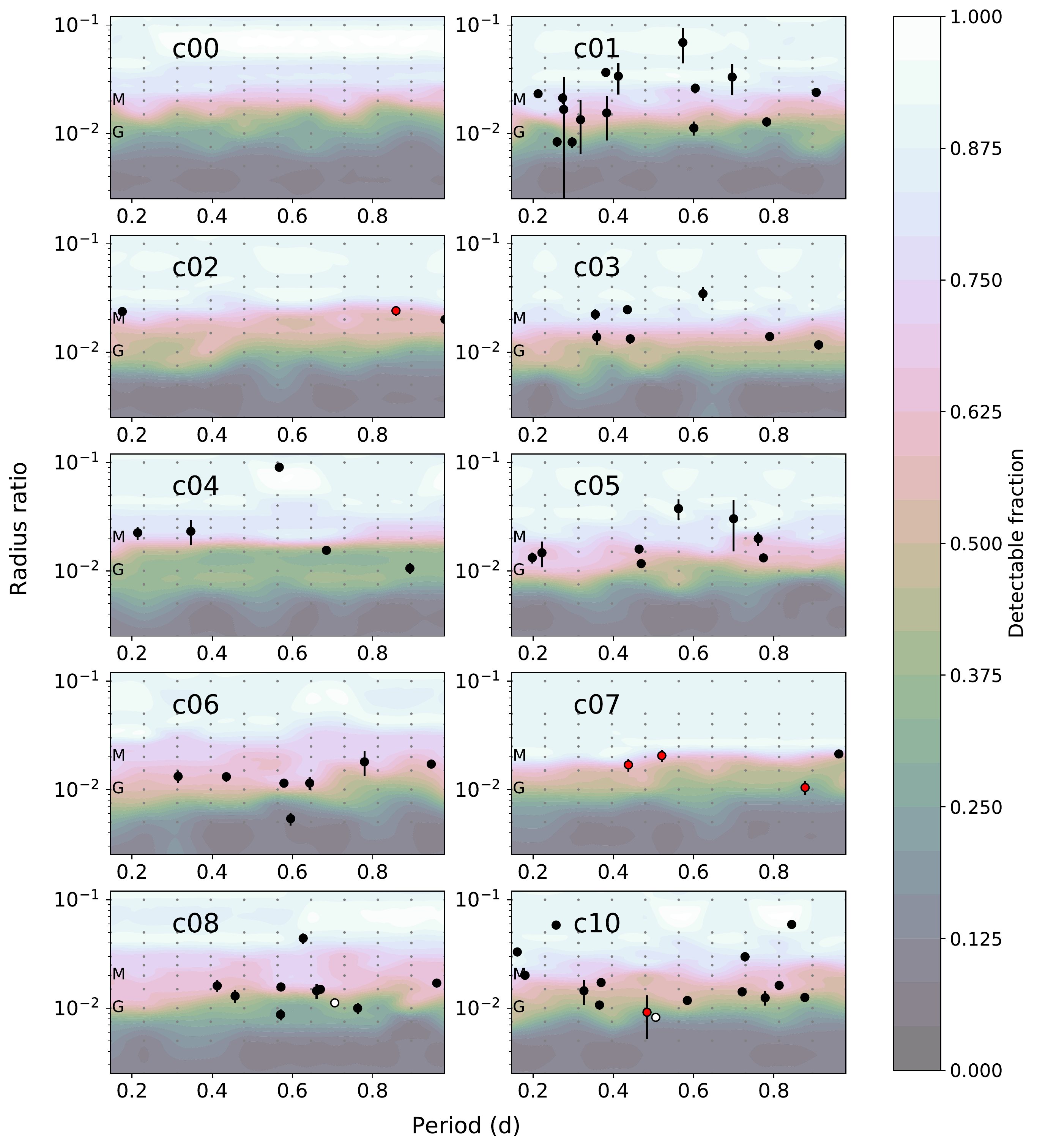}
\caption{Detectability of injected transits by radius ratio and orbital period. Ten light curves at each noise level (the highest, lowest, and each intervening decile) were tested for a suite of injected planets with periods from 3.5-23.5 hours and radius ratios of 0.0025 to 0.25. The grid sampled is shown as small black dots, with the contours showing the detectability interpolated from that grid.} The colorbar is the same for all panels and shades from black (0\% detection) to white (100\% detection), depending on the fraction of samples detected. For size reference, the letters M and G on the left of each plot denote the radius ratio of a 1-$R_{\oplus}$ planet around a typical M and G dwarf. We plot our candidates as large black circles with error bars; candidates with high false positives (FPP$\ge90\%$) are shown in red, and white circles mark the two candidates we initially missed in our survey (EPIC 220554210 in C8 and EPIC 228721452 in C10, see discussion in Section \ref{sec:missed}). 
\label{fig:completeness}
\end{figure}

To test the completeness of our survey, we injected a grid of synthetic transits into a representative sample of light curves, as in \citet{2016AJ....152...47A}. We sorted all light curves from a given campaign by the median noise of the detrended light curve and took ten light curves at each decile (median noise ranging from 238-477 ppm),
as well as the least noisy ($\approx 1$ ppm) and most noisy ($\approx 10^6$ ppm)  curves for each campaign. Into each of these representative time-series we inserted transits with radius ratios from 0.0025 to 0.25 and with periods between 3.5 and 23.5 hours. For computational simplicity, we fixed $i=90$ deg and the limb darkening parameters to those of the Sun. With typically shallow depths and only 1-3 points in any given transit, the average USP has poorly constrained inclinations and limb darkening parameters even after stacking tens of transits together. The significant uncertainties on light curve shape means estimating limb darkening parameters would have been difficult, and so we did not. Each synthetic light curve was run through our EEBLS detection pipeline, and a transit was declared detected if it was recovered within 0.1\% of the injected period and within a factor of 2 of the assumed transit depth. The completeness as a function of the period and radius is shown in \autoref{fig:completeness}. Note that two of the three campaigns closest to the Galactic Plane, and thus with the most potential for false positives due to crowding, also have the most candidates with high false positive probabilities; by distance from the Galactic Plane, they are: Campaign 0 (no candidates), Campaign 7 (3 likely FPs), and Campaign 2 (1 likely FP).

\subsection{Comparing EVEREST 2.0 to k2sff}
\label{sec:comp_k2sff}

To investigate the importance of the improved signal to noise of the EVEREST 2.0 photometry pipeline, we compared the detectability for this survey to previous work by the SuPerPiG collaboration which used a similar detection and vetting program but with k2sff photometry. \citet{2016AJ....152...47A} identified 19 candidates in C0-5; here, using EVEREST, we have identified twice as many candidates (38) in the same six campaigns (and rejected some of the prior discoveries as false positives). Many of the newly identified planet candidates in C0-5 are small: in \citet{2016AJ....152...47A} there were 3 sub-Earth-sized candidates in k2sff data (EPIC 201264302, 201650711, and 211357309), compared to 19 sub-Earth-sized USPs here; there were seven k2sff candidates between 1-2$~R_{\oplus}$, compared to 15 here.

Detectability is a strong function of the radius ratio. In \citet{2016AJ....152...47A} using k2sff data, only 10\% of light curves in C0-2 would have been sensitive to a 1-${\rm R_{\oplus}}$ planet around a G-star (radius ratio of $\sim0.01$), and in C3-C5 it was 30-65\%. In this work using EVEREST data, those numbers are universally higher and much less variable by campaign, ranging from 56\% (C0) to 71\% (C3, C5). The improved detectability fractions agree with our discovery of many additional Earths and sub-Earths, particularly in earlier campaigns (especially C1) that had low sensitivity using k2sff photometry. The fraction of light curves unsuitable for discovering even the largest USPs is also lower: we estimated that 15-20\% of k2sff light curves were unsuitable to detect transits of any size, whereas with EVEREST that value is about 10\% across all campaigns.

\section{Putting the new systems in context}
\label{sec:context}

\subsection{Partial reporting of multi-planet systems is common}
\label{sec:missed}

Examining results across multiple papers has turned up a common theme: additional candidates in systems hosting ultra-short-period planets are often missed if one or more of the planets are close to the detectability thresholds. Due to limitations on labor, surveys for new planets often don't detect additional planets in a system unless they are well above the detection limits, to the detriment of smaller companion planets. Since false positive rates are high for ultra-short periods, we specifically designed our survey's filters and tests to find and vet planets with periods under a day, but a consequence of running a more labor-intensive vetting program is that our targeted survey perforce under-detects longer period planets -- and this is a real problem if the dominant planet signal in a system is at a period greater than one day. At the same time, general searches for planets at all periods tend to perform worse at detecting the shortest-period planets than our dedicated search, as discussed in detail in Section \ref{sec:othersurveys}. We identified several cases where multiple works have reported planetary candidates for the same system at different periods. Sometimes one or more of these reported candidates is in error (e.g., an aliased period), and sometimes they represent a multi-planet systems that was found in pieces.

\subsection{Population parameters for USPs with companions}

Systems of tightly-packed inner planets have been commonly observed in \kepler\ and \ktwo\ data, and it has been estimated that 30\% of all stars have $\sim3$ planets with $P<400$ d, with most more closely packed than in our solar system \citep{Zhu_2018}. 
Such systems have been referred to as ``peas in a pod", since in addition to being closely spaced (typical separation between adjacent worlds is $<20$ Hill radii), the planets in a single system also tend to have similar sizes, though outer planets appear to be a bit larger than inner planets \citep[][though see Section \ref{sec:size}]{Weiss_2018}. Ultra-short-period planet systems are of course rarer \citep[0.5\% of G stars,][]{SanchisOjeda2014}, and for the few with multiple companions that have been studied, it has been noted that while the outer planets are tightly packed and have low mutual inclination, the USP subverts the formula by being further separated and having a higher mutual inclination compared to the rest of the system \citep[e.g., K2-266,][]{2020arXiv200910745B}. 

We will now discuss the observed parameters of the \nummultisystems\ USP-hosting systems with 2+ known planets, among which are six systems with three known planets and two systems with four known planets. In Section \ref{sec:theory}, we will compare how these properties agree with the predictions of different theoretical works.

\begin{figure}\includegraphics[width=0.5\textwidth]{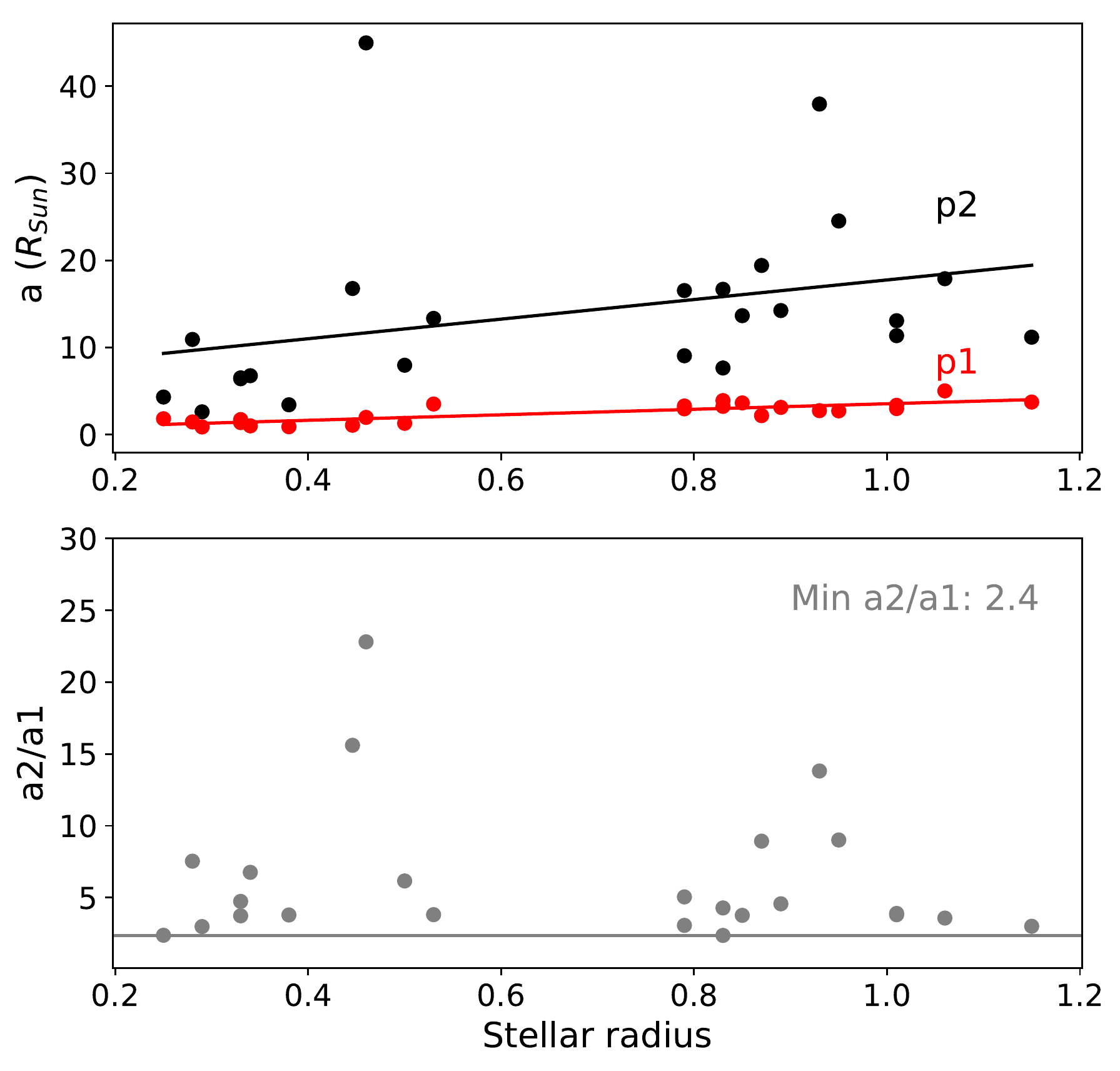}\caption{Comparison of stellar separations of USP and nearest companion in multi-planet systems. Top: a trend toward larger semimajor axes around larger stars is seen for both the USP (p1) and its nearest companion (p2). Bottom: the ratio between semimajor axes has a minimum value (2.4)} that is constant across all stellar types.
\label{fig:separations}
\end{figure}

\begin{figure}\includegraphics[width=0.5\textwidth]{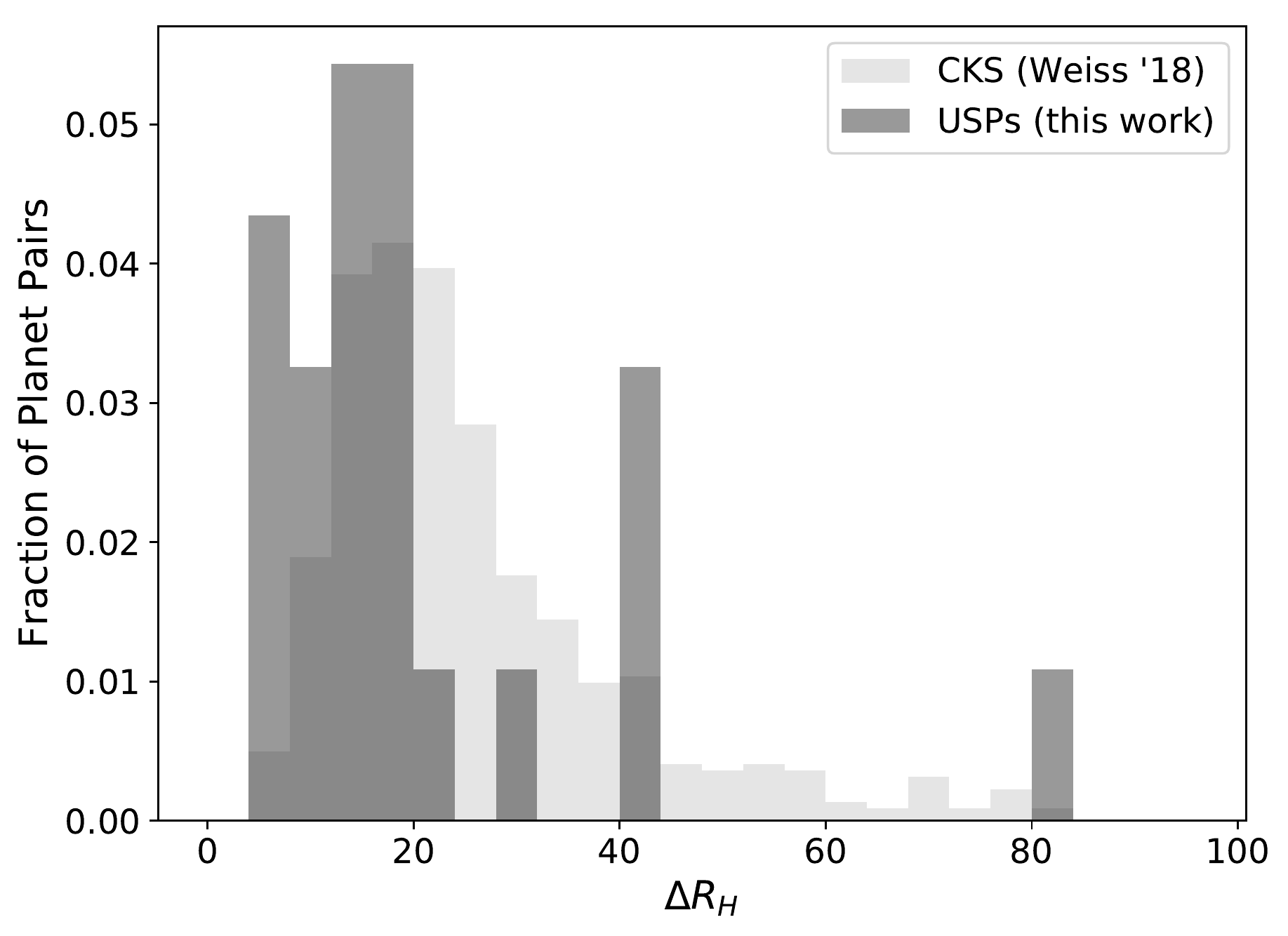}\caption{Separations between planets in terms of the mutual Hill radius ($\Delta R_H$) for adjacent planets in multi-planet systems. Dark grey: 24 pairs of planets in 21 systems containing an ultra-short-period planet (USP) from this work (three pairs with $\Delta R_H=254-2371$ are not shown). Light grey: 554 pairs of planets from the California \kepler\ Survey (CKS), \citet{Weiss_2018}, none of which had $\Delta R_H>100$. }
\label{fig:hill}
\end{figure}

\subsubsection{Semimajor axes}

We first examined the distance of USPs in multi-systems from their stars. We found that the planetary semi-major axis follows a trend of increasing distance from the star with stellar radius, so that USPs around smaller stars are closer to their stars. This trend holds for both the USPs and for their nearest companions, although the USPs follow a shallower trend. The ratio of semimajor axes, however, $a_2 / a_1$, has a minimum value of 2.4 that holds across all stellar radii, as shown in \autoref{fig:separations}.

We next examined how closely USPs in multi-systems are spaced to each other. We compared our sample to the work of \citet{Weiss_2018}, which examined 909 planets in 355 multi-planet systems from the California \kepler\ Survey (CKS). We followed their formulations to estimate the planetary masses from the radius values, and used that to calculate the Hill radius, which is defined following \citet{1993Icar..106..247G} as
\begin{equation}
R_H = \left( \frac{ m_j+m_{j+1}}{3M_\star} \right)^{1/3} \frac{(a_j+a_{j+1})}{2},
\end{equation}

where planets of masses $m_j$ and $m_{j+1}$ orbit a star of mass $M_\star$ at semi-major axes $a_j$ and $a_{j+1}$. The mutual separation between two planets, in units of mutual Hill radii, is

\begin{equation}
\Delta R_H = (a_{j+1}-a_j) R_H.
\end{equation}

We found a similar average separation to \citet{Weiss_2018}, with a median $\Delta R_H = 18$ for USP systems vs. 22 for CKS sample, but a higher fraction of USP systems are more closely spaced, with 22\% of the USP sample less than 10 Hill radii apart, compared to 7\% in the CKS sample (\autoref{fig:hill}). The minimum separation found is almost identical ($\Delta R_H = 4.5$ for USPs and $\Delta R_H = 4.9$ for the CKS sample). We note that \citet{1996Icar..119..261C} found separations of less than l0 Hill radii to be always unstable for systems with three or more planets. We also found that the USP sample had a larger fraction of planet pairs with very large separations ($>50$ Hill radii): 19\% of USPs vs. 8\% in the CKS sample; in our case these are all small, Mercury-sized planets with very low estimated masses and hence very small Hill radii. What our sample is missing, compared to the CKS sample, is planet pairs at intermediate separations. Whether this is a selection bias effect (\kepler\ targets were observed for much longer than \ktwo\ targets, making longer period detections possible) or an artifact of different formation histories is left for future research.

\subsubsection{Mutual inclinations}
\label{sec:inclinations}

\begin{figure}
\includegraphics[width=0.8\textwidth]{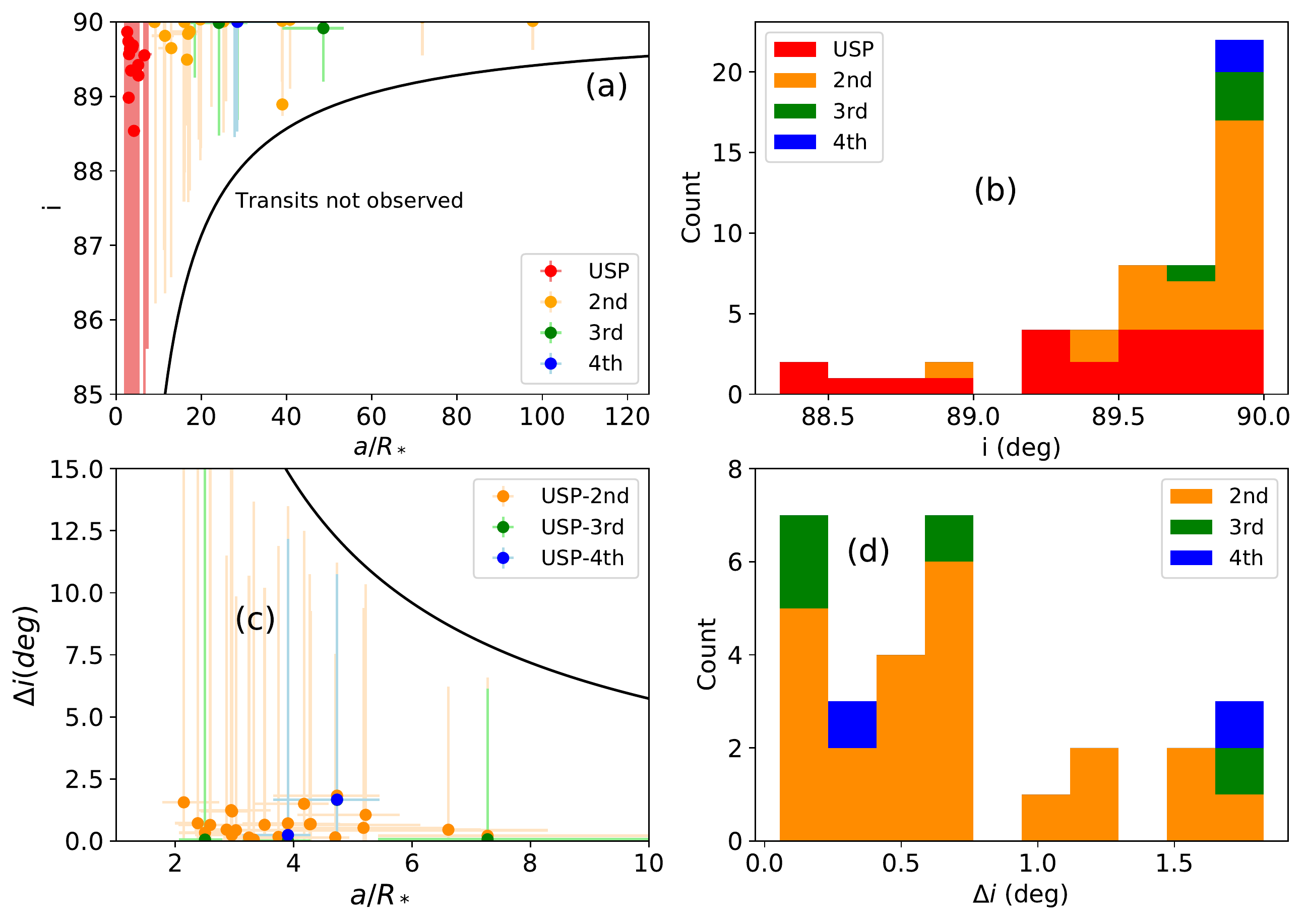}
\caption{Inclination and semimajor axes for multi-planet systems hosting an ultra-short-period planet. (a) $i$ vs.~$a/R_*$ for the USP (red) and the second (orange), third (green), and fourth (blue) companions, by distance from the star. The black line shows where transits geometrically not observable. (b) Histogram of individual inclinations. (c) Mutual inclinations $\Delta i=\lvert i_x -i_1 \rvert$} for the USP minus the second (orange), third (green), and fourth (blue) companions. All mutual inclinations are well below the black line, where the USP would not transit even if the companion had $i=90^\circ$. (d) Histogram of mutual inclinations.
\label{fig:inc}
\end{figure}

The inclination distribution of \nummultisystems\ systems with one USP and one or more transiting companions are shown in \autoref{fig:inc}. We find that USPs have smaller inclinations than their companions, which is consistent with their geometrical advantage, since more distant companions need to be more tightly aligned to be observed to transit (see Section \ref{sec:transit_prob}). We calculated the mutual inclination as the difference between the two fitted inclinations, and calculated the joint error using the maximal values of the asymmetrical errors (corresponding to the lower error bar, after requiring that $i\le90^\circ$ from the fitted values). We also find that the mutual inclinations of systems with USPs are quite low (mean $\Delta i=1.2\pm1.2^\circ$), with few cases significantly different from zero, and none with $\Delta i>8$ degrees. The mutual inclinations presented here are much lower than those in the sample reported by \citet{2018ApJ...864L..38D}, which comprises a different set of objects with little overlap, as discussed in Section \ref{sec:theory}. It is possible that our sample is more strongly biased against objects with higher inclination dispersion, since we include many small candidates at moderate to low SNR, and would not easily detect grazes, which can result in even shallower transit depths. We also note that many of our objects have higher errors than the difference in inclinations.

\subsubsection{Orbital periods and lack of apparent mean-motion resonances}
\label{sec:period}

The distribution of periods and radii for all \numusp\ USPs is shown in \autoref{fig:period_radius_hist}. There is no difference in the observed period distribution of ultra-short-period planets with and without companions, which is essentially flat between 0.2-1 days, with a median of 0.58 d. For companion planets, the median period is 4.5 d for the second planet in the system and 18.6 d for the third or fourth planet, with a maximum of 31 d for transiting companions (WASP-47 c, at 588 d, is one of only two companions known that does not transit). 

Little evidence is seen for current mean motion resonances involving USPs. We examined period ratios between candidate planets in multiplanet systems. If the ratio between the periods is within $\pm0.1$ of an integer or half integer ratio (i.e., $0.0\pm0.1$ or $0.5\pm0.1$), it was considered near resonance. This was intended as a first look, since actual resonance or past resonance membership would need to be determined with dynamical modeling.  For period ratios up to a factor of 10, just four USPs meet that criteria, which is fewer than if the period ratios were randomly distributed (40\% of numbers end in those digits, which would be  $0.4 \times 30$ combinations = 12 planets). None of the USPs are very near low-order resonances, with the closest being: (1) EPIC 206024342/K2-299, where the period ratio of the USP to the second closest planet is within 5\% of a 5:1 ratio; (2) EPIC 220554210/K2-282, where the period ratio of the USP to the second closest planet is within 9\% of a 6:1 ratio; (3) EPIC 201239401, where the period ratio of the USP to the second closest planet is within 1\% of a 15:2 ratio; and (4) EPIC 228721452/K2-223, where the period ratio of the USP to the second closest planet is within 2\% of a 9:1 ratio.

Some additional near-resonances can be found among the outer companions, though with 4 out of 9 combinations flagged it is consistent with random chance. (4) In EPIC 212157262/K2-187, the second and third planets are within 1\% of a 5:2 ratio. Two pairs are within 10\% of a 2:1 ratio: (5) EPIC 211562654/K2-183, second and third planets, and (6) EPIC 212157262/K2-187, third and fourth planets. (7) EPIC 220554210/K2-282, has the second and third planets are within 3\% of a 13:2 ratio.

Only six out of \nummultisystems\ systems have any members near a resonance, as broadly construed here. Clearly, if dynamical processes required mean-motion resonances to emplace the USPs, they have long since decoupled from the resonances in all of the systems we examined. A more detailed dynamical study would be required to determine if any of the near-resonant period ratios actually correspond to potential current or past resonances.

\subsubsection{Planetary radius}
\label{sec:size}

We did not confirm any ultra-hot Jupiters (UHJ, $R_p \gtrapprox 10 R_{\oplus}$), and found just one UHJ candidate with moderately high false-positive probability; however, by optimizing our search to discover small planets it is possible that some of the deepest transits might have been sigma-clipped, so we do not consider our UHJ count to be complete. Medium to large planets ($R > 3  R_{\oplus}$) also appear to be rare among planetary companions to USPs, with just one known among \nummultisystems\ systems (WASP-47 b). We also did not identify any compelling candidates in the short period desert ($\sim3-10~ R_{\oplus}$), as discussed in Section \ref{sec:desert}.

Excluding the lone UHJ candidate, EPIC 210605073, and focusing on the smaller population, we find a few correlations of note. First, USPs with companions are marginally more likely to be larger (median: $1.36~ R_{\oplus}$, sample $N=\nummultisystems$) than those without (median: $1.07~ R_{\oplus}$, sample $N=50$)), with KS test p-value$=0.02$. This is unlikely to be a detection effect since there may, if anything, be a slight bias toward finding a smaller USP in a system with a larger known companion than to find one on its own. 

Second, we do not find evidence that the intrinsic population of USPs must have smaller radius values than their companions. Although it is true that the observed median radius of all USPs ($1.14~R_{\oplus}$, $N=\numusp$) is different than that of their known companions (second planets: median = $1.5~R_{\oplus}$, $N=24$, p-value=$0.04$; third/fourth planets: median = $2.6~ R_{\oplus}$, N=11, p-value=$3\times10^{-4}$), this difference is entirely consistent with different detection efficiencies based on stacking shallow transits. At the median orbital period for a USP, a second planet, and a third planet (at 0.58 d, 4.5 d, and 22.6 d respectively), a \ktwo\ campaign of 80 days would yield a maximum of 138, 18, and 4 transits (ignoring gaps). Assuming that stacking yielded improvements proportional to the square root of the number of transits, the USP would be detectable at 2.8 and 5.9 times shallower depths than the second and third planets; since the radius is proportional to the square root of the depth, the detected companions would have larger radius values by a factor of 1.7 for second planets and 2.4 for third/fourth, very similar to the observed differences in median values. We do not account here for any of the other detection biases that are also at play, including inclination effects (making more distant and smaller planets less likely to transit) as well as sampling effects (making USPs with the shortest periods harder to detect, since each transit has only 1-3 points per transit with \ktwo's 30 minute observing cadence). 

\subsection{Transit detection probabilities and a rough estimate of sample completeness}
\label{sec:transit_prob}

To calculate the transit probabilities, we ignore grazing transits (which would push most of these small transits into being too shallow to detect) and eccentricity (negligible for tidally locked USPs but might be important for more distant companions) and use $P_{tr}=R_*/a$ \citep{1984Icar...58..121B}. For USPs, the probability of transit is quite high: EPIC 203533312, with $a/R_*=1.6$, has a 38\% chance of transit, while even the lowest-probability USP (228814754) has a 5\% transit probability. The \textbf{longer period} companions in our sample have transit probabilities ranging from 0.5-6\% (some with large uncertainties due to $a/R_*$ being poorly constrained).

One takeaway is that the sample of USPs that we detected is much more complete than that of companions. As a first approximation (ignoring all other detection biases), we add the inverse of the transit probabilities to estimate the total number of planets and find an estimated total 535 USPs, of which our sample of \numusp\ would represent 14\% of the total population. For the companion planets, we find that the estimated total from inverse transit probabilities is 1396, of which we only have \numcompanions\ in our sample, or 2\%. Considering only the second planet in each system, the detected fraction is similar (1015 estimated total second planets, of which we have found 4\%). One possible conclusion is that, if our sample of USPs is seven times as complete as our sample of companions (14\% / 2\% $\sim2$), and companions have been found around one third of all USPs, then every USP could be in a multi-planet system.

\subsection{Comparison to theoretical predictions}
\label{sec:theory}

\subsubsection{Petrovich, Deibert \& Wu 2019}
The first paper to provide a mechanism for the origins of USPs involving high-eccentricity migration was \citet{2019AJ....157..180P}. In particular, they found that USP planets should have spin-orbit angles and inclinations relative to outer planets in the range of 10-50 degrees. They also predicted that most USPs should have companions that are either more distant than $P > 20$ d or with high mutual inclination (i $> 20$ deg). As seen in \autoref{fig:inc}, we see very low mutual inclinations in multi-candidate systems (albeit with large error bars). We also see many companions at shorter orbital periods, although our sample is far from complete and transit detectability decreases rapidly for more distant companions.

\subsubsection{Pu and Lai (2019)}

\citet{2019MNRAS.488.3568P} also presented a model which requires the presence of additional companions to emplace planets on short-period orbits. In this scenario, the future USP starts out as the innermost in a chain of planets, with a modest orbital eccentricity ($\sim0.1$), and encounters resonances that increase the mutual inclination with respect to the exterior planets. This theory predicts that: (1) planets exterior to the USP must orbit close enough that secular interactions can maintain a non-zero eccentricity for the USP for a long enough time that tidal interactions can emplace the USP; (2) the USPs should be less massive than the exterior planets; and (3) the USP may have an orbit that is significantly misaligned with the rest of the planetary system.

Our results have broad agreement with \citet{2019MNRAS.488.3568P} on point 1, qualified agreement on 2, and qualified disagreement on point 3: 

\begin{enumerate}

\item Planets must orbit close enough: we have found that a third of our sample of \ktwo\ USPs has a transiting companion within $P<31$ d, and our results are consistent with most of the apparently singleton systems hosting nearby planets that do not transit (Section \ref{sec:transit_prob}). The separations between planets are small both physically (\autoref{fig:separations}) and dynamically (\autoref{fig:hill}), with an average mutual Hill radius of $\Delta R_H = 18$; nearly a quarter have $\Delta R_H < 10$.

\item The observed sample of USPs have smaller radii than their companions, but this is consistent with detection biases toward finding small planets with more frequent transits (Section \ref{sec:size}). A further caveat is that smaller radii may not correspond to smaller masses. Measured masses are unavailable for more than a handful of planets.

\item While we observe that USP inclinations span a larger range than those of the companions (\autoref{fig:inc}), it's not clear how significant this finding is, due to high errors on most of our inclinations. All of our systems are statistically consistent with zero mutual inclination, in contrast with \citet{2018ApJ...864L..38D}, one of the works that \citet{2019MNRAS.488.3568P} drew on, who found an inclination dispersion of 6.7$^\circ$ (using a sample that extends to longer periods than our own, although their USPs also have a higher range of mutual inclinations). Panel (c) of \autoref{fig:inc} reproduces their Figure 3 for our sample, which has a mean $\Delta i=1.2\pm1.2^\circ$. One possible cause for the discrepancy could be sample size and/or quality; their sample comprises few \ktwo\ candidates and has little overlap with ours. Many of our objects are shallow and/or have low to moderate SNR, meaning that a planet in a grazing orbit (i.e., significantly different from $i=90$) might be more likely to be missed. We also don't account for the possibility of oppositely-aligned orbits; a pair of planets at 88 and 89 degrees will have the same $\Delta i$ in \autoref{fig:inc} as a pair at 92 and 89 degrees. Since however most of our nominal values hew closely to 90 degrees, this is unlikely to account for the full difference. A more in-depth analysis of the inclinations of USPs from \ktwo\ is recommended before drawing conclusions on this point.

\end{enumerate}

\subsubsection{Millholland and Spalding (2020)}

The model of \citet{2020ApJ...905...71M} is obliquity-driven tidal migration, where it is the planetary obliquity (the angle between the planet's orbit and its rotational spin) that drives migration of the innermost planet with at least one close companion, using the quick passage through Cassini states, configurations in which precession of the planet's spin axis matches precession of its orbital plane, which can result in substantial tidal dissipation. Five testable predictions are made by that work; we broadly agree with points (1) and (5), disagree with (3) and (4b), and leave the corroboration of points (2) and (4a) for future work.

\begin{enumerate}

\item USPs should have close companions ($P\lesssim10$ d), which may or may not transit. This agrees well with \autoref{fig:period_radius_hist}, which finds the closest companion has a median $P=4.5$ d. About a quarter (6 of 24) of the closest companions have $10<P_2<15$ d, still broadly compatible with this theory.

\item No trends are expected for eccentricity, whereas the \citet{2019MNRAS.488.3568P} eccentricity-driven tidal migration model would expect some remnant eccentricity in the companion planets. Unfortunately, we cannot test this with our current fits because we assumed zero eccentricity for all planets to avoid fitting extra parameters to light curves that are often near the signal to noise limit. Some of the companion light curves are of sufficient quality that they could be examined for non-zero eccentricity.

\item USPs should have non-zero mutual inclinations. In particular: (a) they should have modest misalignments with their planetary companions, in agreement with \citet{2018ApJ...864L..38D} but not our results in \autoref{fig:inc}, and (b) they should be aligned with the stellar spin (which is mostly unknown). Measuring the stellar obliquity of the USPs in multi-planet systems is an excellent goal to constrain this theory.

\item In this model, the USP rapidly breaks out of the Cassini states. This implies that (a) few USPs should be currently undergoing rapid tidal decay; long-term monitoring of the transit timing of these systems would be needed to address this question. Another prediction is that (b) planets will pile up near 1 day, with occurrence rates decreasing at shorter periods, as found in the earlier work of \citet{SanchisOjeda2014}. As noted in Section \ref{sec:period} we instead observed an even spread in orbital period for both USPs in known multi-planet systems and USPs that are apparently single. 

\item Finally, USPs are predicted to be more common around smaller host stars as in \citet{SanchisOjeda2014}. Although we do not have debiased our discoveries to produce a detection rate by stellar type, we found that only 4 of \nummultisystems\ multi-planet host stars have $R_* > 1 R_{\odot}$.

\end{enumerate}

\section{Summary and Conclusions}
\label{sec:summary}

The importance of ultra-short period planets has long been recognized as potential laboratories for exploring the effects of extreme proximity to a star on planet formation and stability. However, there have to date been few large-scale population studies to determine the properties and occurrence rates of such systems. This work provides a new search of 10 campaigns of \ktwo\ data to systematically discover ultra-short period planets and any companions in the same system. We present \numusp\ candidates, \numvalidusp\ of which have been validated and/or confirmed as planets. Almost half (43\%) of our USPs have radius values smaller than Earth, with the smallest similar to Mercury in radius. Twenty-four are in systems with at least one other transiting planet candidate, with \numcompanions\ total companions detected. We note that \numhighfpp\ USPs have FPP$>50\%$, including \numfp\ that are nominally false positives with FPP$>90\%$, and should be treated with caution. It is important to note that \marginalsnrusp\ of our USPs and \marginalsnrcomp\ of the companions are presented with SNR $<10$, a deliberate decision to increase the yield of these rare-but-important objects, which will require future observations to fully clarify their status.

At this point, it should be clear that (a) USPs are commonly found in multi-planet systems (\multirate \% of the USPs in this paper have at least one known companion); (b) such systems often have 3+ planets (33\% of USPs in multiplanet systems have 3 or more transiting candidates), meaning that finding one companion does not mean that you have found them all; and (c) the sample of USPs is much more complete (based solely on transit probabilities) than the sample of companions to USPs. Our results are consistent with every USP system also having at least one nearby non-transiting companion (within $P\sim30$ d). 

We recommend that all systems with known planets or candidates should be searched specifically for USP companions, which tend to be smaller and easy to miss in non-targeted searches. Furthermore, USP systems should be monitored for longer-period transits and for the radial velocity signatures of non-transiting companions, which are expected to be common. Understanding the population of ultra-short period planets is key to understanding the forces that shape planetary evolution right at the edge of stability. 

\section{acknowledgments}

Data products produced as part of this work, including the transit light curves and fitting images, are available at our repository at \url{https://github.com/elisabethadams/superpig-public}.

Thanks to Rodrigo Luger, Tim Morton, Katja Poppenhaeger, Joe Renaud, Ethan Kruse, and Dan Foreman-Mackey for helpful discussions during the preparation of this manuscript. The two anonymous reviewers also greatly improved the quality of this paper and their thorough reviews are much appreciated.

This work has made use of data from the European Space Agency (ESA) mission {\it Gaia} (\url{http://www.cosmos.esa.int/gaia}), processed by the {\it Gaia} Data Processing and Analysis Consortium (DPAC, \url{http://www.cosmos.esa.int/web/gaia/dpac/consortium}). Funding for the DPAC has been provided by national institutions, in particular the institutions participating in the {\it Gaia} Multilateral Agreement.

This research made use of Lightkurve, a Python package for Kepler and TESS data analysis (Lightkurve Collaboration, 2018);  Astropy, a community-developed core Python package for Astronomy \citep{astropy:2013, astropy:2018}; Scipy \citep{2020SciPy-NMeth}; PyMC \citep{Salvatier2016}; Batman \citep{2015PASP..127.1161K}; and Matplotlib \citep{Hunter2007}.

This study is based upon work supported by NASA under Grant No.~NNX15AB78G issued through the Astrophysical Data Analysis Program by Science Mission Directorate and under Grant No.~NNX17AB94G through the Exoplanets Research Program. M.E. and W.D.C. were supported by NASA K2 Guest Observer grants NNX15AV58G, NNX16AE70G and NNX16AE58G to The University of Texas at Austin.  

This research was supported in part by the National Science Foundation under Grant No. NSF PHY-1748958 under the auspices of UC Santa Barbara's Kavli Institute for Theoretical Physics.

Some of the data analyzed in this paper were collected by the \ktwo\ mission, funding for which is provided by the NASA Science Mission Directorate. The data were obtained from the Mikulski Archive for Space Telescopes (MAST). STScI is operated by the Association of Universities for Research in Astronomy, Inc., under NASA contract NAS5-26555. Support for MAST for non-HST data is provided by the NASA Office of Space Science via grant NNX09AF08G and by other grants and contracts. 
Some of the data presented herein were obtained at the W.M. Keck Observatory, which is operated as a scientific partnership among the California Institute of Technology, the University of California and the National Aeronautics and Space Administration. The Observatory was made possible by the generous financial support of the W.M. Keck Foundation. The authors wish to recognize and acknowledge the very significant cultural role and reverence that the summit of Maunakea has always had within the indigenous Hawaiian community.  We are most fortunate to have the opportunity to conduct observations from this mountain. 


{\it Facilities:} \facility{\ktwo} \facility{Keck:II(NIRC2)}  
\facility{Gemini:South}
\facility{Gemini:Gillett}
\facility{SOAR}
\facility{Hale}
\facility{WIYN}

\software{Batman 2.1.0 \citep{2015PASP..127.1161K} \url{http://astro.uchicago.edu/\~kreidberg/batman/};
pymodelfit 0.1.2 \citep{2011ascl.soft09010T} \url{https://pythonhosted.org/PyModelFit/core/pymodelfit.core.FunctionModel1D.html};
PyMC 2.3.4 \citep{2015ascl.soft06005F} \url{https://pymc-devs.github.io/pymc/};
PyLightCurve 2.3.2 \citep{2016ascl.soft12018T} \url{https://github.com/ucl-exoplanets/pylightcurve}; 
Emcee 2.2.1 \citep{2013PASP..125..306F} \url{https://emcee.readthedocs.io/en/stable/};
Kea \citep{2016PASP..128i4502E},
Uncertainties 2.4.6.1 \url{http://pythonhosted.org/uncertainties/)},
vespa 0.4.9 \citep{2012ApJ...761....6M} \url{https://github.com/timothydmorton/VESPA};
astropy 3.1.1 \citep{astropy:2013, astropy:2018} \url{http://www.astropy.org};
scipy 1.2.1 \citep{2020SciPy-NMeth} \url{https://scipy.org/};
matplotlib 3.0.2 \citep{Hunter2007} \url{https://matplotlib.org/}
}

\bibliography{k2_paperIII}

\clearpage



\end{document}